**Title:** A computational framework for evaluating an edge-integrated, multi-ramp construction model of the Great Pyramid of Giza

**Author:** Vicente Luis Rosell Roig

**Correspondence to:** vicente.rosell@gmail.com

**Address:** L'Alcúdia. Valencia. Spain.

**Education:** PhD in Pattern Recognition, Artificial Intelligence and Computer Graphics. Universitat Politècnica de València (UPV).

**Affiliation:** Independent Researcher.

**ORCID ID**: 0009-0003-8857-9706

**Abstract:** *Despite decades of study, a quantitative, integrated framework to evaluate minute-scale throughput, geometric control, and a zero external footprint for Khufu's pyramid has been lacking. We test the Integrated Edge-Ramp (IER) model—a helical path formed by omitting and backfilling perimeter courses—using a unified, end-to-end pipeline coupling parametric geometry, discrete-event logistics, and staged finite-element analysis (FEA). An adaptive multi-ramp strategy can sustain 4–6-minute dispatches and yields a median on-site duration of 13.8–20.6 years (95% CI); including quarrying, river transport, and seasonal pauses gives 20–27 years. FEA indicates that stresses and settlements remain within plausible limits for Old Kingdom limestone under self-weight. The model's geometry is also consistent with internal voids identified by muon imaging (a hypothesis-generating result). The IER helps reconcile throughput, survey access, and zero-footprint closure, and produces falsifiable predictions (edge-fill signatures, corner wear). Our study provides a transferable, open-data/code framework for testing construction hypotheses for ancient megastructures.*

## Introduction

The Great Pyramid of Giza, built c. 2560 BCE as Khufu's tomb, rises ~146.6 m over a ~230 m-per-side base and comprises ~2.3 million blocks, mainly limestone with granite elements. Finishing it within Khufu's ~27-year reign implies ≈1 block every ~3 min on average across the project—an extreme throughput whose logistical and structural feasibility remains unresolved and motivates this study.

Evidence from the Wadi al-Jarf Papyri, the Ahramat branch near Giza, and reconstructed Nile levels indicate an operable waterway that limits construction to a 20–27-year window [1–3], within the uncertainties of the historical record. Proposed ramp methods involve trade-offs in material volume, site footprint, and survey control observed in Fourth-Dynasty works at Dahshur and Giza [4–7]. Accordingly, this study tests the hypothesis that an adaptive, multi-ramp Integrated Edge-Ramp (IER) system could achieve the documented construction rate within the Old Kingdom's material and technological constraints, and derive testable field predictions.

Old Kingdom technology precluded iron tools, wheeled heavy transport, and compound pulleys, but allowed copper chisels, water-lubricated sledges, ropes, levers, earthen works, and Nile barges; accordingly we bound ramp slope, lane width/clearance, and friction (μ) and evaluate the dispatch headway (time between placing successive blocks) required to satisfy the 20–27-year window, encoding these constraints as model parameters.

We situate the IER against the main ramp families' well-documented constraints. Straight or frontal ramps require prohibitively large volumes of material and leave no attested footings at Giza. Zig-zag or side ramps obstruct survey control and block perimeter access. External spiral designs impede corner visibility and turning, while internal spirals assume undocumented internal corridors and struggle to match muography results. Finally, terrace or micro-ramp schemes add handling complexity and create crew interference [4–14]. Descriptive details are in Supplementary S1 and a qualitative comparison in Supplementary S13. These shortcomings expose a gap: no model achieves minute-scale throughput, edge-line geometric control, and a zero external footprint. We therefore evaluate whether an Integrated Edge Ramp (IER) can meet these requirements under Old Kingdom constraints using a unified computational pipeline. In contrast, the IER seeks to reconcile footprint, survey control, corner access, and throughput without requiring massive external works or continuous internal voids, and we test this claim quantitatively below. These limitations motivate the need for a new approach — one that can adapt to shifting logistical demands as the structure rises and can accommodate heavy elements. Computational archaeology enables parametric simulation and finite-element analysis to test these criteria under Old Kingdom constraints and yield falsifiable predictions.

Building on prior ramp theories and recent computational heritage studies [15–17], we integrate parametric geometry, discrete-event logistics, and FEA into a single testable framework to test construction hypotheses with transparency, repeatability, and systematic what-if analyses, while acknowledging dependence on assumed parameters. We adopt a recursive parametric algorithm that updates the build state stepwise (state $_t$ → state $_{t+1}$) and exposes key inputs for reproducibility. Informed by Landreau et al. (hydraulic lifting) [18], Brichieri-Colombi (spiral ramps) [19], Hirlimann (energy budgets) [20], Fajardo et al. (Petri-net workflows) [21], and Manière (least-cost paths) [22], we test an Integrated Edge-Ramp (IER)—a helical, zero-footprint path formed by omitting then backfilling perimeter courses—coupling the algorithm with physics-based logistics and finite-element analysis to evaluate its feasibility.

To assess feasibility, we simulated construction duration under varying ramp configurations, delays, and seasonal calendars. Our scheduling model includes on-site assembly, quarrying, river transport, and seasonal pauses, yielding total durations suitable for comparison with historical benchmarks. In particular, the overall 20–27 year envelope is consistent with the ~27-year reign window reported in the Wadi al-Jarf papyri; this reconciliation is examined quantitatively in the Results section.

Recent muography reports a Big Void, the North Face Corridor (NFC), cavities and notches [12], which argue against a continuous internal helical tunnel and require compatibility with localized cavities and step-core stratigraphy. We therefore evaluate the IER's geometric consistency with these findings, showing that its helical path plausibly coincides with several reported cavities and notches, main entrance and NFC, and thickness change series.

We evaluate whether the IER can sustain minute-scale headways while remaining materially efficient, structurally safe, and geometrically precise. The analysis reveals a throughput bottleneck that motivates an adaptive multi-ramp strategy; our reproducible framework generates falsifiable archaeological predictions and, unlike prior helical proposals, provides an integrated computational evaluation of a testable, archaeologically plausible construction method. In addition, we apply the same open framework to three reference geometries under common assumptions— a straight external ramp, a near-edge spiral ramp and the dual-phase Houdin model. This like-for-like comparison reports per-course and aggregate haul distances, mechanical work, single-lane cadence, and external earthworks, clarifying which geometries can plausibly meet the 20–27-year historical window while preserving survey control.

Accordingly, we test the IER model by coupling a parametric course-by-course generator with a discrete-event/queuing model for on-site duration and staged self-weight FEA, supporting feasibility under our modeling assumptions within the attested 20–27-year window and yielding falsifiable archaeological predictions under our modeling assumptions.

The main aim of this study is to move beyond predominantly qualitative assessments by developing and applying a unified computational framework to quantify IER feasibility. We test whether (i) an adaptively parallelized IER sustains minute-scale headways within a feasible portion of Khufu's ~27-year reign, (ii) temporary edge channels remain structurally plausible under self-weight, and (iii) the model yields falsifiable archaeological signatures distinct from other ramp theories. The novelty lies in the transparent, reproducible integration of geometry, logistics, and structural mechanics for heritage-scale hypotheses.

## Methods

### Fundamental principles of the integrated edge-ramp (IER) model

In the integrated edge-ramp (IER) model, a temporary open-air helical haul lane is created by omitting a 3×6 mask of perimeter blocks along the active edge, forming a ~3.8 m-wide, 4.26 m-high corridor (Fig. S2.1). Owing to the step-core profile, only 12 block positions per longitudinal slice are omitted—a conservative engineering trade-off that preserves structural integrity and provides safe supervisory passage while minimizing cost. The stepped floor is locally leveled with compacted adobe or short planking and sand to maintain a ~7° target grade (6–8° tested). As the path advances to the apex, the omission is backfilled top-down, restoring smooth faces and leaving, in principle, no persistent external footprint. For reproducibility we define $W_{mask} \approx 3.8$ m and $H_{mask} \approx 4.26$ m; the geometry is compatible with the ~51.8° face angle and step-core (Stufenbauweise), and the exported path polylines/turning radii feed the logistics simulator and staged FEA (see Supplementary S3). All symbols use SI units; parameter ranges are in Table S3. We present the IER as a case study within a general framework that integrates parametric geometry, logistics, and staged FEA for repeatable hypothesis testing.

### Construction and Decommissioning Sequence

The IER construction process follows two phases: ascent for core construction and descent for decommissioning and finishing.

During ascent, the pyramid rose course by course via a helical edge channel as the primary haul path: limestone and casing blocks ascended to the working elevation, then crossed the wide terrace. As work advanced, the channel climbed with the structure, preserving edge visibility for survey control and direct access to the highest level. The stepped floor was locally smoothed with compacted adobe or short planking and sand to maintain ~7° grade (~6–8° tested), requiring only minor maintenance.

During the descent phase, after setting the pyramidion, construction reversed. Ramp decommissioning followed the stepped closure shown in Fig. S2.2; rubble-and-mortar packing and final casing were applied per Old Kingdom practice (details in Supplementary S2). Filler and facing stones were pre-quarried and dressed to match masonry. Temporary materials used to smooth the ramp were removed as filling advanced.

Tura limestone casing was installed concurrently on faces free of active channels, maintaining survey baselines and geometric control. After decommissioning, the remaining faces were clad top-down, merging seamlessly into the final volume. Tooling and materials followed Old Kingdom practice (survey lines, plumb/square control, copper chisels, temporary bedding/planking).

## Methodology: Algorithmic and Physical Modeling

IER credibility rests on a parametric, recursive framework that models the build sequence precisely. This algorithm formalizes the construction sequence as a recursive process, allowing reproducible simulation of geometry, logistics, and stress evolution across all stages. The recursion reduces "build a pyramid of height H and base B" to "place one course, then update $H \rightarrow H - \Delta h$ and $B \rightarrow B - \Delta b$," formalizing the "each course supports the next" logic. Given inputs $\{H, B, \beta\}$ and parameters ($\theta_r$, $\mu$, $W_{mask}$, $H_{mask}$), the procedure is deterministic through the logistics layer; the emitted stagewise geometry supplies path lengths, turning states, and terrace distances for the haul/queuing modules and FEA. Full algorithmic details are in S4.

## Software & Reproducibility

Code and inputs (Unity C# generator; Python 3.11 queuing; Code_Aster v15.6.10/MUMPS v5.2.1 FEA) are archived with fixed seeds and mesh definitions; exact versions and parameter files are provided via the Zenodo DOI (see Data/Code Availability). Run-critical parameters (geometry, $\mu$ ranges, slope, headways, corner-delay distributions, speeds) are listed in Supplementary S3 (geometry) and S9 (logistics) with ranges and sensitivity notes. This workflow-level reproducibility is central to our contribution: any scenario can be re-run end-to-end—from geometry to logistics to FEA—under controlled assumptions. The next subsection applies these settings to compute per-stage geometry and the block-placement recursion used in Results.

## Single Integrated Edge Ramp Theory: Block Placement Logic and Termination

We simulate construction with a recursive algorithm that places blocks course by course and subtracts the ramp channel; the geometry is updated stagewise until the apex or until the base can no longer accommodate the channel. Full algorithmic details, equations, and termination criteria appear in S4.

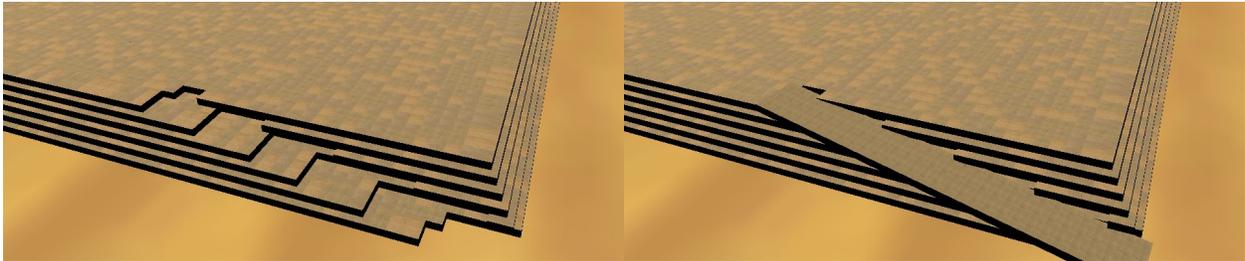

Fig. 1. Early steps in the edge-integrated ramp sequence. (a) Fifth course showing omitted perimeter blocks. (b) Fifth course with the edge-ramp corridor established: a 3×6 mask of temporarily omitted blocks defines an open-air lane; a thin leveling course maintains a ~7° target grade. Dimensions (SI): $W_{mask} \approx 3.8$ m; $H_{mask} \approx 4.26$ m; $\theta_r = 7°$; block module $h_b = 0.71$ m × 1.27 m. Omitted positions are backfilled during decommissioning.

Reserving a 3×6 window of temporarily omitted perimeter courses forms an open-air haul corridor on the active face; a thin leveling course maintains the target grade while edges and corners remain exposed (Fig. S5.1), preserving survey lines and geometric control. Omitted units are backfilled after decommissioning, leaving no external footprint (Fig. 2).

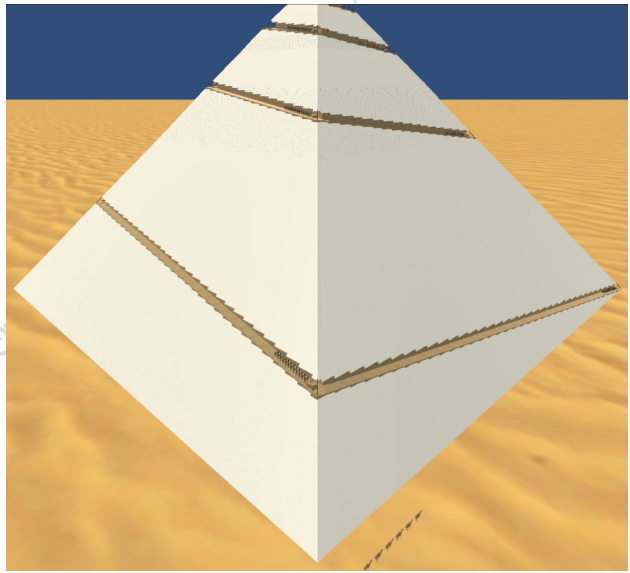

Fig. 2. Face change with an edge-integrated haul corridor (3D render). An open-air edge channel ($W_{mask} \approx 3.8$ m; $H_{mask} \approx 4.26$ m; $\theta_r = 7°$) rotates 90° at corners. Perspective view shows the lane ascending the near edge while adjacent faces remain stepped and fully visible for survey control. The corridor is formed by temporarily omitted blocks; as construction advances it shifts edge-to-edge at corners and, after decommissioning, is backfilled top-down, restoring smooth faces with no external footprint. Geometry exported directly from the parametric recursive model.

During single-ramp phases, the ~3.8 m lane supports counterflow—one loaded ascent and empty sled/pedestrian returns—regulated by headway, with passing limited to corner platforms. Published wetted-sand/trackway experiments support feasible traction ranges. A two-row fan-pull team (shoulder ~0.6–0.7 m/person) occupies ~2–3 m. Each edge-integrated lane is initialized at a corner so crews approach in a straight line, deferring the first turning maneuver to higher courses and reducing early headway penalties. We computed per-course block counts by recursive tiling of each square course with the baseline block unit and face angle, then subtracting the active-edge channel mask (rotating 90° at corners); see S4 for the discrete scheme and parameters.

A detailed time-lapse of the single-ramp sequence is provided in Supplementary Video 1.

**The Physical Model: Quantifying Work and Effort**

To quantify hauling effort, we use a physical force model that computes required force and per-block work under two conditions: inclined transport on the ramp (overcoming gravity and friction) and horizontal transport on the terrace (friction only). Total work per block follows from these forces and the travel distances from the geometric model. Full formulations and equations are in Supplementary S4. These calculations size hauling equipment and feed logistics and schedule simulations. Model validity rests on transparent, justified parameters—especially ramp steepness ($\theta_r$) and kinetic friction ($\mu$).

Ramp inclination ($\theta_r$). A 7° slope was chosen as a compromise: 6° lowers force by 5.58% but lengthens the ramp by 14.3%, whereas 8° raises force by 5.26% while shortening the path. Thus 7° balances path length and force within the 6–8° (≈10–14%) sledging envelope; the implied 23–25 pullers at 300 N per worker for a 2.27-t block match Table S9.2 and experimental ergonomics by Smith [23] and Brier [24].

Friction coefficient ($\mu$). We adopt $\mu$=0.20 as the baseline, matching the experimental optimum at ~5% water saturation (Harrell & Brown [25]; Fall et al. [26]) and implying feasible wetting logistics given nearby water sources on site such as the Ahramat Nile branch. Robustness was checked in our simulations via sensitivity analyses over $\mu$=0.15–0.30, bracketing plausible field conditions for the haul model assumptions.

The physical model couples directly to the geometric algorithm: per-block work (W) and force estimates feed the logistics module to size teams, estimate energy, and compute throughput, enabling a comprehensive feasibility assessment under Old Kingdom technological constraints.

This formulation provides, under the stated assumptions, a transparent, SI-consistent mapping from grade, friction ($\mu$), width, mass and path length to per-block work and crew effort, with exposed defaults/bounds for replication and sensitivity, and couples directly to the stagewise algorithm and logistics/FEA modules without ad hoc assumptions.

**Design refinements**

The baseline IER haul path is feasible, but for Old Kingdom constraints and replication, we add three refinements: (i) a low parapet for safety; (ii) enlarged corner platforms for turning/staging; and (iii) a folding-beam to maintain survey baselines. These adjustments preserve the ramp grade and zero-footprint premise and are encoded as explicit rules/parameters (see Supplementary S6).

To improve safety at height, the ramp channel is widened to a 4×8 aperture, reserving the outermost block as a continuous stone parapet (Fig. S6.1). This yields a corridor $W_{lane} \approx 3.8$ m and $H_{mask} \approx 5.68$ m with a 0.71 m height barrier, safeguarding against falls. The parapet marginally shortens the ramp (1,129→1,089 m) and reduces ascent work by ~3.54%. Additional impact attenuation (e.g., straw sacks) can be placed at turns. This four-block-wide configuration is used on upper courses to evaluate effects on work and safety.

The 90° turns are operational bottlenecks. To provide maneuvering, staging, and survey clearance, corner blocks are temporarily omitted to form 4×4×9 turning bays (Fig. S6.2). In plan, adding the inter-course setback on both inner edges yields a 5.64×5.64 m platform (≈31.8 m²). The inter-course setback is the per-course horizontal recession; for β=51.84° and $h_b$=0.71 m, $h_b/\tan\beta \approx 0.5579$ m (~0.56 m) (geometry from Lehner [7]). These platforms act as operational buffers where teams can pass and execute pivots safely, with minimal impact on the construction sequence; they are backfilled after decommissioning.

External spiral ramps can obscure edge sightlines for survey control; to preserve alignment, the IER uses a dual-purpose corner post (Fig. S6.3). A wooden post aligned to the edge serves dual roles: in surveying it accepts plumbs for backsight references; in hauling it acts as a fairlead, allowing a ¼–½ rope wrap to manage 90° turns. Although no Old Kingdom source attests such a device, the concept accords with documented survey practice and rope handling (Dunham [10]) and is presented as a period-plausible inference.

These refinements were encoded as explicit rules and parameters in the algorithmic model. The parapet adjustment modifies the channel geometry, the corner platforms are instantiated at turning nodes, and the survey system ensures continuous geometric control without affecting the core logistics calculations.

### The Four-Ramp Integrated Edge-Ramp Model: A Paradigm of Parallel Construction

While a single helical IER is feasible, meeting Khufu's timeframe requires parallelization. We therefore instantiate identical edge channels on multiple faces under the same rules, creating a concurrent process without altering the core algorithm or adding auxiliary works. In the 4-ramp system (one channel per face; Fig. 3), geometry dictates an adaptive sequence as height increases: courses 1–183 use 4 ramps in parallel; 184–198 use 2 ramps on opposite faces to preserve access; 199–203 reduce to a single ramp to ensure safe passage and geometric control.

An adaptive 4→2→1 sequence follows the diminishing base: mutual-exclusion rules prevent channel interference as courses narrow. Transition points at courses 183 and 199 arise directly from the parametric geometry when required clearances no longer permit multiple active channels. In the 4-ramp phase, dedicated lanes (3 ascent, 1 descent) minimize counterflow.

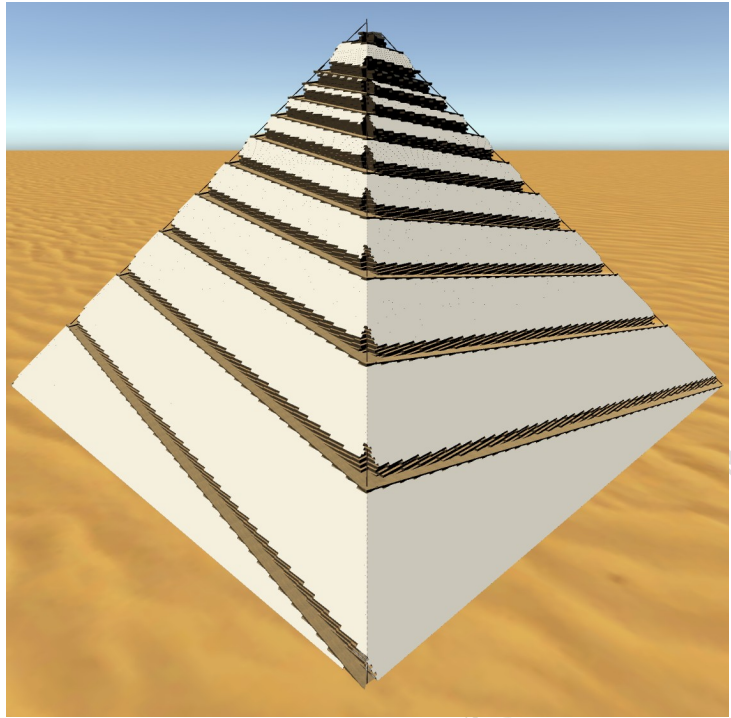

Fig. 3. Four-ramp parallel IER configuration (3D render). Parametric 3-D reconstruction with four edge-integrated helical channels (one per face) using the baseline aperture ($W_{mask} \approx 3.8$ m; $H_{mask} \approx 4.26$ m; $\theta_r = 7°$). Corner platforms enable turning/staging while edge arrises remain visible for survey control. Lanes: 3 for loaded ascent, 1 for empty descent. During decommissioning, all channels are backfilled (zero-footprint) following the same turning/fill rules.

Beyond the 4-ramp case, we consider two independent ramps on opposite faces operating counterflow (loaded up, empty down); being geometrically decoupled, their aggregate capacity is the sum of both channels. We also evaluate a macro-terrace IER variant; full definitions and algorithmic rules appear in Supplementary S8.

The parallel configuration directly feeds into the logistics model, where aggregate capacity scales with the number of active ramps. The adaptive sequence $4 \rightarrow 2 \rightarrow 1$ is triggered automatically by the recursive geometry algorithm when base dimensions fall below safety thresholds.

A detailed time-lapse of the 4-ramp sequence is provided in Supplementary Video 2.

**The Adaptive Sequential Optimization Model**

Building on the capacity and timing analyses, we test an adaptive sequential schedule that adjusts the number of active edge channels to sustain throughput as geometry tightens. It minimizes on-site duration and stabilizes headways by selecting phase boundaries under the baseline geometry (7°; μ≈0.20) and rules (corner mutual-exclusion, buffers). We retain the core IER algorithm and evaluate transitions with the logistics simulator, reporting medians with percentile bands.

Building on the 4-ramp efficiency, we adopt a phased strategy that adapts ramp use to operational demand: the principle is to maximize throughput on wide lower courses with multiple straight ramps along base edges, avoiding corner turns. Phase transitions at courses 9 and 20 follow a quantitative heuristic: choose the largest integers $k_1, k_2$ satisfying $B(k)$ $-2s(k)≥W_{lane}+2p$ and $R_{turn}(k)≥R_{min}$ (no overlap, admissible 90° turns). These thresholds maximize early throughput when block demand and terrace spans are greatest, with no-turn straight-runs ≥200 m (through course 9) and ≥135 m (through course 20) at 7°. Geometry and $R_{min}$ follow S5.3.

The adaptive phased construction hypothesis suggests that builders began with high-throughput edge ramps (16 and 8) to overcome the initial horizontal bottleneck and then shifted to the helical 4-ramp system (Fig. S5.3) for most of the vertical ascent. This constitutes our most complete construction timeline.

In phase 1 (courses 1–9). 16 ramps (4/face; Fig. S5.3a) maximize throughput on wide lower courses; no corner turns, edges clear for surveying. Allocation: 12 up / 4 down (3:1) to minimize interference. In phase 2 (10–20). 8 ramps (Fig. S5.3b) maintain high flow without crossings or turns; discontinued lanes are backfilled top-down to restore solid masonry. Allocation: 6 up / 2 down. In phase 3a (21–183). Shift to 4 helical ramps (Fig. S5.2): 3 up / 1 down. In phase 3b (184–198). 2 ramps on opposite faces (geometric limit). In phase 3c (199–203). Single ramp, alternating ascent/descent to ensure safe passage and geometric control. This phased 16→8→4→2→1 schedule reflects diminishing base width and preserves survey access throughout.

We built a ground-level traffic model to verify multi-ramp logistics: routing rules move blocks from supply points to concurrent ramp entrances along a perimeter path. For plausible quarry and harbor nodes, perimeter supply runs in parallel with hauling without bottlenecks; thus critical path is governed by vertical ascent and placement times from the core model.

For each policy, we run a discrete-event simulator with Monte Carlo explained below.

A detailed time-lapse of this method sequence is provided in Supplementary Video 3.

**Workflow Management: Headways, Buffers, and Supervision**

Operating a multi-ramp system required explicit workflow management. Headways—the interval between blocks on an active ramp—were set to a 4-minute baseline per ramp, derived from safe-separation constraints between hauling teams (see Results). Buffers (enlarged corner platforms) acted as waiting areas and turning points. Corner delays were modeled stochastically from Stocks' ~3–4 min 90° sled pivots [27]; using the folding post as a corner bollard allows controlled rope wraps that further shorten set-up/hold times, while supervisors enforce headways and resolve contingencies on site.

Schedule analysis uses conservative operating assumptions. A standard 24-man pull team, delivering ~7,100 N (≈300 N per person), is sufficient to move a block up a 7° wet-sand ramp. To regulate flow and maintain safe separation, we impose a reference headway of one block per ramp every 4 minutes. For on-site duration, the model assumes a working year is 10-hour day, six days per week. This cadence is maintained because the Giza Plateau is above the annual flood level; therefore seasonal pauses are excluded from the on-site calculation and treated separately when estimating the overall project schedule. This headway is applied per active ramp. These assumptions reflect ergonomic limits and hydrologic constraints.

To model realistic throughput, each ramp and corner platform was discretized into equal-length cells with one team per cell (no overtaking) to enforce separation. Speeds were set conservatively and kept constant by surface type—0.15 m/s on active ramps (grade+footing+lubrication) and 0.20 m/s on terraces [5]—and varied in sensitivity tests. Dispatch headway used a ~4-min baseline per active ramp, with 2–13-min alternatives. Corner maneuvers were treated as stochastic delays because the task comprises multiplicative sub-steps (approach, alignment, lever/pivot, short drags, reset), which we approximate with a lognormal law. In the absence of peer-reviewed Old Kingdom timing data, we adopted the lognormal form common in operations research. Parameters were anchored empirically to experimental archaeology—especially Stocks [27] ~90° sledge pivots on planked, wetted footing—yielding a median 2.8 min and shape $\sigma=0.35$.

This enables calculation of key queueing metrics $\lambda$, $\mu$, and $\rho=\lambda/\mu$; $\rho>0.95$ flags saturation (full utilization with queues/delays). Performance outputs include ramp capacity (blocks/h), corner queue length, and waiting/blocking probabilities. Each ramp and corner platform is discretized into equal-length cells, $L_{cell}=L_{team}+L_{sledge}+L_{buffer}$, with $L_{team}$ from double-file spacing (1.5 m/man), $L_{sledge}=3$ m, and baseline $L_{buffer}=15$ m; discipline: FCFS, no overtaking. In a cell-transmission scheme, one team occupies one cell at a time. Arrival rate per ramp is $\lambda=1/H$ (headway H, min); service rate $\mu=1/S$ with $S=T_{ramp}+T_{corner}$. Corner time is lognormal (median 2.8 min, $\sigma=0.35$), consistent with lever-and-drag sub-steps. Headway is mapped to ramp length within 2–13 min to enforce one-team-per-cell separation at 0.15 m/s (ramps) and 0.20 m/s (terraces). Seeds, parameter files, and code are archived (Zenodo); Monte Carlo details in Supplementary S9.

This module directly consumed geometric outputs (ramp lengths, platform locations) from the recursive algorithm and force requirements from the physical model, enabling end-to-end simulation of the construction logistics.

**Specialized Transport: The Case of Granite Megaliths**

Granite beams for the King's Chamber (5–9 m, up to ~80 t) are incompatible with external spiral ramps: their plan length exceeds the clear width and available turning radius, so 90° turns would require major corner widening; even then, drawbar forces and concentrated reactions at tight corners jeopardize platform stability.

Granite transport is a separate task in the IER. Megaliths are positioned early and lifted course-by-course (~0.71 m) on short 3–4° slips with staged levering (Fig. 4), isolating this low-frequency work from the high-frequency limestone stream. Terrace-mounted wooden bollards provide capstan wrap control ($T_{line}/T_{tail} = e^{\mu\theta}$, the exponential friction law that allows a rope around a post to amplify holding force), with $\mu_{capstan} \approx 0.25$–$0.35$ and $\theta \approx \pi$–$2\pi$ (½–1 turn), yielding ~2–9× reduction in tail holding force; two bollards in series (angles add) and 2–3 parallel lines further distribute load and maintain cadence. Required line tension remains ~90–200 kN, but holding/regulating crew drops by ~36–170 across 1–3 lines, making granite handling a decoupled, parallel operation. To securely anchor the terrace-mounted wooden bollards against the immense reaction forces (~180 kN) from megalith hauling, a plausible engineering solution is proposed. During terrace work, a core block was omitted to form a deep socket for a wooden post anchored with wedges and mortar, transferring haul loads into the surrounding masonry. This "omit–use–backfill" practice fits the zero-footprint logic and predicts localized vertical density anomalies detectable by GPR.

This strategy isolates the granite task and spreads effort over years. Granite monoliths of 50–80 t match estimates for the King's Chamber beams (5–9 m) [7]. To probe heavy-haul limits for long members, we evaluated a dedicated 3–4° ramp—a locally leveled, conservative path—while limestone throughput uses 7°. When needed, two bollards in series (angles add) and 2–3 parallel lines distribute load and maintain cadence. We thus model granite handling as a decoupled, low-frequency operation—short 3–4° slips with staged levering—in parallel with the high-frequency limestone stream, improving safety and feasibility within Old Kingdom constraints. Confining operations so that the limestone delivery stream faced only minimal and localized disruption, which could be managed by temporarily rerouting traffic on the wide terrace.

Another advantage of this method is that it can be worked in parallel with the granite blocks, so each team could prepare a ramp and raise a group of blocks without affecting the other group's position, as they are in parallel positions. All the materials used would be recycled, thus fulfilling the zero-footprint principle.

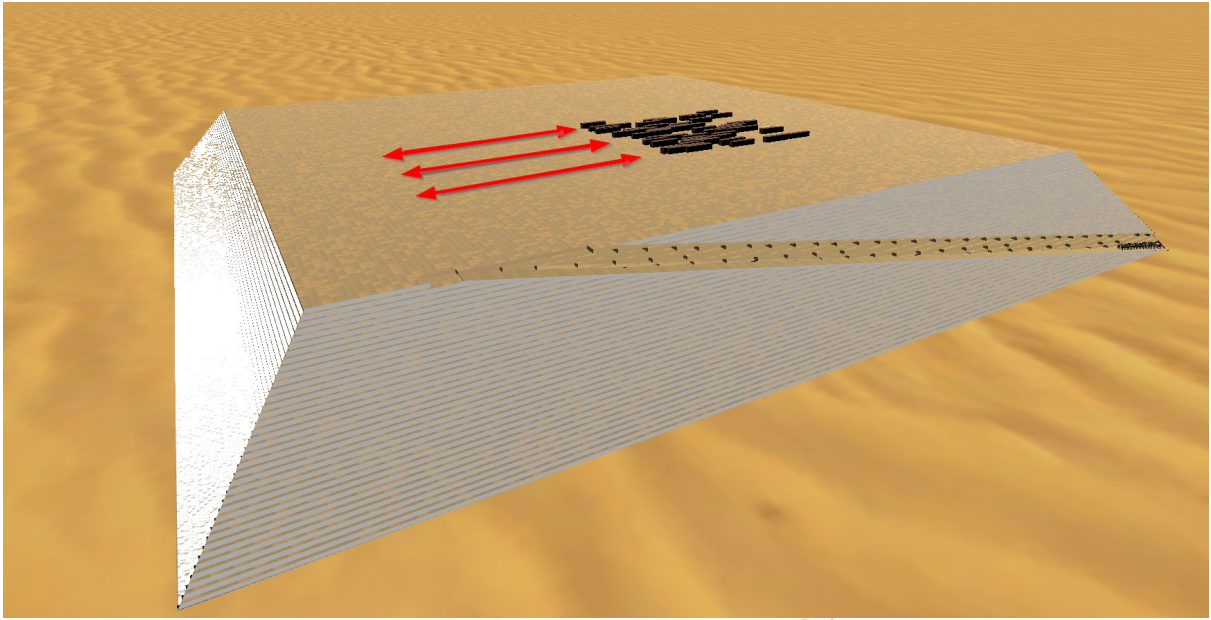

Fig. 4. Granite block transfer across terraces, 3D render. Side-to-side moves between adjacent benches using short slips; baseline $\mu_{track} \approx 0.20$ (lubricated timber/stoneway). Granite beams 50–80 t on short ramps $\theta \approx 3°$ (tested 3–4°). Wooden bollards provide capstan control ($\mu_{capstan}$=0.25–0.35, $\theta = \pi$–$2\pi$). Assumed footing $\mu_k$=0.15–0.30 (wetted clay/timber). Forces: drawbar 90–200 kN (200–300 N/puller); capstan wrap lowers holding force without changing $T_{line}$. Operates in parallel with limestone delivery; excluded from the main throughput timeline. Teams could prepare terrace ramps in parallel for granite blocks

Millimetric placement at the King's Chamber uses pre-pointing with water level and datum marks. The block arrives on a short 3–4° ramp to the working terrace and is wedged; cams and screens lift by millimetres, while opposing wedges allow fine vertical/lateral inching until bedding lines coincide. Final seating is by sequential wedge release, verified with feelers/ochre transfer. The Grand Gallery may have served as a staging or safety zone, but it is not essential to the project.

This specialized transport model is compatible with the safe handling of the largest elements within Old Kingdom constraints and offers a testable operational scenario.

**Structural Validation via Finite Element Analysis (FEA)**

We use finite-element analysis (FEA) because the staged, evolving geometry requires 3D stress and settlement resolution that simpler models cannot provide; The Discrete Element Method (DEM) is computationally impractical. We assume a staged linear-elastic limestone continuum with targeted edge/corner refinement, verified by convergence tests, balancing accuracy and runtime for reproducible parametric sweeps.

The model used Code_Aster 15.6.10 / MUMPS 5.2.1 (a parallel sparse matrix solver) via SimScale, meshed with TET10 (volume) and TRIA6 (surface). Limestone was linear-elastic [13]. The base was fully fixed ($U_x=U_y=U_z=0$), yielding upper-bound stresses and lower-bound settlements. Global equilibrium checked base reactions against $\rho g V$. Stress statistics (Point Data, MPa) reported nodal mean and 95th percentile from a 512-bin histogram. Displacements were apex cap-averaged $U_z$ ($r=0.5$–$1.0$ m) for mesh-objective comparison.

Cap-averaged apex displacement is our primary convergence metric because it reflects global stiffness under self-weight and is far less sensitive to local meshing artefacts. In heritage FEA, convergence is best tracked by global displacement rather than local stress maxima. Peak $\sigma$vM increases near corners or ramp gaps with refinement, misrepresenting bulk behavior; thus we adopt the displacement metric, suitable for large masonry where long-term stability depends on global deformation.

To complement the analytical estimate $\sigma \approx \rho g h/3$, we solved a 3D static FEA of the pyramid (base $\approx 230$ m, height $\approx 146$ m). Limestone was linear-elastic, isotropic ($E=35$ GPa, $\nu=0.25$, $\rho=2600$ kg·m$^{-3}$), with self-weight ($\rho g=-25{,}506$ N·m$^{-3}$) on a fully fixed base. Second-order TET10 meshes at two densities assessed convergence: fineness 3 (539,666 elements; 958,872 nodes) and fineness 5 (897,069 elements; 1,323,383 nodes). We ran coarse–medium convergence tests; once stress/settlement changes were within tolerance, the medium mesh was adopted for parametric sweeps. Fine-grained refinement was not feasible given computational limits. These verification steps—global equilibrium, cap-averaged displacement, and stress percentiles—establish mesh-objective reliability for the staged analyses reported in Results.

**Alignment of the IER geometry with Observed Structural Features**

We evaluate whether IER twist levels are followed by increases in course thickness at the studied ramp angles $\theta_r=6.0$–$8.0°$ (step $0.1°$). For each $\theta_r$, the 10 modeled turn elevations $T_\theta=\{t_i\}$ were compared with observed amplitude-change events $\{z_k\}$, defined as consecutive-course jumps > 0.25 m. A match occurs when an event follows a turn within $z_k-t_i < 1.5$ m ($\approx$ two courses). To assess chance coincidence, we ran 2,000 Monte Carlo simulations of 203 courses with thicknesses drawn from a truncated normal $N(\mu=0.722, \sigma=0.235)$ bounded to [0.495, 1.50] m.

**External validation against ScanPyramids muography**

We conducted a pre-registered external validation against ScanPyramids muographic anomalies [12], registering anomaly centroids (cavities C1–C2; notches N1–N3) using published survey tie points (First conclusive findings with muography on Khufu Pyramid, *HIP Institute press release*, 2016, available at: http://www.hip.institute/press/HIP_INSTITUTE_CP9_EN.pdf, accessed 2025/10/26). For each scenario in our 6–8° slope grid (IER parallel), we generated model-predicted, stagewise internal features—temporary edge channels, corner platforms, and post-backfill residual gaps—for quantitative comparison.

For each scenario, we defined a quantitative match rule per scenario: geometric consistency was recorded when a predicted ramp-channel centerline lay within 2.5 m of an anomaly centroid, an error margin reflecting positional uncertainty. The procedure was fully automated and applied uniformly. Critically, no parameters were tuned; the identical pipeline ran across all scenarios to ensure a fully objective comparison.

**Comparative framework**

We benchmark four implemented geometries under common assumptions—straight external ramp (4° or ~7%) [6,7], near-edge spiral ramp (4°, 5 m offset, 6.35 m lane) [6,7,19], Houdin's dual-phase model (straight 4° to ~40 m, then internal spiral 5° or ~9%) [11], and the single-lane IER (7°)—using the same parametric–logistic pipeline ($\mu$≈0.20, 0.15 m s$^{-1}$ on ramps, one-cell separation rule, working year). For straight and the external leg of Houdin, auxiliary earthworks are computed as clipped fills supported against the pyramid face (single outer talus, $\alpha$=60° from the ground) and the fraction that would overlap the pyramid body is corrected. Material efficiency (ME) is reported as ME=1$-V_{aux}/V_{pyr}$. Outputs are generated per course and in aggregate: ramp/terrace distances, integrated mechanical work F=mg(sin$\theta_r$+$\mu$cos$\theta_r$), single-lane cadence (team size, cell length, headway), and external earthworks up to target heights. For unimplemented families beyond these four, we retain a qualitative/semi-quantitative rubric (Supplementary S13) covering (i) survey visibility, (ii) expected traces, (iii) compatibility with ScanPyramids, and (iv) architectural advantages. For parity, the benchmark uses a single-lane IER; multi-lane/adaptive results are reported separately in Results.

All implemented runs (geometry → logistics → work) are under identical parameters and reproducible from the Zenodo dataset/code; the same workflow combines geometry, forces, and logistics into a single, testable framework that enables like-for-like comparison and quantitative evaluation against archaeological evidence.

**Results**

Having defined the IER geometry and the governing forces, our computational model provides quantitative outputs that characterize construction logistics. For each analysis, we state the objective and method context, then present data and conclusions. We report stagewise geometry, block counts, headways, queueing and duration, FEA structural envelopes, and parallel-operation schedules. Baseline assumptions follow Methods.

**Material Dynamics: Block Distribution Analysis**

Within the framework, this section establishes baseline logistical demand and checks that creating the edge channel does not itself create an intractable workload. Blocks per course (Fig. 5) decrease quadratically with height—from 33,124 at the first course to 8,836 by the 100th (Eq. 3, Supplementary S4). Skipped blocks are ≈10,600 for a single ramp (~0.5%) and ≈42,000 for four ramps (~1.8%) (Eq. 4, S4). These fractions are structurally insignificant, and minimal omissions in lower courses (<0.05%) show that ramp creation did not exacerbate the early horizontal bottleneck. In the adaptive configuration, skipped blocks are ≈8,200 (1.6%) with sixteen ramps and ≈4,700 (1.5%) with eight; the proportion then declines with height. Even in 4-ramp runs, omitted cells remain <~2% (details in Supplementary S4). All values derive from the simulation algorithm.

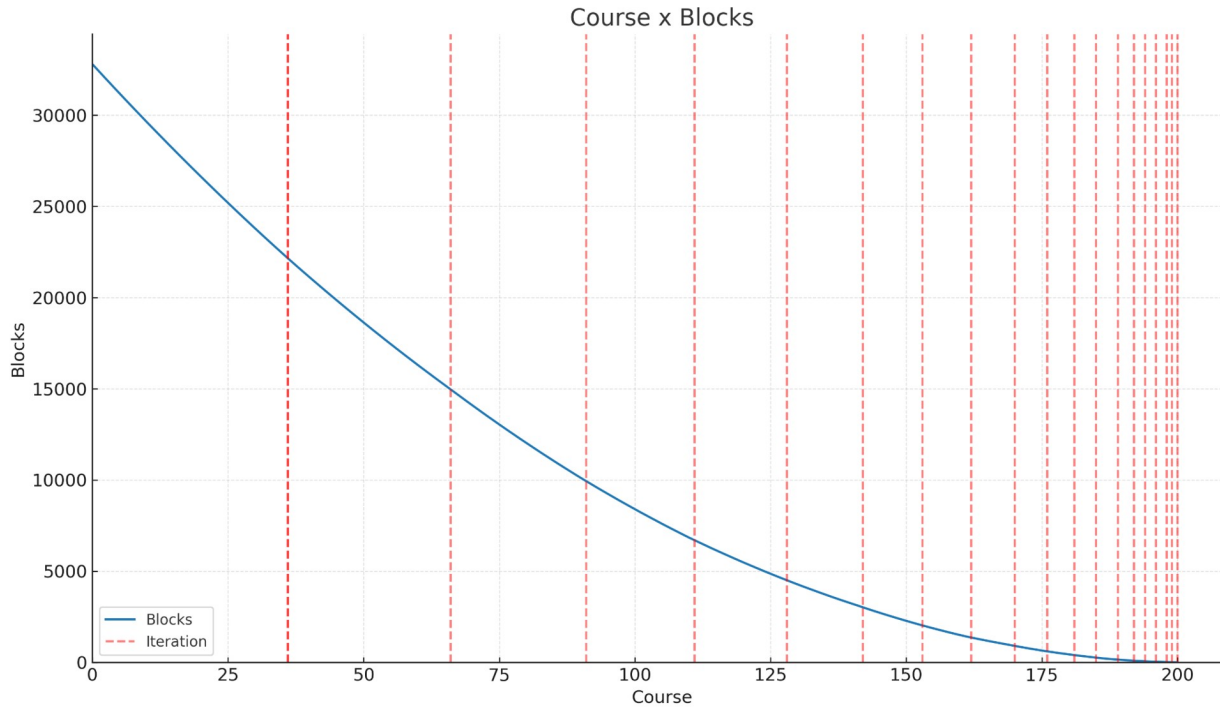

Fig. 5. Simulated blocks per course (baseline geometry). Parametric recursion yields per-course totals while subtracting the active-edge ramp channel. Vertical dashed red lines indicate stage transitions where the helical channel is re-instantiated with baseline $\theta_r=7°$, $\mu\approx0.20$. Axes: x = course number; y = blocks placed (count).

The first four algorithm iterations formed the initial spiral, reaching ~80 m (111 courses) and accounting for ~90% of the 2.3 million blocks. This indicates that block demand was front-loaded, concentrating quarrying and hauling effort in the lower half. Consequently, logistical demands peaked early and progressively eased during upper-level construction.

**Logistical Effort Analysis: Distance and Work Expenditure**

Horizontal transport dominates early; by course 18 the vertical component exceeds it, they balance near course 36, the total workload peaks at course 43, and by course 65 roughly half of cumulative work is completed. Effort concentrates in the monument's lower half, with post-peak ramp turns occurring after the workload maximum, which may help avoid cadence disruptions in the model (Fig. 6).

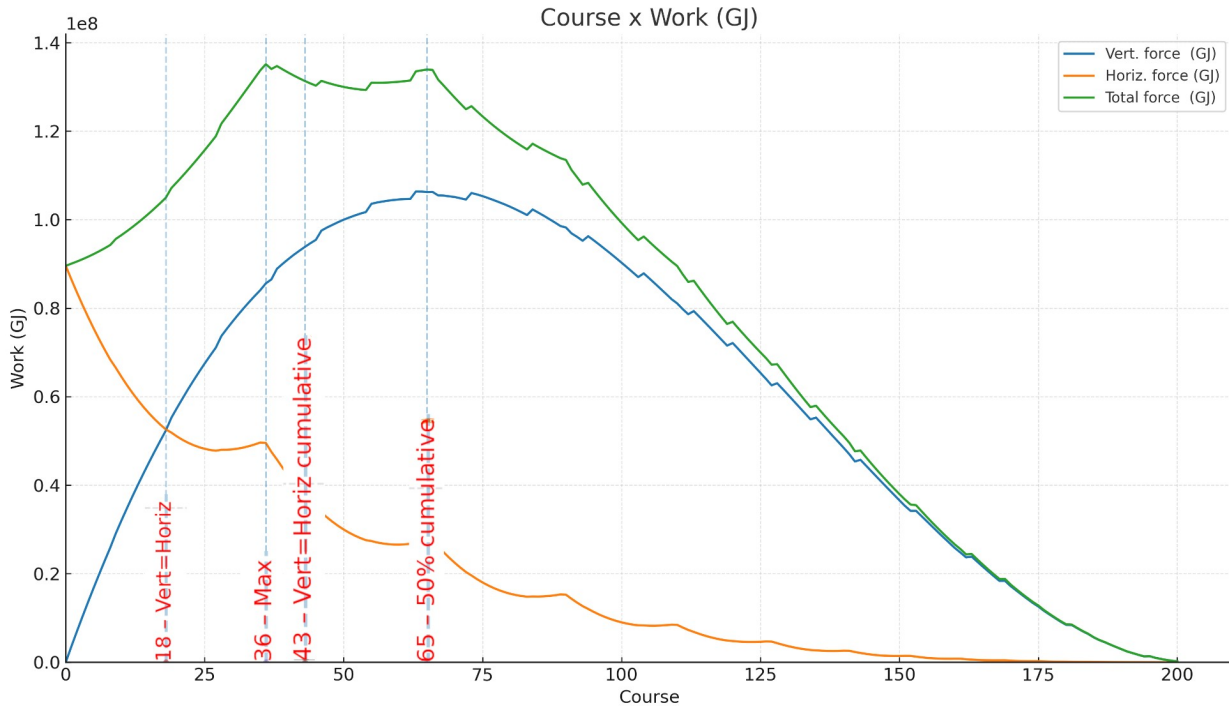

Fig. 6. Cumulative work per course (GJ) with vertical (up-channel) and horizontal (terrace) components, computed from per-block travel distances using the baseline haul model ($\theta_r$=7°, $\mu\approx$0.20). Axes: x = course number; y = cumulative work (GJ). Lines: vertical (blue), horizontal (orange), total = vertical + horizontal (green). Vertical dotted lines mark key transition courses; Note the crossover at course 18, where vertical work (blue line) surpasses horizontal work (orange line), marking the shift in the primary logistical bottleneck.

The horizontal work bottleneck is geometry-driven, as terrace transfer distance varies. Corner placements require the longest lateral moves versus the shortest at mid-side. As the active edge turns 90° at corners, this shift creates a sawtooth horizontal-work curve. This is an expected geometric modulation, not a capacity failure.

The analysis of work expenditure reveals two distinct phases. In phase 1 (<18 courses), >50% of effort is horizontal, distributing tens of thousands of blocks across broad lower courses. In phase 2 (>18), shrinking platforms reduce horizontal work and vertical haul becomes dominant: the primary constraint shifts from width (distribution) to height (longer ramps). This shifting bottleneck quantifies the transition in system control.

Additionally, turning was deferred: no block was turned before course 36 and only once up to course 66, thresholds at ~45% and 69% of cumulative volume (Fig. 5) that postpone turning bottlenecks through the most demanding stages. Beyond course 66 a second turn is required, but by then workload is already declining. Per-course effort reveals a lower-half bottleneck: vertical haul overtakes horizontal at course 18 and peaks near course 43 (pre–mid-height), identifying when ramp cadence is most at risk. At height, vertical transport dominates, reinforcing the case for multi-ramp adaptation.

To quantify this bottleneck, we next assess whether parallel edge channels could alleviate vertical transport constraints while maintaining structural safety.

**Traffic and Queuing for the Multi-ramp Model**

Monte Carlo trials showed early phases (16–8 ramps) below saturation, whereas later phases (4–2–1) ran near capacity yet avoided gridlock (Supplementary Fig. S9.1). Median timelines (95% CIs) stayed within ~27 years, with variability mainly tied to adaptive and 4-ramp strategies. Parallel flows preserved throughput, and the shifting bottleneck was quantified.

The simulations indicate a median on-site duration ranged from ~12.6 years ($\mu$=0.15, $\theta_r$=7°) to ~15.9 years ($\mu$=0.30, $\theta_r$=8°), with 95% CIs of ~10.2–19.2 years. Thus, even with higher friction and slope, the modeled on-site timeline remains within ~27 years, given our assumptions and parameter ranges. Adding planning, quarrying, transport, and seasonal pauses yields a total consistent with the 20–27-year Wadi al-Jarf window. Lower phases (16–8) stayed below saturation, upper phases (4–2–1) ran near capacity yet remained feasible (Fig. S9.2).

Across the six scenarios, phase capacities vary only modestly, with P10–P90 bands substantially overlapping. Throughput is driven chiefly by the adaptive 16→8→4→2→1 sequence, not by minor changes in $\mu$ or $\theta_r$, indicating robust performance. Queuing results show early phases well below saturation and later phases near capacity yet feasible.

**Detailed Construction Time Analysis for the Multi-ramp Model**

Having established system capacity, we then compared single and multi-ramp schedules against the historical 20–27-year window. In our simulations, only the adaptive multi-ramp and 4-ramp strategy fits Khufu's reign, requiring 13.67 and 16.50 working years respectively, whereas a single-ramp scenario would take 49.51 years.

The 4-minute baseline headway (≈15 blocks/hour) follows from ramp length and turns by course. Courses 1–20: high block counts and short ramps (~135 m) permit ~2-minute headways (~18 m spacing at 0.15 m/s) buffered by wide terraces. Up to course 36: ramps ~206 m with no turns allow ~3-minute headways (~27 m spacing). Courses 36–66: one turn on ~376 m ramps yields ~4-minute headways (~36 m spacing); ~69% of volume is placed without over-saturation. Beyond course 66: longer hauls and more turns raise headways to ≥10 minutes (~90 m spacing), but lower block counts limit impact. Headway mapping uses a continuous ramp-length function bounded at 2–13 min (Supplementary S9, Fig. S9.4); bootstrap tests confirm robustness to ±20% variability.

Above the baseline $\mu$=0.2, crews may reach ~32 pullers at $\mu$=0.3 (Table 1; Table 2), sharply increasing team size. On a 3.8 m lane this risks congestion and slower dispatch, even if well-coordinated groups can sometimes haul faster. Managers must balance force needs against spatial/organizational limits to prevent bottlenecks and maintain flow.

Table 1. Crew size as a function of ramp grade ($\theta_r$) and friction ($\mu$).

Entries give the estimated number of pullers per block needed to sustain the baseline dispatch headway under the haul model.

| Ramp Inclination | $\mu = 0.1$ | $\mu = 0.2$ | $\mu = 0.3$ | $\mu = 0.4$ | $\mu = 0.5$ | $\mu = 0.6$ | $\mu = 0.7$ | $\mu = 0.8$ |
|---|---|---|---|---|---|---|---|---|
| 6° | 16 | 23 | 30 | 38 | 45 | 52 | 60 | 67 |
| 7° | 17 | 24 | 32 | 39 | 46 | 54 | 61 | 68 |
| 8° | 18 | 25 | 33 | 40 | 47 | 55 | 62 | 69 |

Table 2. Headway variability under ramp conditions. Per-ramp headway required to maintain one-team-per-cell separation at baseline speeds falls within 3.9–4.7 min, supporting the 4-min baseline. Safety distance budget: team length + ~3 m sledge + 15 m dynamic buffer on long ramps, ensuring no overtaking and corner clearance.

| Ramp angle | Friction (μ) | Team size (number of workers) | Team length (m) | Total min. safety distance (m) | Calculated avg. headway (minutes) |
|---|---|---|---|---|---|
| 6° | 0.2 | 23 | 17.0 | 35.0 | 3.89 |
| | 0.3 | 30 | 21.5 | 39.5 | 4.39 |
| 7° | 0.2 | 24 | 17.0 | 35.0 | 3.89 |
| | 0.3 | 32 | 23.0 | 41.0 | 4.56 |
| 8° | 0.2 | 25 | 18.5 | 36.5 | 4.06 |
| | 0.3 | 33 | 24.5 | 42.5 | 4.72 |

Under the tested conditions, realized headways ranged between 3.9 and 4.7 minutes, which is consistent with a ~4-minute baseline in our model. This finding quantitatively supports the 4-minute average baseline as realistic and grounded in workforce space constraints rather than as an arbitrary choice. A detailed time analysis, under conservative assumptions, was performed to quantify the efficiency of the multi-ramp strategy.

Analysis of ramp configurations revealed large timeline differences. A single-ramp model appears operationally unfeasible under the tested settings: long horizontal average haul (~135 m) and 10.3 min/block imply 49.51 years at a 4-min headway—well beyond Khufu's reign. In contrast, a four-ramp system is viable: shortening the horizontal average haul to ~57 m (5.42 min/block) and enabling parallel delivery reduces the total to 16.50 years at the same headway, a roughly threefold speedup that makes the timeline historically plausible. The most efficient strategy is the adaptive approach, which reduces the number of ramps with height; this further optimizes the schedule to ~13.67 years.

Efficiency can be increased by tailoring the ramps used in each phase and addressing the shifting bottleneck. The use of straight ramps addresses width-related issues; then, a helical system can be used during the vertical ascent (Table S9.3).

The construction spans five phases, with the number of active up-ramps decreasing from 12 to 1. Each phase's duration is determined by dividing the total work (blocks × headway) by the number of ramps used. Meeting the 27-year construction timeline requires a speed of ~0.46 blocks/min, or one block every 2.18 minutes.

Our conservative model assumes placing the pyramid's full theoretical volume from a flat base. However, geological studies indicate a natural bedrock knoll forms a substantial part of the lower core (order-tens of percent; ~first 17 courses; Hemeda & Sonbol [28]), reducing early block placements, mitigating the "horizontal bottleneck," and shortening the schedule. Thus, while reported timelines use the full block count, the bedrock core acts as a conservative buffer supporting feasibility within Khufu's reign. To contextualize on-site construction, we summarize overlapping off-site phases (Supplementary S9): Planning ≈1.0 y; Quarrying ≈15.7–21.5 y; River transport ≈6.5–8.8 y; Seasonal pauses ≈3.3–4.5 y. We aggregate these with on-site Monte Carlo outcomes under non-overlap rules; these values provide realistic envelopes rather than a simple sum, consistent with the Wadi al-Jarf horizon. Sensitivity tests in S9 show that varying each phase by ±20% does not change the central result: the combined timeline remains within ~20–27 years.

Table 3. Comparison of construction times (working years) per average headway and method. This table quantifies how different ramp configurations and block dispatch intervals affect the estimated on-site construction duration for the Great Pyramid. $\mu=0.2, \theta_r=7°$

| Method | 2 min | 3 min | 4 min | 5 min | 6 min | 7 min | 8 min |
|--------|-------|-------|-------|-------|-------|-------|-------|
| 1 ramp | 24.75 | 37.13 | 49.51 | 61.89 | 74.26 | 86.64 | 99.02 |
| 2-Ramp | 12.29 | 18.43 | 24.57 | 30.72 | 36.86 | 43.00 | 49.15 |
| 4 ramp | 8.25  | 12.38 | 16.50 | 20.63 | 24.75 | 28.88 | 33.01 |
| Adaptive | 6.84 | 10.25 | 13.67 | 17.09 | 20.51 | 23.92 | 27.34 |

The model's robustness was tested across inputs, especially headway. A conservative 6-min headway covers accidents/slowdowns; at 8 min the 4-ramp build exceeds 27 years, whereas the adaptive schedule remains near the upper bound (27.34 y). An optimistic 3-min headway (10.25 y) is not sustainable (Table 3; Fig. 7). Thus, with a 6–7 min average, on-site assembly still fits Khufu's 27-year reign, indicating feasibility under realistic variability and allowing delays and off-site margins; together these results bound practical cadence choices.

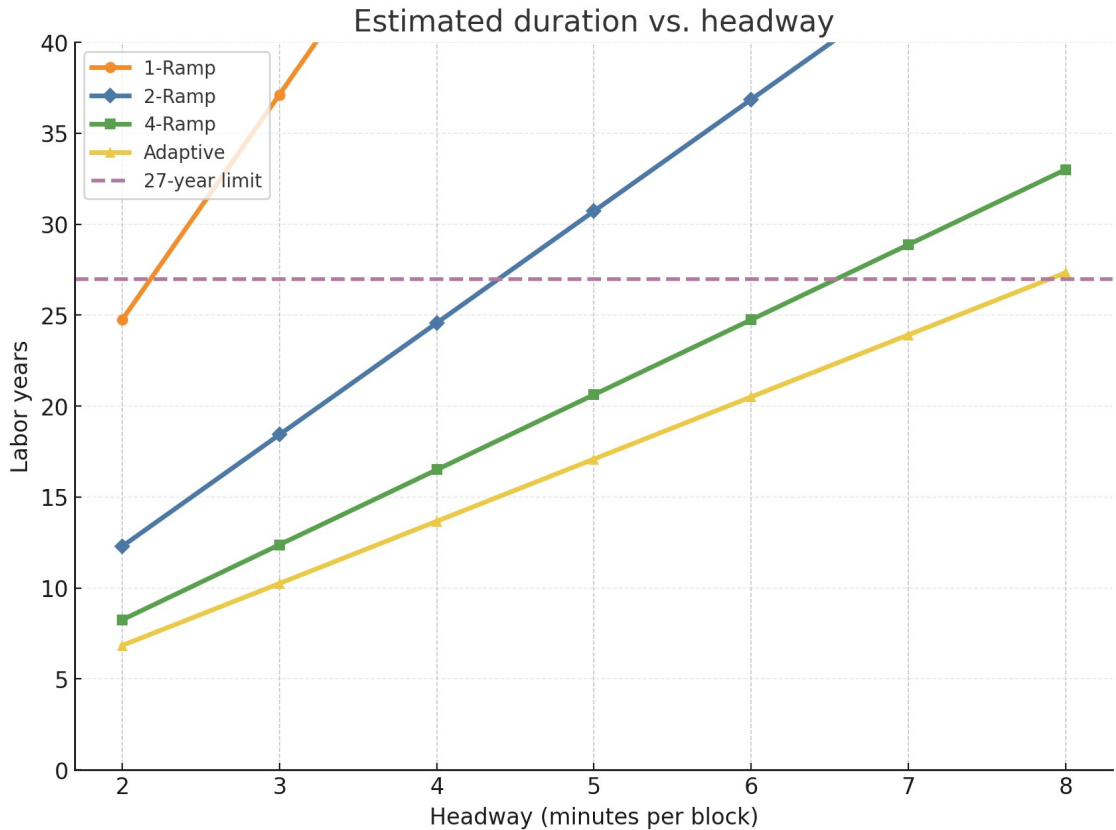

Fig. 7. Construction timelines (working years). Project durations for the schedules summarized in Table 3; bars report onsite time (and integrated totals where noted). The horizontal purple dashed line marks Khufu's ~27-year reign as a reference limit. Baseline settings ($\mu$=0.2, $\theta_r$=7°) follow Methods.

In addition, sensitivity tests show that raising friction from $\mu$=0.2 to 0.3 ($\approx$32-man crew; Table 1) lengthens on-site time by 1.91 years, while increasing slope to 8° adds 0.21 years. Conversely, $\mu$=0.15 shortens time by 1.85 years, and a 6° ramp reduces it by 0.37 years. Despite these parameter shifts, timelines remain feasible within Khufu's reign.

To test robustness, we bootstrap-resampled the dispatch plan, varying headway by ±20%, corner-delay variance, and adding random stops (MTBF ~6 h; repair 5–12 min). Across 10,000 trials, the adaptive strategy's median was 13.8 y (95% CI 12.54–15.18) at $\mu$=0.20, $\theta_r$=7°, 4-min headway. With 6-min headways and slower speeds the median rose to 20.6 y (95% CI 18.53–22.90), still within 27 years (Fig. 8). Results are summarized by medians and 95% bootstrap intervals.

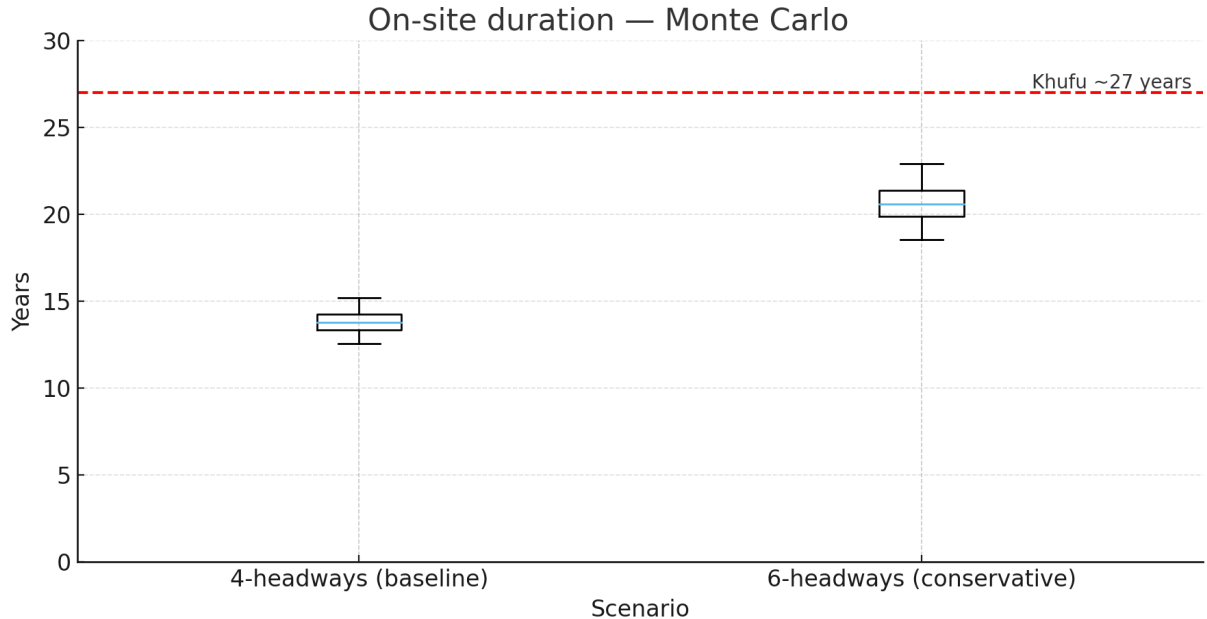

Fig. 8. On-site construction duration from Monte Carlo (N=10,000; seed=42). Boxplots show median (center line) and interquartile range; whiskers are P2.5–P97.5. The red dashed line marks Khufu's ~27-year horizon. Baseline speeds: 0.15 m/s (ramps), 0.20 m/s (terraces); corner delay is lognormal (median 2.8 min, σ=0.35). Settings per Methods.

Monte Carlo simulations with μ=0.15–0.30 and $\theta_r$=6–8° (Fig. S9.3a) yielded median on-site durations of ~12.2 years (95% CI 9.1–15.8) under low-friction μ=0.15, $\theta_r$=6°, and ~15.9 years (95% CI 12.9–19.2) under conservative μ=0.30, $\theta_r$=8°. Across scenarios, 95% CIs remained <20 years (overall range ~9.1–19.2), using 4-minutes headway as baseline. Including quarrying, transport, and pauses aligns totals with the 20–27-year Wadi al-Jarf window.

To quantify how phase configuration controls throughput, we derive realized capacities per phase. Median capacities decrease as ramps consolidate, with variability bounded by friction/slope sensitivity (Fig. S9.1–S9.2). These distributions underpin the on-site duration (Fig. S9.3) benchmarked against Khufu's 27-year horizon.

The model is robust to input variability. Larger pulling teams can offset losses from higher friction or steeper ramps, but this trade-off requires greater headway to avoid congestion, slowing cadence. Managers must balance team size against ramp throughput. Crew footprints, quantified by team length, range from ~11 m to >51 m in double-file organization (1.5 m spacing). Even under demanding conditions the hauling footprint remains significant yet manageable, reinforcing the need for regulated headways.

The time analysis shows on-site duration scales chiefly with the number of active ramps: the adaptive schedule removes the early bottleneck. Reasonable variations in μ (0.15–0.30) and grade (6–8°) shift medians but not scenario ranking. On-site placement remains within the

median of 13.8 (95% CI 12.54–15.18), and 20.6 (95% CI 18.53–22.90) years overall with 4–6-min headways per ramp. Monte Carlo simulations at 6-minute headways show on-site medians ranging from about 18 to 24 years across the μ–θ grid (Fig. S9.3b), all remaining below Khufu's ~27-year historical window.

This on-site logistics model is integrated into an overlapping project schedule (e.g., quarrying continues while river transport runs during Akhet high-water windows), so phase totals do not sum to project duration. We assume a conservative river season, fixed festival/low-water pauses, and a front-loaded planning year for survey/path conditioning (Supplementary S9). Table S9.1 lists parameter baselines, ranges, and ±20% sensitivities with literature justifications. Including planning, quarrying and transport keeps the integrated timeline within the historical 20–27-year window; throughput envelopes thus indicate completion is plausible even under conservative crew assumptions.

**Granite megaliths project**

While the multi-ramp system efficiently handles standard limestone blocks, the 50–80 t granite megaliths for the King's Chamber pose a distinct challenge that requires a dedicated analysis.

Granite handling used short, reusable low-grade ramps built from on-site materials over a prepared surface, enabling batch moves and material recycling consistent with the IER's zero-footprint principle. Staging all megaliths occupied ≈355–601 m² (≈1.3–2.3% of the terrace), confining operations so ordinary limestone delivery proceeded uninterrupted. For a representative 70 t beam on a 3.5° slip with μ=0.20, the steady drawbar is ≈180 kN; the haul model yields ~90–200 kN drawbar. This implies ≈600–900 pullers; for 60–80 t blocks the modeled band is ≈300–700, distributed over 6–16 parallel ropes, keeping the footprint ≤ 3.3% of the terrace. Terrace-mounted wooden bollards (capstan effect) do not lower line tension (≈90–200 kN) or puller count, but reduce the regulating/holding force to ≈30–50 kN, requiring only ≈100 regulators. Calculations are summarized in Supplementary S10.

Whereas completing a single terrace of limestone took months early on, our analysis illustrates the entire batch of granite beams could be positioned in a few days of concentrated work (S10). Conservative upper bound: the first 60 courses require 30 working days; the 60–85 horizon adds 13 working days; thus the total maximum is 43 working days. The early and latest courses likely contained fewer granite elements, so these figures are maxima. Even under an upper bound of ≈45 working days, the granite lifting is spread over time and largely decoupled on terrace slips; against the cumulative ~8.3 on-site years to course 80 in the baseline (or ~12.4 years in the conservative headway), it is clearly not rate-limiting. Moreover, these beams constitute <0.01% of the pyramid's volume, and the short-duration workforce could be reallocated or supplemented from the flexible labor pool at Heit el-Ghurab, as evidenced by that site's scale and organization.

Across conservative ranges of friction and team size, modeled haul forces, bearing stresses, and maneuvering clearances remain within admissible limits without ad hoc mechanisms. These outcomes follow from the haul model, lane geometry, and the route shown in Fig. 4. We therefore treat granite handling as a short-duration, decoupled operation—short 3–4° slips with staged levering—scheduled during natural pauses or in parallel, with negligible impact on the 27-year timeline. In brief, the granite module is not rate-limiting for limestone throughput. We next set these findings against model sensitivity and external anomaly evidence.

**Structural Validation via Finite Element Analysis**

To validate the framework's core structural assumption (temporary edge-channels under self-weight), we perform staged FEA with mesh-objective convergence checks. Mesh-objective convergence was achieved and stress/settlement envelopes remained low: global reactions closed within tolerance, apex settlements were stable across meshes, and stress fields showed only localized bands with ample safety margins (Table 4; Fig. 9).

The mesh convergence analysis confirmed the model's numerical stability (Table 4): (i) base-reaction equilibrium against $\rho gV$ (mismatch ≤0.60% at fineness 3; ≤0.16% at fineness 5); (ii) cap-averaged apex settlement, differing by only −0.008% between meshes; and (iii) point-data stress statistics in MPa (nodal mean and p95 from a 512-bin histogram). Solver residuals were ≤$10^{-10}$, with reported components along the vertical (Z).

Over the full solid, nodal von Mises stresses average 0.530 MPa (fineness 3) and 0.521 MPa (fineness 5), with p95($\sigma_{vM}$) of 1.159 and 1.057 MPa, consistent with the order-of-estimate $\sigma \approx pgh/3$ near the base. Localized bands along ramp zones remain modest (upper envelope ≈3 MPa), far below limestone compressive strength (≥60 MPa), consistent with safety factors ≥50 in bulk and ≥20 at hot spots under linear-elastic assumptions. The fine-mesh apex point probe gives $U_z$=−0.0489 mm, whereas the cap-averaged apex settlement is 3.596 mm; we use the latter as the global metric because point probes at geometric singularities under-represent global settlement.

Two systematically refined meshes were assessed: fineness 3 (coarse) and fineness 5 (fine). Global equilibrium ($\Sigma Rz$ vs $\rho gV$) closed within 0.60% at fineness 3 and 0.160% at fineness 5. Cap-averaged apex settlement changed by only −0.008% (3.59629→3.59600 mm). Stress statistics were stable: nodal mean −1.8% (0.530→0.521 MPa) and p95($\sigma_{vM}$) −8.8% (1.159→1.057 MPa). By contrast, local extrema are mesh-sensitive (e.g., single-node apex probes). We therefore adopt global, mesh-objective metrics—equilibrium, cap-average displacement, and stress percentiles—for convergence rather than $\sigma_{vM}^{max}$ or single-node values.

The stress field matches self-weight expectations: compression concentrates mildly where material is locally reduced (ramp bands, corners), while bulk masonry carries load efficiently, with no gravity-only failure mechanism. With verified meshes (fineness 3–5), nodal mean and p95($\sigma_{vM}$) remain ~0.52–0.53 MPa and ~1.06–1.16 MPa—well below limestone capacity (≥60 MPa).

Table 4. Two-level mesh convergence (fineness 3 → 5). $\Delta$ = |Xf3−Xf3|/Xf5×100. Displacements are cap-averaged apex settlement over a fixed top patch (r = 0.5–1.0 m; downward positive). We report Point-Data von Mises stress (nodal mean, p95 in MPa) instead of $\sigma_{vM}^{max}$, which is mesh-sensitive. Global equilibrium mismatch is |$\Sigma Rz$−$\rho gV$|/($\rho gV$). Convergence is evidenced when all $\Delta$ fall below the preset tolerance (Methods) and equilibrium mismatch is near zero, confirming reliability of the reported FEA fields.

| Mesh level | Fineness | Elements (TET10) | Nodes | Apex cap-avg $U_z$ (mm) | Δ apex (%) | Mean $\sigma_{vM}$ (MPa) | p95($\sigma_{vM}$) (MPa) | Equilibrium mismatch (%) |
|---|---|---|---|---|---|---|---|---|
| Coarse | 3 | 592,877 | 958,872 | 3.59629 | – | 0.53009 | 1.159 | 0.60 |
| Fine | 5 | 839,877 | 1,323,383 | 3.59600 | 0.008 | 0.52055 | 1.057 | 0.160 |

Pointwise von Mises peaks are governed by geometric singularities at corners/ramp interfaces and do not converge with refinement. The stress map (Fig. 9) accords with this: blue tones dominate the core, with only small localized bands at ramp interfaces, all far below limestone capacity. Consistent with Weiss et al. [29] on cumulative effects and stability, our FEA indicate the IER design remains within safe stress margins for Old Kingdom limestone, with no gravity-only failure mechanism.

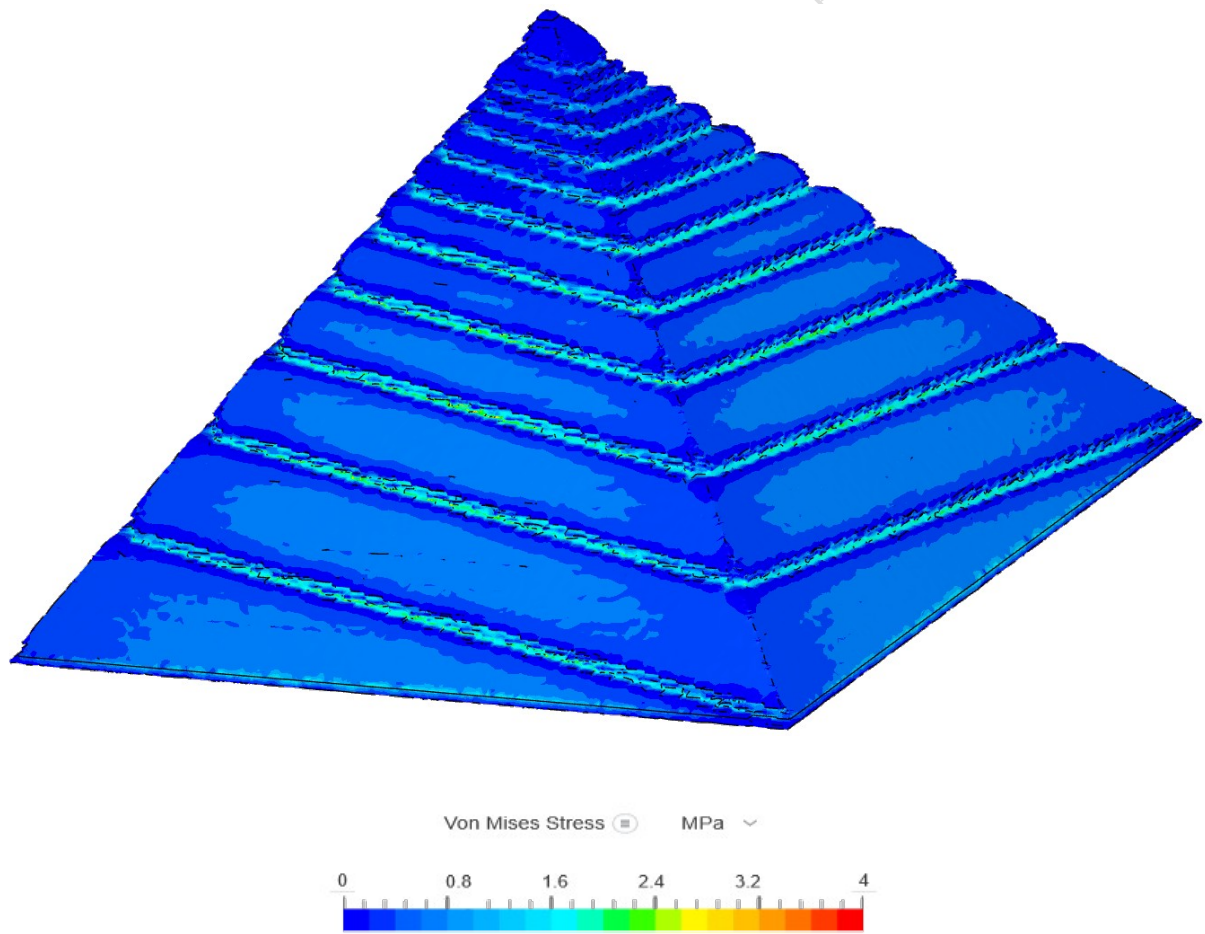

Fig. 9. FEA under self-weight: von Mises stress (MPa) and apex settlement (mm). Stresses are low and localized along the temporary ramp channel; no critical core concentrations at the representative stage. Color bar: von Mises (MPa). Model: linear-elastic limestone, self-weight only; see Methods (FEA) for properties and convergence. Mesh: fine 3-D TETRA10 with local refinement at edges/corners/chamber zones. Solver: Code_Aster 15.6.10 / MUMPS 5.2.1. Interpretation uses global displacement and percentile stress summaries (95th). Material: E=35 GPa, ν=0.25, ρ=2600 kg·m⁻³; body force ρg=−25,506 N·m⁻³. Full meshes, input decks, and seeds are archived on Zenodo.

Peak stresses along the edge-channel path are well below typical limestone strengths; together with mesh-objective convergence, this supports structural plausibility under the linear-elastic assumptions tested. These results do not address dynamic loads or jointed-rock behavior.

Finally, the framework extends to external geometric checks, comparing predicted turning tiers and access heights with independent datasets (course thickness and muography) as hypothesis-generating tests.

**Alignment of the IER geometry with Observed Structural Features**

Another possible external geometric alignment merits study: a pre-specified comparison between IER-predicted turning tiers and published course-thickness series (McKenzie [30]) shows a non-trivial co-alignment. The correlation between predicted turning tiers and course-thickness changes showed a narrow maximum at a ramp angle of 7.4° with 5/10 post-turn matches (Fig. 10), whereas adjacent angles are ≤4 (Table S11.1). This result is statistically significant ($p < 5 \times 10^{-4}$), as Monte Carlo simulations found no instances of ≥5 matches. Our study shows raising $\theta_r$ from 7° to 7.4° adds only ~0.1 years and no extra workforce. We interpret this as a geometric correspondence under our pre-registered test, hypothesis-generating rather than confirmatory; higher-resolution data would be required to discriminate mechanisms; full tier-by-tier counts and window/detrending sensitivities are in Supplementary S11. Higher-resolution muography and endoscopy would be needed to resolve morphology and fill signatures.

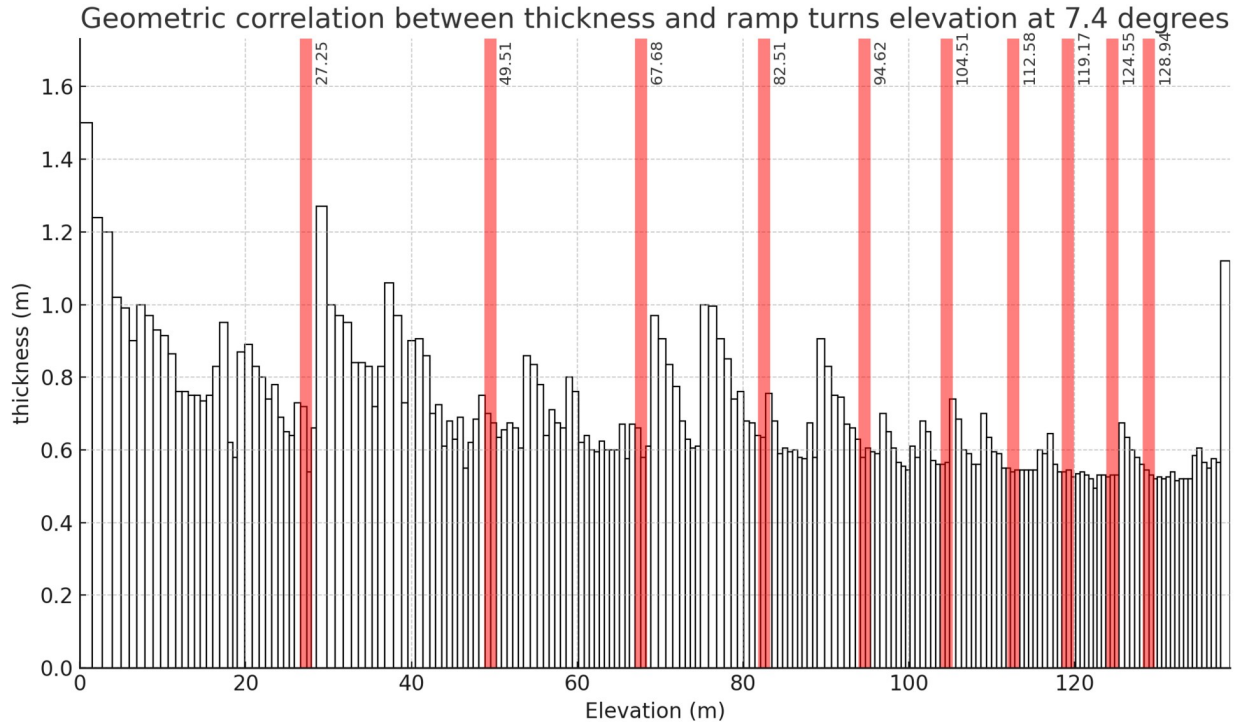

Fig. 10. Course thickness vs elevation (m) ($\theta_r = 7.4°$), by McKenzie [30]. White-filled bars with black borders show thickness per course (m) versus elevation (m). The red bands indicate the turning nodes predicted by the IER scenario with a 7.4° slope in courses 38 (27.5 m), 95 (67.68 m), 116 (82.51), 147 (104.51) and 175 (124.55); The turning elevation positions are highlighted in red. This figure tests geometric alignment between IER turning tiers and observed anomalies; interpretation is hypothesis-generating. Other prominent thickness peaks not aligned with IER turning nodes likely reflect survey-driven leveling/stiffening belts or operational phase boundaries (e.g., platforms near the King's Chamber horizon); see Supplementary S11 for details.

The northeast (NE) corner is uniquely informative: its exposed interior permits a geometric cross-check against model predictions and highlights a key distinction between ramp models. Whereas Houdin's internal ramp requires an entrance within the masonry, the IER's helical path is an open-air channel formed by temporarily omitting perimeter blocks, so its access is the edge aperture itself. Notably, the predicted IER access aligns with a visible zone of structural loss on the face (Fig. S11.2). Although this notch could reflect post-construction erosion or quarrying, its alignment with the void expected from an omitted-block ramp offers a plausible geometric correspondence that a deeper, internal access does not explain. Such zones would be expected to respond differently to seismic loading than solid, interlocked faces. This is consistent with independent geological assessments: Hemeda and Sonbol [28] report corners as areas of "permanent deformation" and "increasing weakness," attributed to "friction and sliding between the facing and backing blocks" during seismic events. While not proof, the localized degradation pattern plausibly matches an edge-integrated method that backfilled corners, a prediction testable via geophysics targeting rubble-and-mortar fills along the helical path.

**The ScanPyramids Mission Findings**

Beyond internal validation through FEA and possible geometry consistency, the IER model must also suggest consistency with external empirical evidence. We therefore conducted a geometric comparison with the muographic anomalies reported by the ScanPyramids mission.

A useful hypothesis-generating check is whether the IER geometry plausibly corresponds to published ScanPyramids anomalies [12]—cavities/notches on the NE edge, the Big Void (BV) and the North Face Corridor (NFC). We compute turn elevations for $\theta_r$=6.0–8.0° (see Supplementary S12, Tables S12.1–S12.2) to be contrasted with the positions of the known cavities and notches. The best fit within the tested grid occurs at 7.4–7.5° with mean absolute error 1.30 m, acknowledging registration/localization uncertainties. Given muographic localization/registration uncertainties, we treat this overlay as hypothesis-generating, not confirmatory. In the 4-ramp model, the 7.4–7.5° helical path lies near the reported anomaly centroids (Fig. 11). This consistency warrants further investigation but is not evidentiary: current muography cannot distinguish a backfilled ramp from other heterogeneities by centroid proximity alone. Sensitivity shows raising $\theta_r$ from 7° to 7.5° adds only ~0.11 years and no extra workforce, so geometric consistency does not compromise the studied logistical feasibility. The alignment is compatible with density anomalies from backfilled construction infrastructure, but alternative explanations remain viable at present resolution. Non-detection of internal spirals or specific helical voids must be interpreted within method sensitivity and coverage limits; current data neither confirm nor exclude edge-integrated backfill.

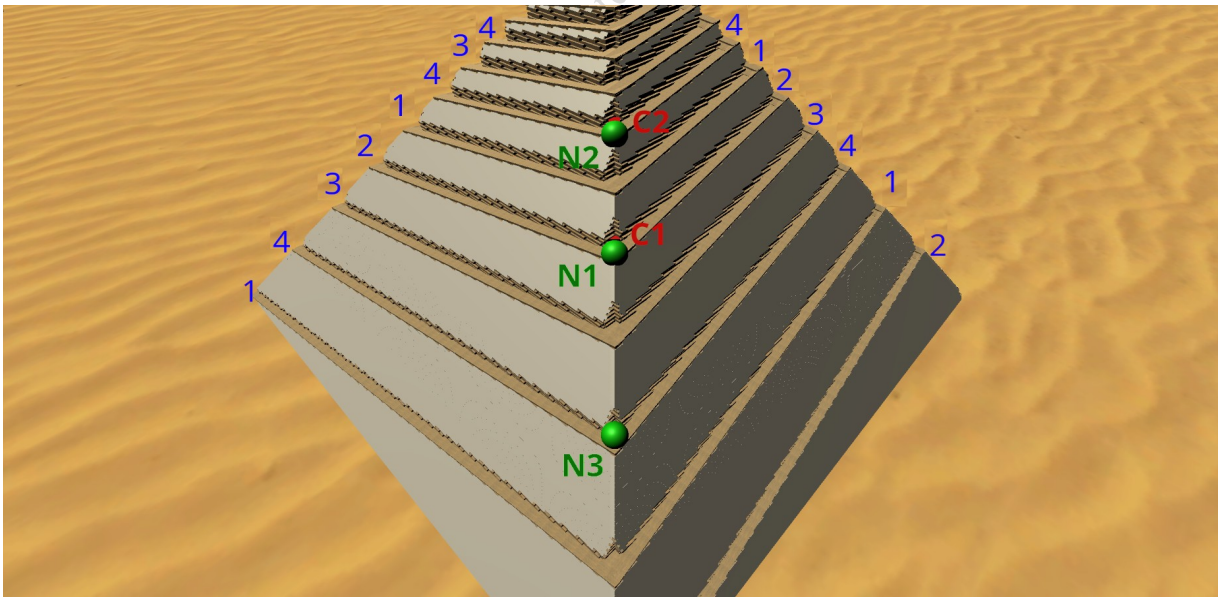

Fig. 11. Geometric consistency of the IER with ScanPyramids anomalies. 3-D parametric reconstruction of four edge-integrated helical channels at 7.5°; channels are numbered by start edge. Reading guide: numbered helices = predicted 7.5° edge channels; green spheres =

notches (N1–N3); red cubes = cavities (C1, C2). The overlay illustrates proximity, not identity; values reflect model–data registration within meter-scale uncertainties and are not diagnostic of process and should be interpreted as geometric consistency rather than proof.

If Khufu used an integrated edge ramp, Khafre's near-contemporary pyramid may have employed the same method. A falsifiable test is to compute a Khafre-specific list of predicted edge-band elevations from $\theta_{Khafre}$ and $h_c$ (Table S12.4), assuming a ≈7.5° grade, and to target these heights in muographic/endoscopic surveys. These results are compatible with muographic signatures but are not evidence of a specific void geometry. We retain $\theta_r=7°$ as the logistics/FEA baseline (force–path balance); 7.5° is used only for external geometric-consistency tests against ScanPyramids anomalies. The apparent 7.5° alignment (Table S12.3) suggests possible geometric correspondence, but our model is a parametric reconstruction rather than a precise survey.

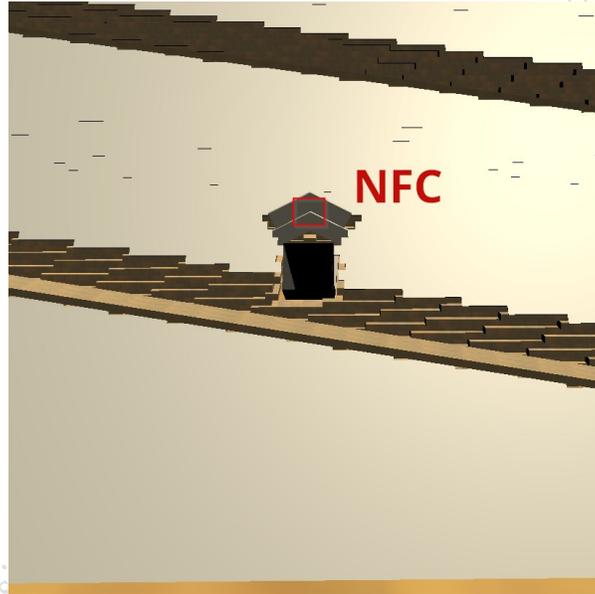

Fig. 12. 3D IER rendering of the north face. The edge lane along the arris reaches ~15–16 m, ~1.6–1.8 m below the NFC (~20 m) and ~7–8 m inside the frontage, for a 7.5° ramp. A short ~10° temporary walkway/stair could link them; the lane is later backfilled, preserving the façade —unlike large external ramps. Interpretation: correspondence only, not diagnostic at current muographic resolution.

While the IER does not predict the BV, the NFC shows intriguing compatibility: its ~20 m elevation and ~7–8 m inside the frontage and 0.8 m behind the chevrons [7,12] closely matches the edge-ramp's ~15–16 m (Fig. 12). A short ~10° temporary ramp/stair could link the main entrance (~17 m) [7]. This suggests a temporary access could have connected the ramp to the NFC and main entrance. The alignment is treated as a geometric consistency and a testable hypothesis, not a proof. Verifying whether the voids are backfilled ramps requires higher-resolution muography or endoscopy to resolve morphology and density.

**Comparative framework analysis**

To contextualize the Integrated Edge-Ramp (IER) within prior ramp hypotheses, we applied the same computational–logistic framework (see Supplementary S13). All were re-implemented under identical parameters to ensure direct comparability of geometry, logistics, and mechanical work.

See results in Table 5, when lowering ramp slope to 4–5° reduces per-block traction but inflates total haul distance, work, and single-lane timelines. Under identical assumptions, straight and spiral variants remain single-lane and distance-dominated, yielding on-site duration well beyond the 20–27-year window. The Houdin split (short external then internal spiral) lessens early earthworks but still behaves as a single-lane system with multi-decade timelines. By contrast, the IER (7°) shortens the ramp paths and cuts the required work by roughly two-thirds. Full per-course tables and sensitivities are provided in Supplementary S13.

Table 5. Aggregate quantitative comparison. Summary of key geometric, logistical, and energetic metrics for the configurations modeled under identical parameters using the same parametric–logistic pipeline (see Supplementary S9). Distances refer to the cumulative haul of all blocks (rows 0–99); ramp share indicates the fraction of total distance occurring on inclined paths. Work total integrates mechanical work. On-site duration assumes single-lane operation. Fill to h≈71 m corresponds to the external adobe volume required for clipped ramps (embankment 60° slope angle from ground). Data illustrate the consistent framework used to benchmark different ramp geometries, allowing direct, parameter-controlled comparison and future export of results for structural (FEA) or sensitivity analyses.

| Parameter | Straight 4° (~7%) | Spiral 4° (~7%) | Houdin 4–5° (~7–8%) | Single IER 7° (~12%) |
|---|---|---|---|---|
| Ramp length (m) | 1007.65 | 1025.80 | 887.28 | 579.93 |
| Total distance blocks ramp (m) | $1.0197 \times 10^9$ | $1.0466 \times 10^9$ | $9.3848 \times 10^8$ | $6.9603 \times 10^8$ |
| Team size | 20 | 20 | 20–23 | 24 |
| Headway (min) | 3.7 | 3.7 | 3.7–4.0 | 4.0 |
| Work (GJ) | $2.46 \times 10^{12}$ | $\geq 2.46 \times 10^{12}$ | $\approx 1.46 \times 10^{12}$ | $8.27 \times 10^{11}$ |
| Fill to h≈71 m (Mm³, α=60°) | 0.75 | >1.0 (multi-face) | 0.25 (ext. ~40 m) | 0 |
| On-site years (~2 M blocks) | 40–50 | $\geq$45–55 | 35–45 | 41.9 |

The parametric IER model is consistent with completion within Khufu's reign under the Old Kingdom constraints modeled here. Under 4–6-minute per-ramp headways, on-site assembly spans ~13.8–20.6 years; including quarrying, river windows, and seasonal pauses yields ~20–27 years overall, consistent with the Wadi al-Jarf horizon—a finding we contextualize against competing theories in the Discussion.

**Discussion**

This study deployed an integrated computational framework to evaluate the feasibility of the Integrated Edge-Ramp (IER) model for the construction of the Great Pyramid. Our principal findings demonstrate that: (i) an adaptively parallelized IER can sustain minute-scale block dispatch, yielding an on-site construction timeline that fits within Khufu's reign; (ii) the temporary edge channels pose no significant structural risk under self-weight; and (iii) the model's geometry shows consistent alignments with modern geophysical surveys and yields distinct, falsifiable archaeological predictions. The following discussion interprets these results in the context of Old Kingdom engineering constraints and prior ramp theories.

This Discussion first synthesizes the main quantitative findings, then situates them within the broader context of Old Kingdom engineering and ramp theory under identical and transparent parameters (geometry, friction, cadence rules, and work model). This homogeneous setup translates disparate proposals into commensurable results (distances per lane, single-lane performance, rock movement, and integrated work), so that past, current, and future methods can be compared on a fair basis. The process is reproducible and extensible: all tables and 3D objects can be rerun with alternative assumptions or exported as limit/load inputs for finite element analysis (FEA) (or other simulators), allowing for structural checks consistent with the logistics used to generate schedules. In this sense, the framework not only generates results for the IER but also constitutes a standardized test for quantitative evaluation across different ramp families.

While qualitative comparison highlights the IER's advantages, its viability is ultimately tested quantitatively against the primary Old Kingdom constraint: the historical timeline. These results confirm the quantitative feasibility previously established; here we focus on their archaeological and methodological implications. This aligns with the Wadi al-Jarf papyri and reconciles Merer's high-frequency logistics with a viable construction model. Under minute-scale headways, on-site construction lasts ~13.8–20.6 years. This alignment supports a coordinated, adaptive IER workflow. Separating on-site placement from off-site supply reconciles pace with reign length. Future chronology should integrate radiocarbon with Nile records to further refine the timeline. This chronological contrast serves as the critical external benchmark: if the simulated schedule did not match the 20–27-year horizon, the IER would have to be reconsidered.

The temporal analysis shows that regulated headways balanced speed and safety: a 4-minute interval limited congestion while sustaining throughput. Crucially, the IER defers corner turns until after course 36 (≈45% of total volume), minimizing bottlenecks in the most demanding phase. This contrasts with earlier ramp theories (Arnold [6]; Lehner & Hawass [8]; Brichieri-Colombi [5,19]; Houdin [11]; Klemm & Klemm [13]; Müller-Römer [14]) that assumed 2–3 min placements; as Müller-Römer noted, at 5 min/block such systems exceed 40 years, revealing acute headway sensitivity. By comparison, adaptive IER remains feasible within the 20–27-year Wadi al-Jarf window even at 6–7 min headways, showing greater resilience. Simulations suggest throughput is fundamentally headway-limited, especially early. Sustaining 4–7 min dispatch requires avoiding "starvation" via early quarry mobilization and base-level stockpiles—consistent with Merer's batched deliveries and proposals by Lehner and Houdin for initial reserves. The novel contribution is a quantified logistics requirement—buffered supply to stabilize headways—reframing reign-scale timelines as a testable coordination challenge.

This reconciliation of the macro-timeline rests on resolving micro-logistical challenges, chiefly the 90° corner turns. Our simulations show 4×4×9 corner platforms provide clearance for sledge pivots, with turn delays and lever-and-drag actions consistent with Stocks' experiments [27]. The brief (~7.8 m diagonal) edge occlusion during turns did not hinder precision: a master builder could hold the line with plumb bobs and sighting rods against exposed step arrises, matching Old Kingdom survey practice (measuring cords, plumbs, edge sighting) documented for Fourth Dynasty works. Future work should include microscopic analysis of corner block surfaces to detect wear patterns from repeated turning operations.

Furthermore, keeping corners and faces clear addresses a key architectural challenge: placing the fine Tura limestone casing. Most ramp theories (straight, zig-zag, external spiral) obscure the faces, forcing post-removal casing that complicates precision alignment and lacks archaeological support. By contrast, both Houdin's internal ramp and the IER keep faces exposed, enabling course-by-course casing. The IER achieves this advantage without unverified internal corridors, while preserving visible arrises as survey baselines. This capacity for progressive casing strengthens the IER's archaeological and logistical plausibility over external-ramp alternatives. Petrographic analysis of casing surfaces could test for expected wear and bedding signatures.

This advantage in block and casing placement underscores a key distinction from Houdin's internal spiral, which requires hauling through narrow, ~200 m enclosed corridors; sustained work in dim, poorly ventilated tunnels would demand impractically high energy for lighting. By situating haulage on the pyramid's edges, the IER provides natural daylight, fresh air and direct supervision, supporting visibility and safety. The edge-ramp is the shortest, most efficient variant of an external spiral, using far less material while preserving surveyable edges. Unlike face-entry zig-zag schemes, the corner-start IER affords a ~25.56 m vertical no-turn run (≈210 m along the ramp), delaying the first turning penalty. A straight frontal ramp to 30–40 m would require ~0.24–0.56 million m³ of fill, taking months to over a year to build and remove, and it would leave clear foundations. to build/remove—and would leave clear foundations. Structurally, the IER appears among the more material-efficient helical variants considered in this study, combining external exposure with a zero-footprint signature.

Beyond structural correlations, the IER addresses the long-standing critique of "no ramp remains" via a zero-footprint system—framed as a falsifiable prediction, not evidence (Howell & Prevenier [31]). A massive external ramp would be expected to leave foundations or debris, yet none are found; instead, an integrated system could have been reabsorbed, leaving only subtle density anomalies. Discovery of large ramp bases at Giza would challenge the model, whereas the absence of such traces at other Old Kingdom pyramids is consistent with a no-residual framework. By contrast, Müller-Römer's temporary mudbrick ramps would have been unstable, required non-existent stone foundations, and lengthened construction. The IER may mitigate these issues under the stated constraints by using and then removing its infrastructure during the build. Although Klemm & Klemm's integral helical ramp [13] shares a conceptual foundation, its original description was brief, lacking detailed geometry, logistical analysis, or quantitative validation—a gap our computational framework now fills. While some indirect features remain ambiguous, the lack of external ramps in early Giza excavations coheres with a zero-footprint IER.

In addition to logistical feasibility, the IER's structural simplicity confers operational advantages —security and scalability. While a single ramp shows baseline feasibility, meeting Khufu's reign requires parallelizing a multi-ramp system. This layout improves efficiency and flexibility: crews can switch lanes and use descending ramps for debris removal, making parallelism a plausible operating regime under these constraints. Archaeology supports this: Klemm & Klemm [13] noted simultaneous quarrying on several fronts near Giza, consistent with deliveries to multiple ramp bases; the parallel workforce structure at Heit el-Ghurab fits such organization; and the Ahramat branch's proximity (Ghoneim et al. [2]) enabled a harbor feeding several entry points.

The multi-ramp system offers qualitative advantages within Old Kingdom constraints. Dedicated lanes (three ascent, one descent) sustain continuous flow and reduce conflicts. Building all four faces simultaneously preserves structural symmetry, lowering settlement risk. The layout adds organizational resilience, allowing crews to redirect when a ramp is blocked with minimal disruption; corner supervisors regulate traffic and enhance safety. Beyond efficiency, parallelism improves overall traffic management, stability, and safety. Finally, as shown in Fig. S7.1, base-level distribution paths speed block delivery to ramp entrances.

Within the adaptive IER framework, plausibility extends beyond simulations. IER lanes—formed by omitting and later backfilling perimeter blocks—require negligible temporary works compared with external ramps. The adaptive sequence quantitatively explains the shift from broad lower platforms to narrower upper courses, numerically confirming what Lehner and Hawass [7,8] and Müller-Römer [14] suggested qualitatively. While Hatnub evidence [32] shows straight ramps in the Old Kingdom, their steep gradients (~20°) and limited capacity preclude them as a sole solution for Khufu; our model formalizes that bottleneck as a measurable threshold. Moreover, the substantial bedrock core documented by Hemeda & Sonbol [28], in the first ~17 courses the demand for initial blocks could be reduced, meaning straight-run ramps could be minimized or omitted so the four-ramp IER regime starts immediately without saturation—reducing corner queues and shortening the initial phase in proportion to the bedrock fraction. Although the 4→2 ramp shift at course 184 is geometry-driven, an earlier transition is plausible: by course 66 (Fig. 5), with ~69% of the volume placed and workload decreasing quadratically (Fig. 6), managers could reduce to two ramps, trading upper-course cadence for manpower reallocation and simpler coordination. This potential shift from throughput-maximization to resource optimization warrants quantitative testing in future simulations.

Having situated the IER among competing theories, we examine the central physical parameter: kinetic friction ($\mu$) between sledge and ramp (Fig. S9.5). Our baseline is $\mu$=0.20, consistent with Liefferink et al. [33,34], where controlled experiments show a non-monotonic $\mu(w)$ with a minimum at ~3–5 % water ($\mu \approx 0.16$–0.22) and higher values in dry or overwet regimes ($\mu \approx 0.25$–0.35). We exclude West's rolling proposal ($\mu \approx 0.06$) given its preprint status and lack of archaeological support [35]. While maintaining optimal wetting is non-trivial, the nearby Ahramat Branch suggests water was available. Thus, we adopt $\mu = 0.20$ as a field-plausible baseline (not a tuned lab optimum). Crucially, results are robust across $\mu$=0.15–0.30, so conclusions do not rely on optimistic assumptions. Moreover, all ramp theories share a similar friction–slope envelope; comparable parameter shifts yield similar travel-time effects.

A key finding is that minute-scale headways are sustainable with modest on-site crews when operations are parallelized (adaptive or four-ramp), concentrating effort in early phases. This pattern aligns with Hirlimann's analysis [20], which found hauling consumes little energy and that fewer than ~2,000 workers per season (~200 active on site) would suffice—figures that necessitate highly parallel operations and contradict single-ramp models. The adaptive IER enables simultaneous block movement via multiple access routes, reconciling energy budgets with the timeline. Crucially, scheduling and concurrency—not raw manpower—limit throughput, directing tests toward predictive signatures of parallel workflows (e.g., phased corner wear) and toward integrating quarry/transport data to refine crew size estimates.

Because throughput is constrained by scheduling and concurrency rather than manpower, river supply must be synchronized with ramp operations to keep minute-scale headways stable. In practice, staggering barge arrivals across harbor nodes feeding different ramp bases, with site buffers absorbing variability, prevents any single entry from overloading. Our IER model suggests that integrating transport windows, staggered deliveries, and buffer limits sustains the 4-min baseline headway in high-demand phases. Coordinated river schedules are therefore essential, linking external supply to on-site cadence. Archaeology illustrates this capacity was within Old Kingdom means, with planning and heavy-haul know-how at Hatnub/Wadi Hammamat (Klemm [36]) and Giza (Lehner [37]). Multiple corner ramps are favored given nearby limestone sources (Fig. S9.6): splitting sources by corner shortens trips, enables local stockpiles, minimizes cross-traffic, and stabilizes headway, yielding higher throughput with lower variance than any single-entry supply. The scheme is testable via geoarchaeological coring/geophysics and perimeter surveys for corner wear and thin stockpile lenses; their absence would argue against a multi-entry system.

While the model centers on ramp logistics, Khufu's ground plan was highly accurate (Clerc [38]). Feasibility also hinges on base-level logistics: the Ahramat branch supported a high-volume harbor (Marriner et al. [39]), and parallel ramps required distribution paths from harbor staging areas to ramp bases—well within Old Kingdom capability at the planned Heit el-Ghurab settlement with specialized crews. This framework of harbor, distribution system, and large labor pool fits Khufu's ~27-year reign yet remains flexible, since base-level phases could lengthen without invalidating estimates. Future work should test these assumptions via geophysics for stockpile lenses, mapping corner-platform wear, and edge surveys. The IER logistics thus suggest concrete, excavation-grounded field tests (Der Manuelian [40]).

The results underscore that work management—not just manpower—may be crucial to sustaining efficiency. Headways and buffers functioned as dynamic risk-mitigation tools, allowing managers to regulate dispatch and absorb delays or accidents; corner platforms served as physical buffers for disruptions. This reflects adaptive uncertainty management and aligns with Daganzo's principle [41] that steady headways maximize throughput, as well as evidence from Old Kingdom quarries (Bloxam & Storemyr [42]) indicating sophisticated supply-chain control. Efficiency also relied on specialized crew coordination: assembly-line teams (loaders, pullers, pivot crews, setters) reduced fatigue and built expertise. Archaeology supports this structure: Heit el-Ghurab sustained a rotating workforce of ~20,000–30,000 (Lehner [43,44]), while crew marks (Tavares [45]) and 10-day cycles in the Wadi al-Jarf papyri indicate formal divisions and rotation. Future work should parameterize rotations and rest cycles to quantify their impact on headways and safety.

Further supporting this interpretation, Dormion's 1980s microgravimetric survey [46] found lower-density anomalies along all four edges (Fig. S12.3), consistent with the IER prediction of backfilled ramps and optimized corner platforms, though alternative explanations exist (e.g., Houdin's small L-shaped chambers). The NE notch at ~80 m (Fig. S12.2) and a small chamber identified by Brier [24] also fit both models. The approaches diverge structurally: Houdin's requires an archaeologically unsupported roofed tunnel, whereas the IER employs open-air platforms consistent with Old Kingdom practice. For muography, a backfilled channel should yield a heterogeneous density signature, unlike the uniform void of a preserved tunnel; thus the IER offers a testable, though unproven, interpretation. Moreover, the IER predicts continuous edge backfill, while an internal ramp expects discrete, corner-focused anomalies—motivating a reanalysis of Dormion's data to determine which geometry better fits the observations.

Complementing geophysical evidence, the ScanPyramids record is central to current debates. Schumacher et al. [47] reported undocumented voids, including the Big Void (BV) and the North Face Corridor (NFC) [48,49]. The NFC is compatible with a 7.5° IER lane that could have linked a temporary access to the main entrance—unlike external ramps that would obscure the north face. Methodologically, the IER and Houdin models diverge: IER sensitivity analyses reveal two tight geometric correlations. First, constant-slope 7.4–7.5° helices align with cavities C1–C2 and notches N1–N3. Second, an independent test finds 7.4° yields a statistically significant correlation with course-thickness changes, suggesting ramp turns were followed by architectural re-leveling. Together, these lines of evidence imply an operational gradient in the 7.4–7.5° range. By contrast, Houdin's 2024 model requires a dual-slope ramp and still presupposes an unobserved continuous internal tunnel. NE-corner evidence (Fig. S12.1) also fits an open-air haul lane better than a face-internal entrance; a rubble-filled edge band is testable, whereas an internal spiral demands features not seen at current resolution. Since muography coverage is weakest near edges, non-detection is not decisive. Consistent with Ivashov et al. [50], we treat muography as a first pass requiring corroboration; our edge-band fill prediction implies rubble-rich, mortar-bearing signatures (lower velocities/resistivity), though moisture can mask them (Sharafeldin et al. [51]). A predictive scan grid is provided (Tables S12.1–S12.2). The IER fit is geometric and predictive, remaining testable via targeted surveys.

Beyond high-frequency limestone flow, a comprehensive model must also solve the low-frequency, high-mass placement of the King's Chamber granite megaliths. Our approach supports feasibility via a decoupled operation on wide terraces—supported by heavy-haul experiments (Ayrinhac [52])—that requires a manageable workforce and is made safer by capstans. Systemically, isolating granite moves from the main limestone stream prevents bottlenecks and reinforces overall logistical plausibility, suggesting that granite handling is not rate-limiting. By contrast, single-lane ramp theories demand massive auxiliary works that block workflows. The IER's stage-lifting method, though longer in path, is decoupled, uses no significant auxiliary materials, and distributes effort modularly—consistent with experimental methods and the King's Chamber layout. While our model does not require the Grand Gallery counterweights proposed by Houdin, comparable incremental methods with short ramps and levers were noted by Arnold, Houdin, and Stocks. We therefore infer that granite handling was unlikely to be rate-limiting under the modeled procedure and suggest targeted field checks (wear patterns, shoring sockets) and controlled trials to refine cycle-time bounds.

In addition to logistical coherence, structural safety during construction is paramount. Our linear FEA illustrates comfortable stress margins under staged self-weight, consistent with Old Kingdom limestone properties: stress envelopes and settlements remain well below conservative bounds, and the temporary edge channel produces only localized, low-magnitude perturbations that dissipate after backfill. Geotechnical studies (Abd El-Hady et al. [53], Fahmy et al. [54], Hemeda & Sonbol [28]) and petrographic analyses (Mahmoud & Abdel-Hady [55]) support material variability and our conservative margins. Consistent with large-scale stability work (Weiss et al. [29]) emphasizing cumulative stresses, within the linear-elastic, monolithic assumptions, we see no progressive failure, reinforcing that the edge-ramp posed no structural risk under the specific assumptions. This aligns with heritage-engineering guidance prioritizing global response measures over local stress peaks at geometric singularities. FEA supports plausibility—not proof—and excludes dynamic loading, jointed-rock mechanics and accident scenarios, which could alter local responses. Because the ramp is peripheral and distant from ScanPyramids voids, our analysis focused on edge stability, unaffected by those features. Unlike internal spirals (tensioned roofs) or large external ramps (retaining works), the IER incurs no such penalties. This analysis assumes an idealized, monolithic limestone body and does not account for jointed-rock mechanics, potential slip-plane failures, or dynamic loading from construction activities, which could significantly alter local stress fields. Future models should integrate these features; targeted non-destructive tests can discriminate rubble backfill from continuous voids.

The model's internal consistency is robust, but its greatest strength is external testability: the IER is falsifiable. The computational model maps predicted edge-band backfill zones, enabling targeted non-destructive tests: (i) geophysics (GPR/ERT) to reveal ramp channels with rubble signatures—recent surveys report anomalies near the base [56,57,58]; (ii) petrographic and microwear analysis of corner blocks for turning wear; (iii) testing the prediction that no large external ramp bases exist; (iv) muon radiography (e.g., SciDEP at Khafre's Pyramid [59]) to detect heterogeneous rubble signatures in C1, C2 and N1–N3, not uniform voids; (v) endoscopy to find maneuvering wear; and (vi) holographic/ground radar, following Ivashov et al. [60], to discriminate debris fills from continuous voids. We treat muography as a first pass needing corroboration: edge-band fills imply rubble-rich, mortar-bearing signatures rather than continuous voids.

In considering alternative mechanisms, Scheuring's [61] counterweight/pulley reinterpretation is stimulating, but several gaps remain relative to the IER framework. Its critique of ramps as uniformly "single-file" overlooks the IER's parallel, multi-lane arrangements with explicit headway control. The proposal also relies on very low friction and high rope tensions through non-rotating fairleads lacking direct Old Kingdom evidence, and it does not specify lane-level cadence or the number of concurrent stations required to meet a ~27-year schedule; the assumed symmetry is not anchored in current anomaly patterns, and completing the apex appears to require additional service spaces that are presently undocumented. By contrast, the IER provides a continuous base-to-apex logistics model with quantified headways, minimal temporary works, and falsifiable edge-fill signatures. While elevator-style hypotheses lack repeatable archaeological signatures, Scheuring's approach could be a compelling solution for granite megaliths; should geophysical surveys confirm the necessary vertical shafts, a hybrid model—internal elevator for megaliths with parallelized IER for bulk limestone—would be coherent with our findings.

Beyond theoretical comparisons, the following limitations define our evidence and set priorities for validation. First, the IER implies a level of logistical planning and engineering acumen often under-emphasized for the Old Kingdom; its adaptive strategies suggest algorithmic thinking and applied physics not typically foregrounded in that period. We also acknowledge the following limitations: absence of direct archaeological traces; dependence on experimental friction data; simplified headway assumptions; logistics-based findings and staged linear-elastic FEA with literature parameters (potential bias/uncertainty); use methods like DEM; verified global mesh convergence but mesh-sensitive local maxima; unresolved nonlinear/jointed behavior; inferred crew sizes/equipment; and generalizability for Fourth-Dynasty step-core pyramids. These limits do not overturn IER plausibility—results are robust across $\mu$/grade ranges and parallelization policies—but they point to concrete tests: targeted geophysics, inspection for corner wear, and improved FEA. Finally, quarrying technology remains a constraint: Old Kingdom arsenical copper tools (Odler et al. [62]) may impose added logistical limits.

Ultimately, the IER framework demonstrates how quantitative modeling can transform debates on ancient engineering from speculation to testable science. The results indicate that the edge-ramp model reproduces observed anomalies while remaining consistent with structural and logistical constraints. These outcomes frame the broader implications for construction logistics and archaeological interpretation explored in the following conclusions.

**Conclusions**

This study advances a testable, data-driven account consistent with the evidence considered for Khufu's construction: an edge-integrated, zero-footprint ramp that preserves survey control, sustains minute-scale headways under adaptive parallelization, and appears structurally plausible under staged FEA assumptions. Our main contribution is a unified, reproducible framework—integrating parametric geometry → logistics → stress checks—that links per-course operations to reign-scale timelines while remaining consistent with archaeological signatures (e.g., absence of massive external works) and guiding targeted field verification. The framework provides a fully reproducible, open-source pipeline for hypothesis testing in ancient construction, yielding falsifiable predictions—edge-band low-density zones and corner wear—amenable to targeted GPR/ERT and muography. Methodologically, nonlinear/jointed FEA, in-situ friction calibration, and crew-rotation modeling will further strengthen IER assessments across Old Kingdom projects.

**Declarations**

**Data availability:** All the raw and processed datasets (Excel/CSV tables, geometries, finite-element meshes, solver logs, Monte Carlo outputs, and 3D visualizations) are archived at Zenodo under DOI https://doi.org/10.5281/zenodo.16732345 (v1.0.10). The deposit includes a README file that documents and software versioning, dependencies, and workflow steps in line with FAIR principles. No new terrestrial laser scanning or photogrammetric surveys were conducted for this simulation-based study. The pyramid's geometry was parametrically generated from canonical dimensions published in the literature. Accordingly, no coordinate reference system (CRS/EPSG), ground control points, or registration error metrics are applicable to this work.

**Code availability:** Detailed software versions, scripts, parameter files, and fixed random seeds are consolidated in Methods — Software & Reproducibility. The recursive algorithm and simulation scripts are archived in the same Zenodo repository under an open license, enabling full reproduction of the figures, tables, and sensitivity analyses.

**Acknowledgments:** Not applicable

**Funding**: This research received no external funding. The author will cover open access fees.

**Author's contributions** V. Rosell conceived the study, developed the simulation code, performed the analyses and wrote the initial draft. The author also prepared the figures, tables, and supplementary code. The author read and approved the final manuscript.

**Competing interests** The author declares that he has no competing interests.

**Additional information:**

- Supplementary Video 1. Time lapse of the single-ramp construction sequence (29 s).
- Supplementary Video 2. Time lapse of the 4-ramp construction sequence (34 s).
- Supplementary Video 3. Time lapse image of the adaptive ramp sequence (34 s).
- Supplementary Material

(CRediT: Conceptualization, Methodology, Software, Formal analysis, Visualization, Writing – original draft, Writing – review & editing: Vicente Rosell)

**Note**: All 3D representations presented in this article are direct exports of the geometric model generated by the algorithm in the Unity 3D engine (v 6000.2). They are presented without photorealistic rendering to accurately and directly depict the simulation results. Some renderings show the pyramid with the Tura limestone casing, whereas others do not, to aid visual comparison with its present-day appearance. LLM tools were used solely for language polishing; no content or data were generated by AI.

# A computational framework for evaluating an edge-integrated, multi-ramp construction model of the Great Pyramid of Giza

Vicente Luis Rosell Roig.

vicente.rosell@gmail.com

L'Alcúdia. Valencia. Spain.

## Supplementary Information

**Table of Contents**

## S1. Construction ramp theories schemes

Prevailing ramp theories, while plausible in principle, encounter critical flaws when scaled to the magnitude of the Great Pyramid.

Straight external ramps, despite their conceptual simplicity and moderate gradient (≈7–8%), are problematic at scale, requiring a ~1.5 km ramp that exceeds the Giza plateau. Hölscher [4] estimated the ramp's auxiliary mass at 20–40% of the pyramid's—monumental in itself—and its removal would have left scars, which are not observed. Brichieri-Colombi's variant [5] cuts the volume to ~6% but still obstructs casing and survey control at upper tiers and creates traffic conflicts. Thus, the straight external ramp is unsuitable for upper construction, failing to reconcile its mass, footprint, lack of removal evidence, and the required minute-scale headways.

Encircling/zig-zag designs wrap the haul path along a single face to reduce the site's footprint, but their repeated ~180° switchbacks with multi-ton blocks are slow, hazardous, and obstruct corner setting and survey control, making the required minute-scale headways unattainable [6]. Despite footprint savings, they prove impractical for construction at height.

While using less material than straight ramps, spiral ramps wind around the pyramid, hiding edges, blocking sightlines, and restricting corner access—issues noted by Lehner [7], Hawass and Lehner [8,9], and Dunham [10]. No archaeological traces of such a vast structure exist at Giza [9]. Consequently, spiral ramps compromise geometric precision and access while lacking evidential support.

Houdin [11] proposed an internal helical tunnel to cut auxiliary mass and keep faces clear, but it demands a continuous load-bearing roof unprecedented in the Old Kingdom and interprets the NE notch's L-shaped cavity as a turning station. Crucially, ScanPyramids muography detected no continuous spiral void, undermining the model's premise [11,12]. The theory lacks archaeological support and imposes structural demands inconsistent with Old Kingdom practice.

Seeking to embed routes while avoiding massive external works, Klemm & Klemm's integral helical route from opposite corners (~6°, ~4–5 m wide) minimizes material and leaves no trace, requires unobserved recesses, and incorrectly assumes horizontal layering instead of the attested step-core technique. Its narrow geometry restricts turning and two-way traffic, making granite transport impractical. With Müller-Römer [14] estimating a >50-year build time—historically implausible—the model fails on evidential, geometric, and temporal grounds. Although method shares a conceptual foundation, its original description was brief, lacking detailed geometry, logistical analysis, or quantitative validation—a gap our computational framework now fills.

As a terrace-based alternative, Müller-Römer's short, tangential ramps lower auxiliary volume but need constant reconfiguration, tight switchbacks, and interrupt precise outer-course seating. They presume dense winch arrays and heavy levering—hardware unattested for Old Kingdom haulage—and lever-only cycles are too slow for minute-scale headways; high-course ramp beds are scarcely evidenced [14]. This makes the method logistically inadequate at the pyramid's scale.

This section illustrates six ramp families reviewed previously. Each schematic shows plan and elevation views with annotated slope, useful path width, and representative turning radii; shaded zones indicate auxiliary footprint and corner/survey obstructions. The panels provide a visual glossary for non-specialists and are intended as reference diagrams, not as evidence of existence (see main-text references for sources).

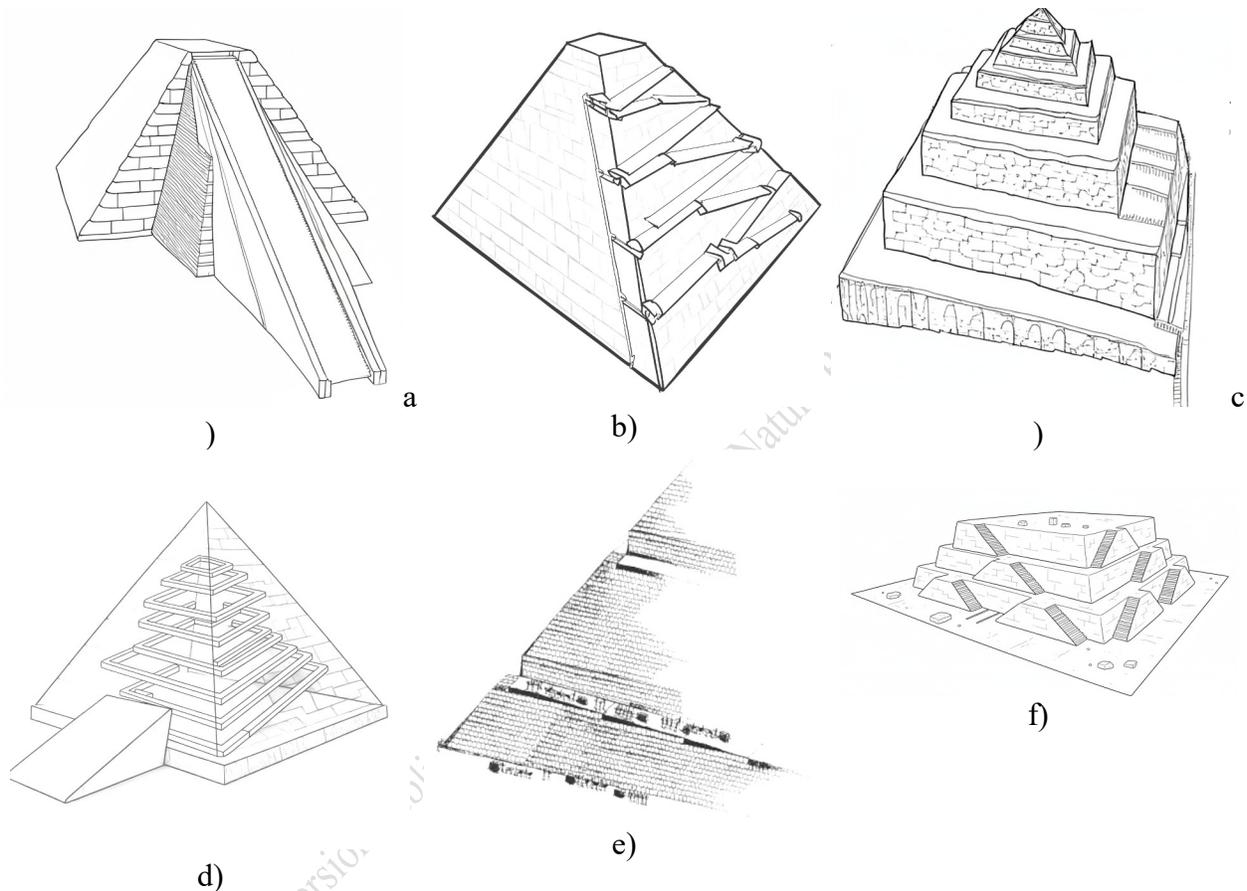

Fig. S1.1: (a) straight frontal external ramp—simple alignment but prohibitive auxiliary mass and footprint at upper courses; (b) zig-zag / encircling ramp—reduced bulk but tight switchbacks that obstruct corners and depress throughput; (c) external spiral ramp—wraps the monument yet obscures edges and survey control, with no convincing traces at Giza; (d) internal spiral ramp—zero external footprint but requires continuous internal voids and turning stations not supported by muon data; (e) integral helical ramp from opposite corners—embedded routes with narrow width and turning issues, lacking sealing evidence; (f) step-core with short/tangential ramps—lower auxiliary volume but slow cycles and minimal archaeological support at high courses

## S2. Basic IER schematics: aperture, closure sequence, and configuration views

This section compiles the methodological schematics moved from the main text: (i) the 3×6 edge-integrated aperture with annotated width ($W_{mask}$), clearance ($H_{mask}$); (ii) the inside-out, top-down closure of the temporarily omitted perimeter courses (panels a–c: interior filler, bench/step blocks, exterior facing); and (iii) configuration views (aerial/frontal) indicating helical direction and face changes at corners. Diagrams are schematic (not to scale), serve as a visual glossary for non-specialists, and mirror the formal rules and parameters defined in Methods; they illustrate the modeling workflow and do not constitute independent archaeological evidence.

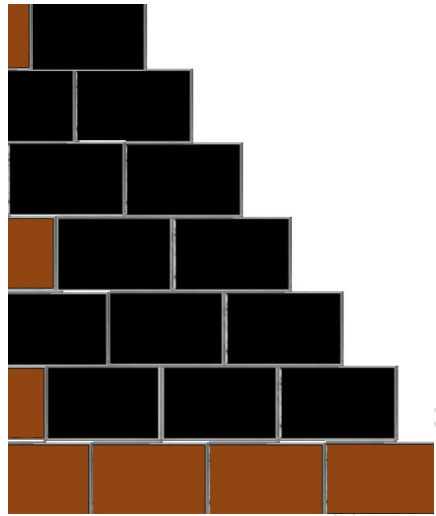

Fig. S2.1. Baseline 3×6 edge-ramp aperture. Black blocks mark the 12 perimeter blocks temporarily omitted during the ascent to form the helical haul channel; brown blocks are placed courses. The aperture yields a corridor of $W_{mask}$=~3.8 m width and $H_{mask}$=4.26 m clearance at ~7–8° grade; omitted blocks are backfilled after reaching the next target elevation (zero external footprint). (see Methods: Fundamental principles of the integrated edge-ramp (IER) model). Schematic not to scale.

Ramp decommissioning:

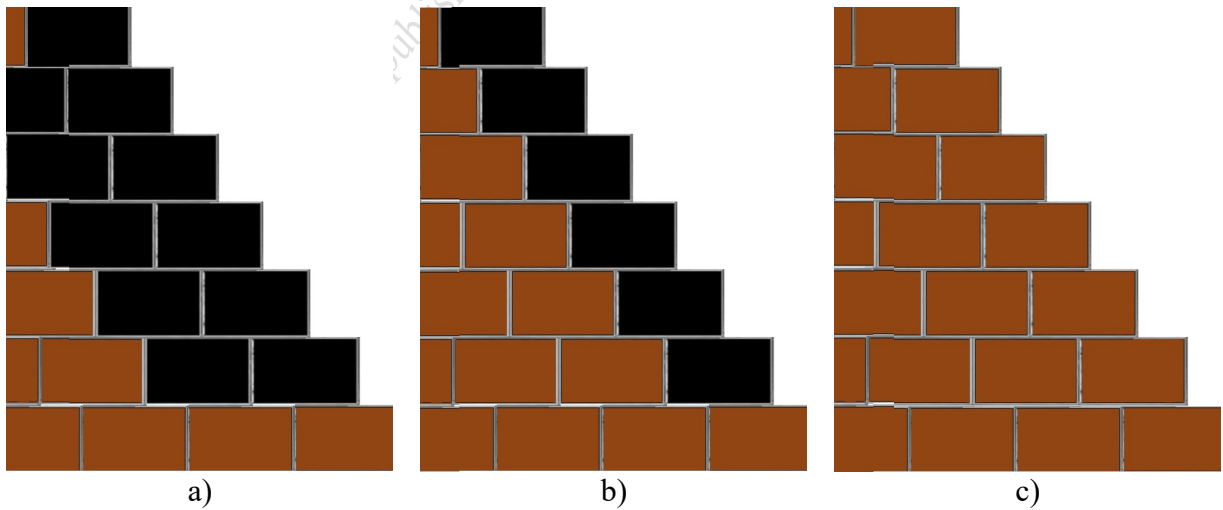

a)                              b)                              c)

Fig. S2.2. Top-down closure of the edge-integrated haul channel, baseline 3×6 edge-ramp aperture. (a) Interior filler placed first to create the bearing surface; (b) installation of step/bench blocks;(c) placement of external facing blocks flush with survey lines. The sequence closes the temporarily omitted perimeter courses, removes the temporary leveling materials, and restores smooth faces with no external footprint. (see Methods: Construction and Decommissioning Sequence). Schematic not to scale.

Larger corner recesses were first packed with smaller stones and gypsum mortar to create stable seating surfaces for facing blocks; temporary leveling materials were progressively removed as backfilling advanced. This rubble-and-mortar packing parallels documented Egyptian methods for closing temporary voids, as discussed by Lehner [7] and other scholars [9,10].

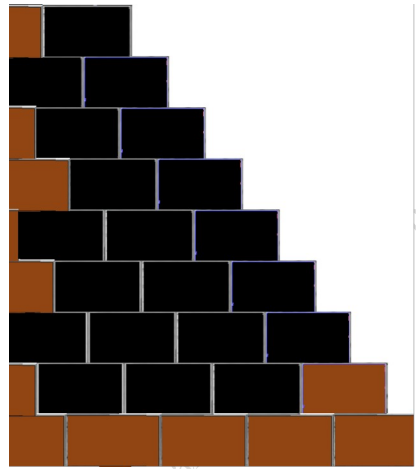

Fig. S2.3. Schematic 4×8 edge-integrated aperture with parapet. This creates a corridor of $W_{lane}$=~3.8 m width and $H_{mask}$=~5.68 m height, with a 0.71 m height protective parapet. Black blocks mark the temporarily omitted perimeter units that define the open-air haul corridor along the edge; grey blocks are placed courses. The brown block is a low parapet intentionally left in place to contain sledges and preserve a safe survey line. After decommissioning, the omitted units are backfilled top-down, restoring smooth faces with no external footprint (see Methods: Design refinements; The Edge Ramp with a Parapet). Schematic not to scale.

## S3. Core geometric inputs (recursive IER algorithm)

This block centralizes every parameter and seed required to reproduce geometry, logistics, Monte-Carlo timelines, and FEA runs. Canonical machine-readable files are archived at the Zenodo record cited in Data Availability.

Global geometric parameters (Great Pyramid) by Lehner[7]:
1. base_size$_{initial}$ (B): 230.363 m
2. Height$_{total}$ (H): 146.50 m
3. pyramid_inclination (β): 51.84°

Block parameters (assumed):

4. block_height ($h_b$): 0.71 m
5. block_width ($w_b$): 1.27 m (square base)
6. block_mass: 2,267.96 kg

The core blocks varied in size and shape and were often irregular. In the model, we used standardized units derived from the work of Lehner [7] for clarity and reproducibility, and we accounted for archaeological variability through tolerances, temporal channel leveling, and sensitivity analysis.

Other parameters:

1. friction_coefficient ($\mu$): 0.2
2. ramp_inclination ($\theta_r$): 7°
3. ramp_width ($W_{mask}$): 3.8 m
4. ramp_width_useful ($W_{lane}$): 3.8 m, denotes the useful haul-lane width (measured between inner parapet and inner curb)
5. ramp_height ($H_{mask}$): 4.26 m, vertical aperture height required
6. sideSlopeAngle: 60° (from ground) for straight and spiral ramp slope angle for the embankment
7. spiralRampSeparation: block separation from the pyramid in the spiral ramp
8. internalRampStraightRampHigh : straight ramp maximum height for the internal ramp
9. Ramp method: straight, external, internal, integrated (default)
10. Method: single / 2-ramps / 4-ramps / adaptive
11. headway_baseline: 4 minutes
12. working_year: 10-hour day, 6 days per week, when Giza was above the flood level, so no seasonal pauses occurred.
13. clearance (p): 0.35 m (is the lateral safety clearance per side)
14. speed: ramp $v\uparrow$ = 0.15 m/s; terrace $v\rightarrow$ = 0.20 m/s
15. Corner delay model: lognormal with median 2.8 min, shape $\sigma = 0.35$ (right-skew, multi-step maneuver).
16. Cell/separation rule: one team per cell; cell length = team length + sledge ($\approx 3$ m) + dynamic buffer (15 m); team length from team size (double file, 1.5 m spacing).
17. Workforce sizing per block: from force $F_{incline} = m\ g\ (\sin\theta_r + \mu\cos\theta_r)$ with $\mu$ scenario (below) and ergonomic 200–300 N/person.

Monte-Carlo / bootstrap.
- Trials N = 10,000; seed = 42.
- Perturbations: speed ±25%; headway ±1 min; random stops MTBF ≈ 6 h; repair 5–12 min.

FEA:
- Material (Eocene limestone): E = 35 GPa; $\nu$ = 0.25; $\rho$ = 2600 kg·m⁻³; load $\rho g$ = 25,506 N·m⁻³.
- Boundary conditions: base fully fixed $U_x = U_y = U_z = 0$.
- Meshes (TET10): fineness 3 ≈ 0.54–0.60 M elems; fineness 5 ≈ 0.84–0.90 M elems;

TRIA6 faces.

Outputs:

- W = work (GJ)
- F = force (N)
- T = time, $T_{ramp}+T_{corner}+T_{terrace}$ (minutes / working years)
- L = block displacement length, $L_{ramp}+L_{terrace}$ (m). $L_{ramp}$ denotes the helical ramp length actually traversed during the edge-ramp phase. It is obtained from the parametric model's exported path at grade $\theta_r$, from the base to the apex. $L_{terrace}$ denotes the horizontal displacement over the terrace until the final block location.
- Monte-Carlo: medians with 95% percentile interval
- $Pyramid_{Volume}$: Pyramid volume in $m^3$
- $Embankment_{Volume}$: Embankment volume in $m^3$ for the straight ramp

**S4. Algorithm's core functions geometry**

The algorithm's core function computes the geometry at each construction stage. At each recursive call, it reads the current base size and remaining height to calculate the dimensions for the next, smaller stage. This formalizes the "each iteration supports the next" logic and is reproducible from the provided inputs.

The two central equations are as follows (Fig. S4.1):
- Height gained per stage (h) - Equation (1): The vertical distance covered in one ramp segment, modeled as a right triangle, calculated as $h = (base\_size\_current - sep) \cdot \tan(\theta_r)$, where base_size is the side adjacent to h, $\theta_r$ is the ramp inclination angle, and sep is the base reduction in each stage.
- Base reduction per stage (sep) - Equation (2): To preserve the pyramid inclination angle with increasing h, the base decreases by $sep = h/\tan(\beta)$. The total base reduction for the next stage is 2·sep and derived from similar triangles preserving face angle β.

The function then calls itself with the updated parameters:
- $base\_size_{(n+1)} = base\_size_{(n)} - 2 \cdot sep$
- $Height_{(n+1)} = Height_{(n)} - h$  (remaining height).

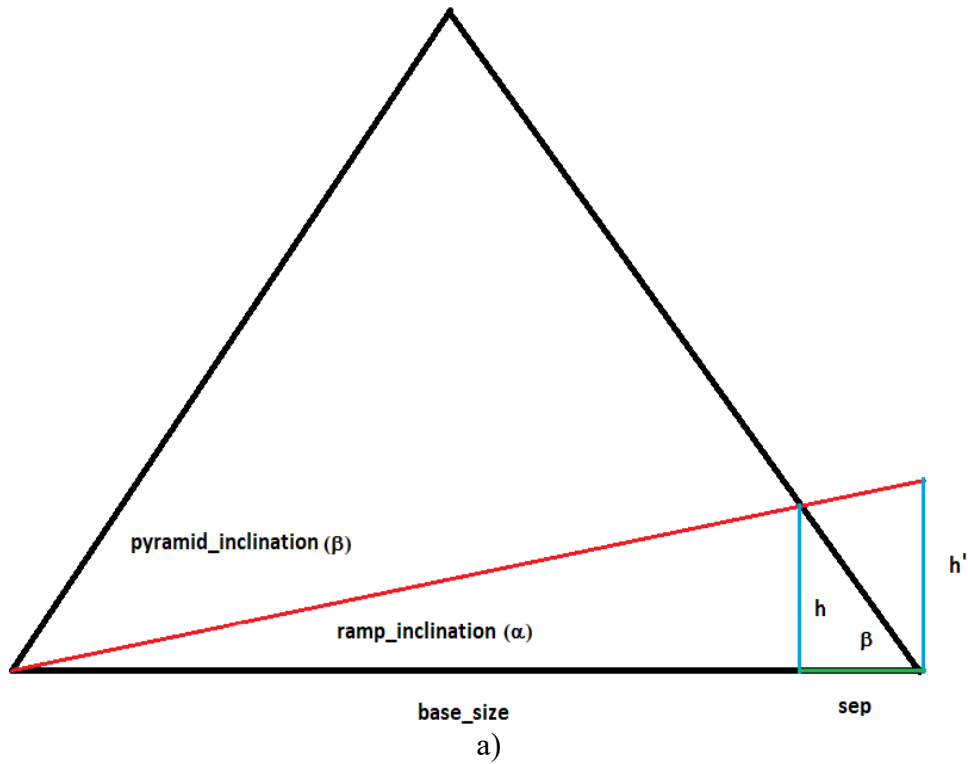

a)

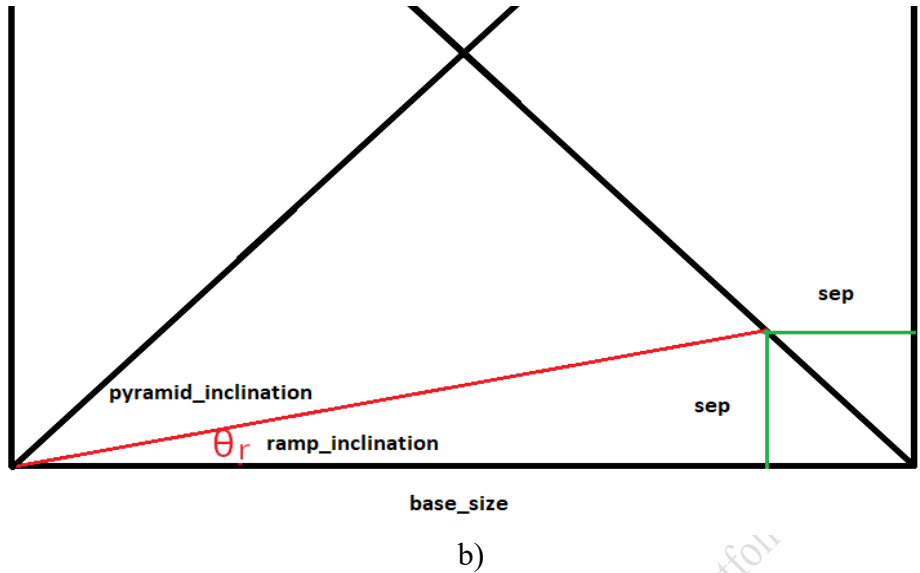

b)

Fig. S4.1. Stagewise geometric relationships used in the parametric update. Schematic not to scale. (a) Height gained per stage (h). One ramp segment is modeled as a right triangle; the vertical gain is h = (base_current − sep) · tan(θ$_r$), where θ$_r$ is the ramp inclination angle and base_current is the adjacent run at the current stage. (b) Base reduction per stage (sep). To preserve the pyramid face angle β as h increases, the base shrinks by sep = h/tan(β) on each side; the total reduction to the next stage is 2·sep.

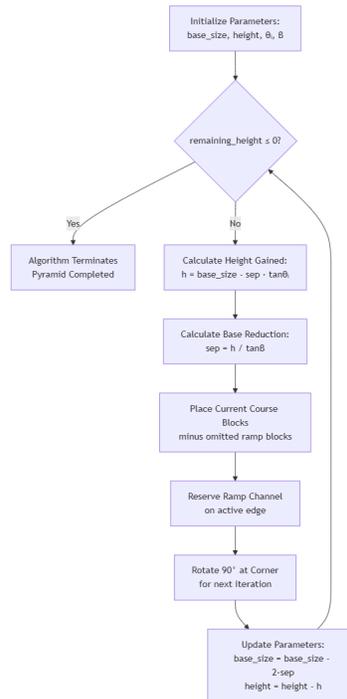

Fig. S4.2. IER algorithm schema

Within each stage (n), the algorithm iterates over the stone courses (c = 1…C(n)). For each course, it computes the full square tiling from input block module and joint parameters, then subtracts the cells intersecting the edge-ramp channel mask (baseline 3×6 blocks; 4×8 with a parapet at height). Stage height h and base setback sep are obtained from Eq. (1–2); courses per stage are C(n) = round(h/$h_b$), where hb is block height. If a non-integer value results, the algorithm creates a block with the remaining 3D dimensions and corresponding mass. The channel mask is fixed within a stage; its total omissions depend on $\theta_r$ via h($\theta_r$) (i.e., how many courses the mask spans), not on a per-course basis.

Block count and mask subtraction (all integer counts):

$n_{side}(n) = floor(B_n /w_b)$, where $B_n$ = base_size at stage n (m); $w_b$ = block_width (m)

$N_{full}(n) = [n_{side}(n)]^2$

$N_p(n, c) = N_{full}(n) − N_{mask}(n, c)$     (Equation 3)

We round to integers before mask subtraction; edge-partial cells are discarded. Eq. (3) is dimensionless. Symbols/units are summarized elsewhere in Methods.

Helical advance and corner turns: The corridor runs along the active edge and turns 90° at each corner; one perimeter cycle ascends by the stage height and reduces the base by 2·sep, after which the recursion advances to stage n+1 with updated {B, H}.

Per-step outputs (units in SI): (i) edge-ramp length $L_r$ (m) for the active edge; (ii) placed blocks $N_p$ from Eq. (3); (iii) travel distance per block d = $d_{up}$ + $d_{platform}$ (m), decomposed into up-ramp and terrace translation; (iv) drag work W = F · d (J), using the haul model in Methods.

We quantify the temporarily omitted ('skipped') blocks forming the open-air haul channel by combining the channel aperture with its helical length. For the baseline 3×6 aperture, step-core geometry means 12 block positions are omitted per longitudinal 'slice' of the channel. With block module $w_b$=1.27 m and helical centerline length L (for 7° we use L≈1,129 m), the single-ramp omission count is 12×[L/$w_b$] (Equation 4).

The model computes the force required to haul a block on a wooden sledge under two primary conditions:

Inclined Transport (Ramp): The force must overcome both gravitational and frictional components:

$F_{incline}$ = block_mass·g·(sin $\theta_r$ + μ cos $\theta_r$),  (Equation 5), where the mass block_mass is ≈ 2.27 tons; g is gravitational acceleration, with a value of 9.81 m/s²;

Horizontal Transport (Platform): The force must overcome friction only:

$$F_{horizontal} = \mu \cdot m \cdot g, \text{ (Equation 6)}$$

The total work per block is calculated as:

$$W = F \cdot d, \text{ (Equation 7)}$$, where d is the total travel distance obtained from the geometric model, comprising both ramp ascent (d↑) and platform transfer (d→) components.

The backfilled ramp channel volume, $V \approx \int A(s)ds$, is approximated by a half-triangular section $A \approx (W_{mask}d)/2$ with $W_{mask} \approx 3.8$ m and $d \approx 0.71$ along a helical length $L \approx 1.2$ km. Since the ramps were created on the edge and subsequently filled, their material trace amounts to $<0.063\%$ ($V \approx 1,633$ m³) per ramp.

Termination criteria: The recursion terminates when either (a) the remaining height $H_{\{n+1\}} \leq 0$ (apex reached), or (b) the channel cannot be safely instantiated, i.e., $B_n < W_{lane} + 2 \cdot p$ or $h < H_{mask}$. (Fig. S4.2).

Safety instantiation check. In (b), $W_{lane}$ denotes the clear haul lane (baseline $W_{mask} \approx 3.8$ m), and $H_{mask}$ the vertical aperture required by the active mask ($\approx 4.26$ m for 3×6). These quantities are reported by stage in the exported geometry to document feasibility at height.

Determinism and I/O: The procedure is fully deterministic from the inputs {H, B, β, $\theta_r$, $w_b$, $h_b$, $W_{lane}$ } and emits stagewise geometry (paths, turning states) consumed by the parametric 3D Scene software and analyzing by FEA (see section S3).

The algorithm can be applied to other Old Kingdom pyramids—Khafre, Menkaure, and the Red Pyramid—by adjusting geometric parameters. For the Bent Pyramid, the slope change is modeled by resetting the ramp angle at the break point. Still, ramp theories must be studied individually for each case. Beyond Khufu, the work contributes a replicable, evidence-linked framework for modelling other Old Kingdom pyramid projects, clarifying how ancient organizations coordinated labor, safety, and metrology under material constraints.

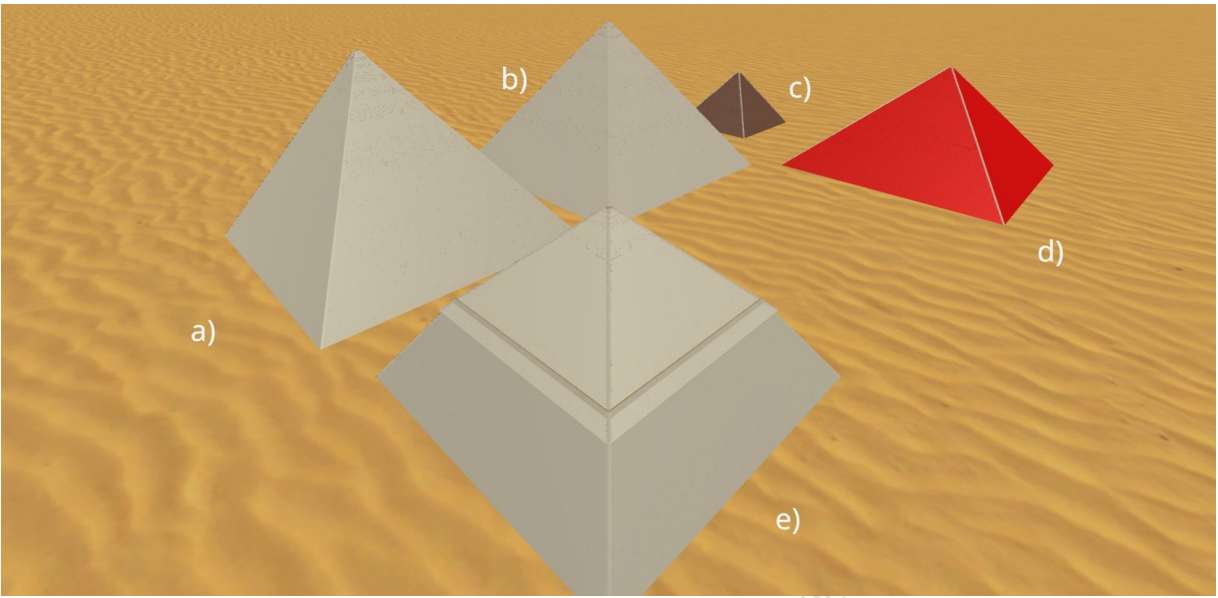

Fig. S4.3. Parametric 3D renderings: Khafre (a), Khufu (b), Menkaure (c), Red (d), and Bent (e). The Bent Pyramid includes a break in slope. Colors are schematic for clarity.

A significant strength of the computational pipeline is its inherent generalizability, a direct result of its parametric design. The entire analytical process—from the recursive generation of the geometry to the discrete-event logistics simulation and the final finite-element analysis—is driven by a set of core inputs. This allows the framework to be readily adapted to model other Old Kingdom monuments by simply substituting the canonical parameters of the target structure. For example, to generate a complete construction analysis for the Pyramid of Khafre, the model's inputs would be updated to a base size of 215.25 m, a height of 143.50 m, an inclination of 53.17°, and block height 0.70 m. Likewise, for the Pyramid of Menkaure, these parameters would be adjusted to its specific dimensions, base size of 108.50 m, a height of 65.50 m, and an inclination of 51.34°, and block height 0.65 m. This parametric flexibility makes the computational model a powerful and reproducible template, capable of generating equivalent logistical timelines and structural assessments for a range of Fourth-Dynasty pyramids, thereby enabling systematic comparative studies of their construction methods. These specific outputs—ranging from the expected elevations of corner turns to the material signatures of backfilled ramp channels—serve as critical, falsifiable points that can directly guide future archaeological and geophysical investigations, moving the debate from theoretical possibility to empirical verification.

## S5. Continuous helical path for IER model

A continuous helical path emerges by reserving a narrow edge channel of temporarily omitted perimeter courses, evident in both aerial and frontal perspectives for each model.

Single ramp model:

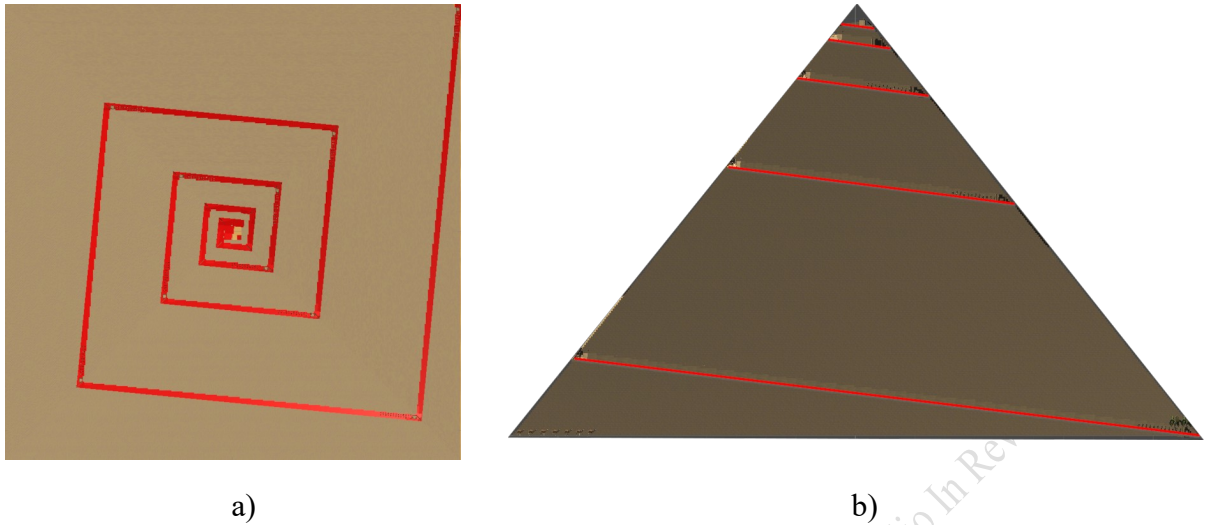

a)                                    b)

Fig. S5.1. Single integrated edge-ramp (IER) configuration—two perspectives. (a) Aerial view: a narrow red edge channel of temporarily omitted perimeter courses forms a continuous helical haul path along the pyramid edge. (b) Frontal view: faces remain exposed for survey and casing while the red channel advances; after decommissioning

Four-Ramp model:

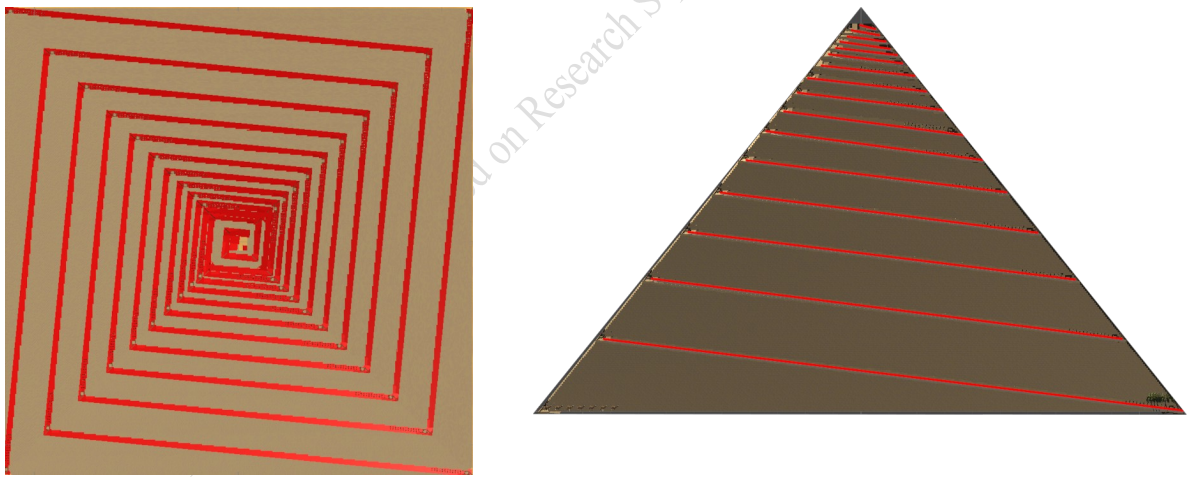

Fig. S5.2. Four Integrated edge-ramp (IER) configuration—two perspectives. (a) Aerial view: a narrow red edge channel of temporarily omitted perimeter courses forms four continuous helical haul paths along the pyramid edge. (b) Frontal view: faces remain exposed for survey and casing while the red channel advances; after decommissioning

Adaptive ramp model:

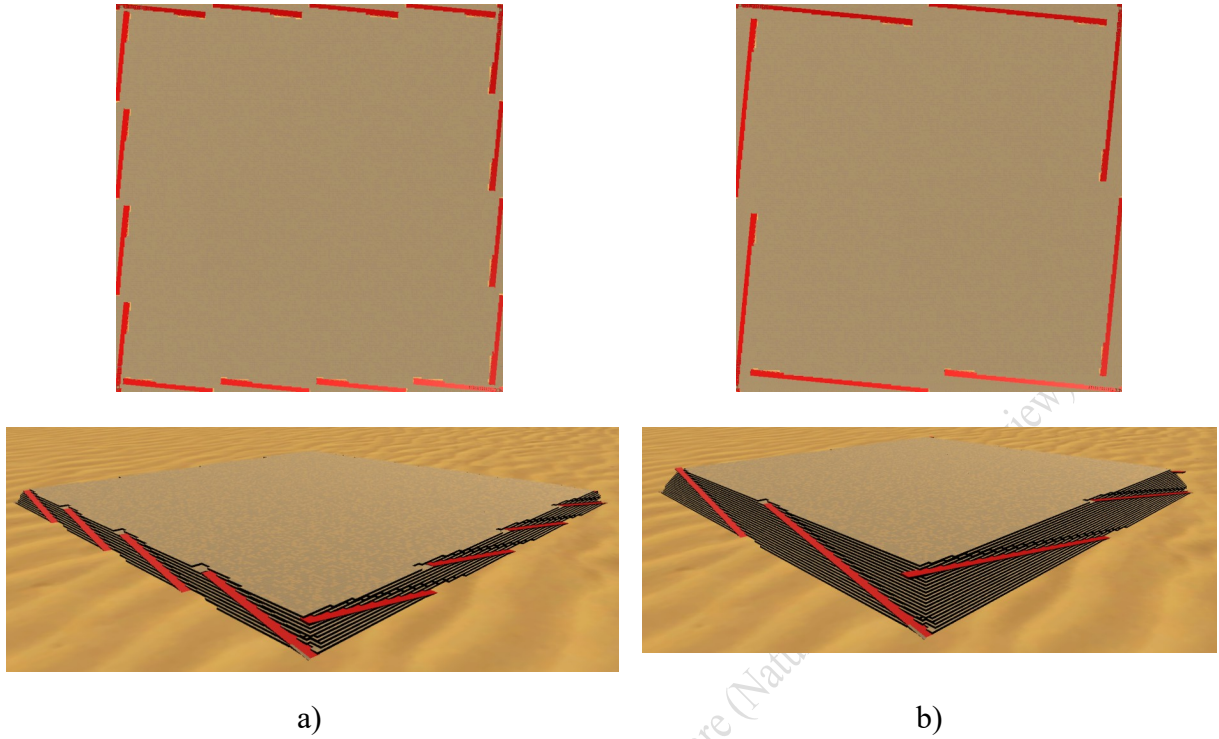

<div align="center">a)             b)</div>

Fig. S5.3. Adaptive Integrated edge-ramp (IER) configuration until course 20— Aerial view and frontal view perspectives. (a) a red edge channel of temporarily omitted perimeter courses forms 16 continuous straight paths along the pyramid edge. (b) a red edge channel of temporarily omitted perimeter courses forms 8 continuous straight paths along the pyramid edge

## S6. Design refinements figures

Edge ramp with parapet

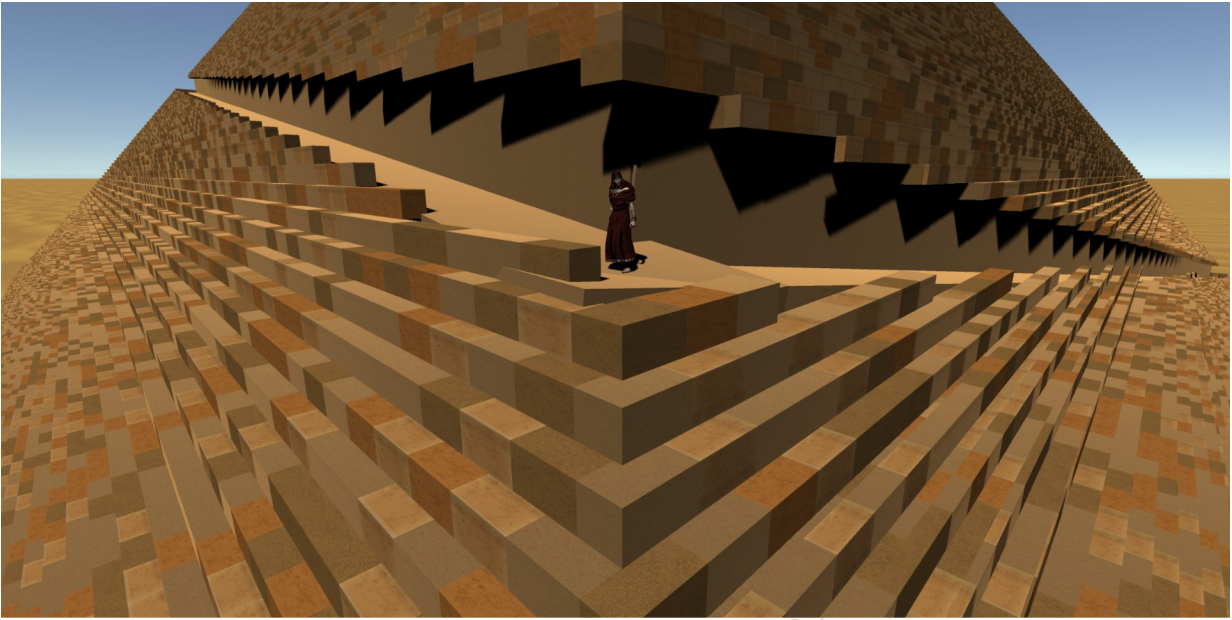

Fig. S6.1. Edge-integrated ramp with parapet (3D render). Open-air haul corridor formed by temporarily omitted perimeter courses along the edge; a low parapet ($h_b$=0.71 m) runs on the outer side to contain sledges and protect the survey line. $W_{mask}$ = 4.5 m, $W_{lane}$= 3.8 m, $H_{mask}$= 5.68 m, denotes the useful haul-lane width and height. Indicate functions (containment, staging, survey control). The parapet remains during ascent and is removed/backfilled during decommissioning, preserving a zero external footprint.

Optimized corner platforms

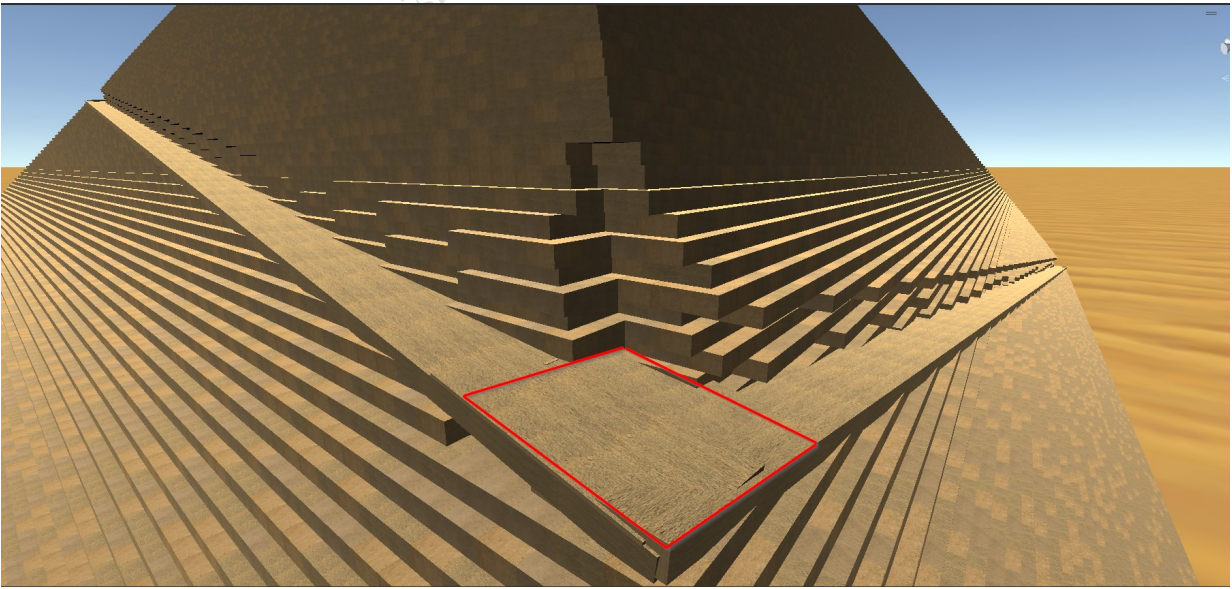

Fig. S6.2. Optimized corner platforms (3D render). Corner blocks are temporarily omitted to create a 4×4×9 platform incorporating the inter-course setback on both inner edges yields (≈5.64×5.64×6 m) platform, providing additional space for 90° turns, staging/queuing of incoming blocks, and survey access at height. The platform can be lowered if temporary shoring is used, and is backfilled during decommissioning to restore the exterior profile.

Geometric precision: the folding beam solutions

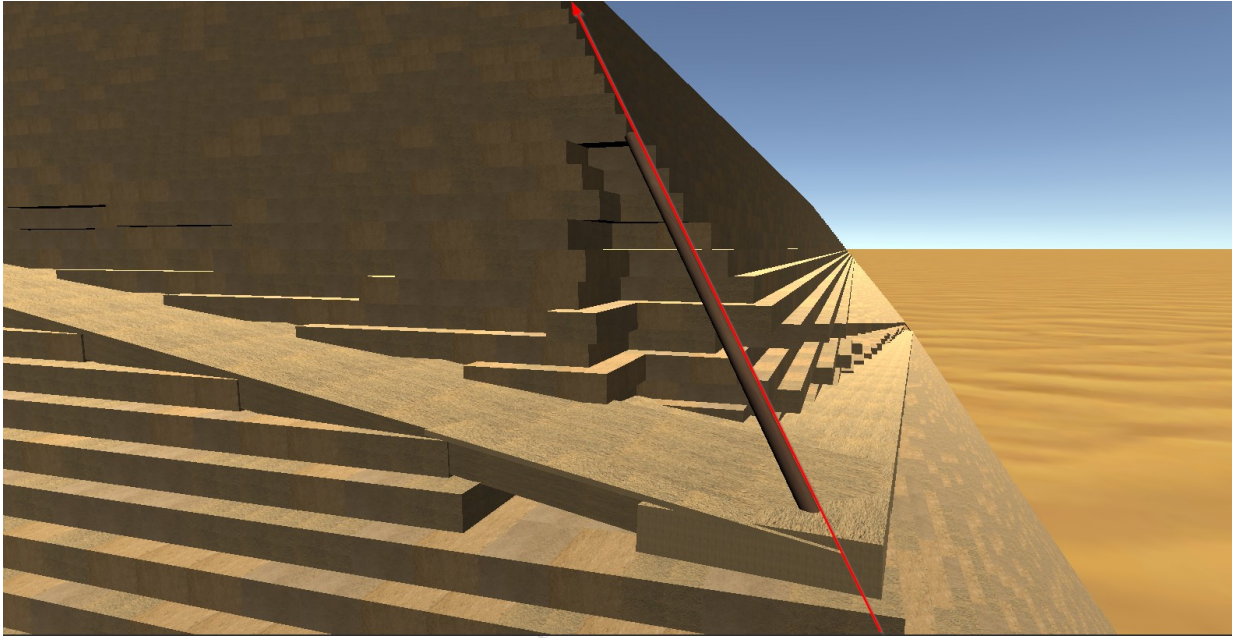

Fig. S6.3. Corner folding beam for survey control (3D render). Straight wooden beam seated in the corner notch and aligned to the true edge provides a continuous reference line while the edge channel is active. The beam is briefly hinged/lifted for block passage and then reset to indexed stops to preserve alignment. During hauling, the post functions as a simple fairlead: the haul rope is given a ¼–½ wrap to control a ~90° turn. Compatible with Old Kingdom sighting tools

**S7. Horizontal displacement improvement with multi ramps system figures**

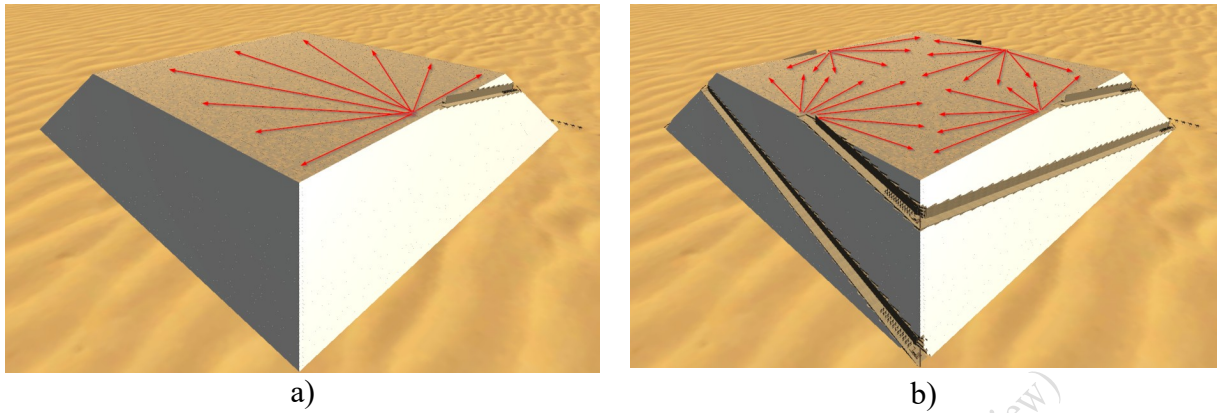

a)                                        b)

Fig. S7.1. Terrace displacement: (a) single-ramp vs (b) four-ramp (≈ 50% shorter mean terrace path optimization). Arrows trace the horizontal transfer on the working terrace from ramp egress to the final placement cell (direction and schematic length; not to scale). In (b), operating one channel per edge reduces the mean lateral path by ~50% relative to (a), as measured by average terrace distance per placed block over the same courses. Corner platforms are used for staging; geometry and settings match the baseline. Reduced terrace travel implies fewer platform crossings and lower local congestion under identical materials and rules.

## S8. Applicability of the Adaptive Edge Ramp to Macroterraces

The adaptive model also applies to macro-terraces—large stepped platforms proposed by Müller–Römer [14]. For Khufu, a first terrace could be ~197 m per side, rising ~11.5 m, with a ~5.75 m setback and ~80° faces, consistent with step-core (Stufenbauweise) interpretations.

The edge-integrated ramp fits naturally: instead of advancing course-by-course, the haul path proceeds along the edge to the terrace elevation. Each terrace functions as a broad working platform, facilitating heavy-block handling and providing additional space for turning, staging, and survey control (Fig. S8.1). The geometry remains compatible with the global ~51.8° face angle and the attested step-core construction.

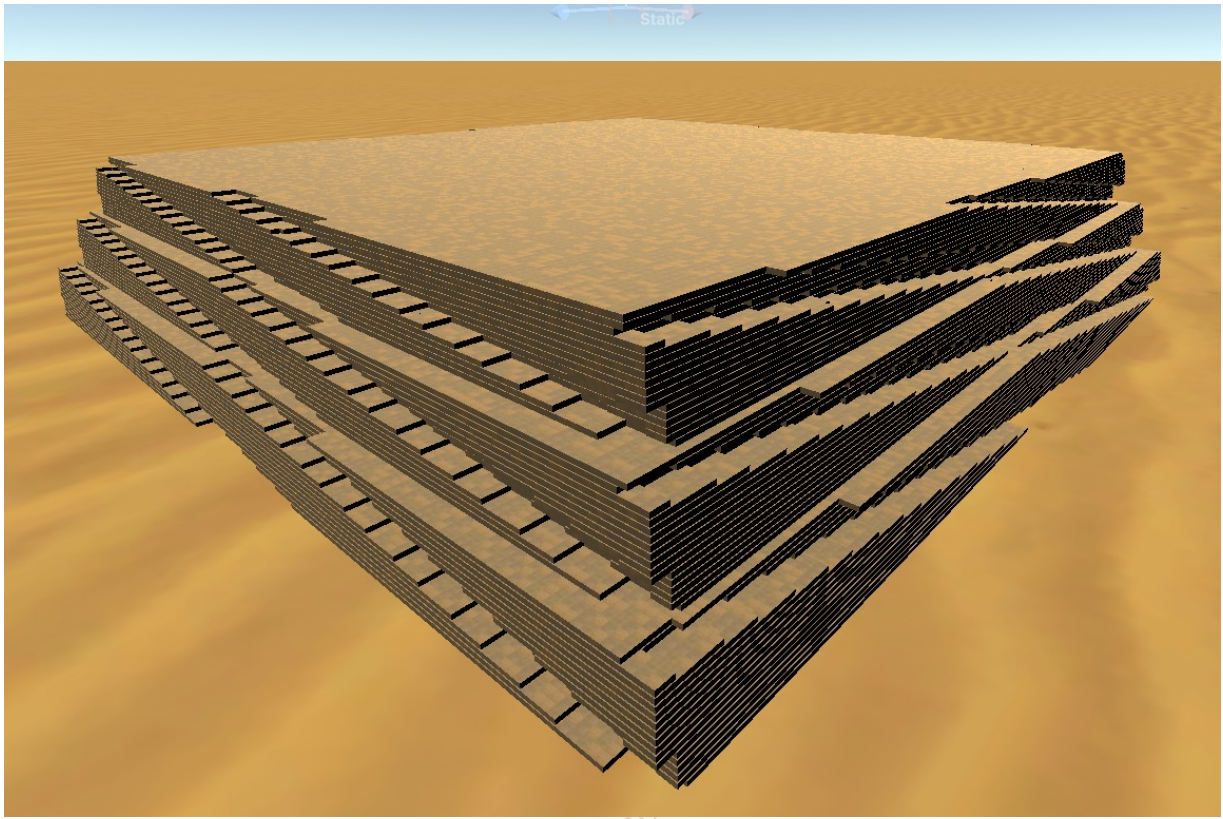

Fig. S8.1. Simulated 3D model of Khufu's pyramid with the adaptive edge-integrated ramp applied to Müller–Römer's macro-terrace concept.

This approach reduces vertical transitions and corner turns but increases intra-terrace travel. The wider platforms likely aided granite handling and improved safety. Ramps still derive from temporarily omitted perimeter courses that are backfilled afterward, leaving no external footprint.

The adaptive IER accommodates both course-by-course and macro-terrace increments, offering flexible sequencing within Khufu's step-core framework. A head-to-head assessment of Müller–Römer's micro-ramp/winch scheme versus the adaptive IER (e.g., throughput and structural margins) lies beyond the scope of this study, which focuses on edge-integrated logistics under Old Kingdom constraints.

## S9. Traffic and Queuing Model and Construction Time Analysis

This section documents the Monte Carlo traffic-and-queuing model used to propagate uncertainty in haul headways, friction, and calendars. We specify entities/events, buffer and corner mutual-exclusion rules, the dispatch policy, and baseline geometry. Two figures report (i) phase capacity distributions and (ii) project-duration ECDFs; we also tabulate phase totals for planning, quarrying, river transport, and seasonal pauses. Exact parameters, and non-defaults match Methods; code and seeds, software, and inputs are archived on Zenodo.

Here are the totals for each phase under the two project-length scenarios, plus the range across them (months → years):
Planning: 12 months (both) → 1.0 year (range: 1.0–1.0 y) (Arnold; Lehner/Hawass [6,7,8])
Quarrying: 188 → 258 months → 15.7–21.5 years (Klemm & Klemm [13])
River transport: 78 → 106 months → 6.5–8.8 years (Tallet, Ghoneim, and Sheisha [1,2,3])
Seasonal pauses: 40 → 54 months → 3.3–4.5 years (Nile seasons, festivals,...)

Detailed phase definitions, calendars, and a ±20% sensitivity table are consolidated in Table S9.1.

Table S9.1. Critical parameters, baselines, tested ranges, justification, and impact of a ±20% change on on-site duration (approximate, from model behavior and reported sensitivities).

| Parameter | Baseline | Range tested | Justification (source) | Impact of ±20% change on on-site duration |
|---|---|---|---|---|
| Headway (per active ramp) | 4.0 min | 3–7 min | Regulates separation for teams; derived from ramp lengths, team footprints, and corner delays; explicitly simulated in Monte-Carlo (N=10,000; seed = 42). | $\approx \pm 20\%$ (near-linear scaling with headway when ramps are not saturated). |
| Corner delay (lognormal median) | 2.8 min ($\sigma$=0.35) | ±25% about median | Calibrated to lever-and-drag trials (Stocks); modeled as lognormal to capture multiplicative sub-steps (align, pivot, chock). | $\approx \pm 1$–3% (small effect except at upper courses with high corner frequency). |
| Ramp travel speed | 0.15 m/s | 0.12–0.18 m/s | Conservative haul speeds from sledge trials/ergonomics; varied in sensitivity. | $\approx \mp 5$–8% (inverse relation where headway is not binding). |
| Terrace travel speed | 0.20 m/s | 0.16–0.24 m/s | As above; terrace runs dominate early courses. | $\approx \mp 3$–6% (largest effect in first ~30–40 courses). |
| Friction $\mu$ (haul) | 0.20 | 0.15–0.30 | Experimental optimum with ~5% moisture; range brackets realistic field conditions (Harrell & Brown; Fall et al. [25,26]). | $\approx \pm 5$–6% for ±20% $\mu$ (+50% $\mu$ adds ~13.8%; linearized to ±20% $\Rightarrow$ ~±5.5%). |
| Ramp grade $\theta_r$ | 7° | 6–8° | Balances force and path length within attested haul angles; both tested. | $\approx -2.7\%$ (to 6°) and +1.5% (to 8°) relative to 7° (reported). |
| Random stops MTBF / repair | 6 h / 5–12 min | Scenario variants | Encodes incidental disruptions; values used in Monte-Carlo. | Sub-percent to few % unless clustered; absorbed by buffers/headways. |

Note: The ±20% entries summarize direction and order-of-magnitude using the reported model sensitivities where available (headway linear; μ from reported +50% ⇒ +1.91 y at baseline 13.79 y; slope from 6–8°).

Table S9.2. Ramp inclination scenarios (μ = 0.2 for wet sand). Comparison of ramp angles $\theta_r$ = 6°, 7°, 8° under the baseline block mass and friction used in the physical haul model. Total helical path length for the baseline stage (m). For each $\theta_r$ we report: (i) ramp length per unit vertical rise, (ii) steady pull force per block on the ramp (N) from the sled-on-slope model at μ = 0.20, (iii) the assumed sustained force of 200–300 N per puller is consistent with ergonomic limits for repetitive pulling tasks and is supported by experimental archaeology replicating Egyptian sledge transport and levering techniques. Reported forces are steady drawbar forces from Equation 4 in S4.

| Ramp Inclination | Total Ramp Length (m) | Required Force (N) | Number of Workers |
|---|---|---|---|
| 6° | 1290 | 6745 | 23 |
| 7° | 1129 | 7121 | 24 |
| 8° | 968 | 7497 | 25 |

Table S9.3. Phase-wise onsite construction time for the adaptive sequential optimization at a 4-minute mean headway (baseline). The schedule is partitioned into five ramp-configuration phases (adaptive helical ramps) aligned with the indicated course ranges. For each phase, the block count comes from the recursive geometry; estimated working years are computed as (blocks × mean dispatch headway) / active ramps, using the 4-min baseline headway, μ=0.2,$\theta_r$ =7°. This estimated duration assumes consistent hauling rates, crew coordination, and uninterrupted seasonal work.

| Construction Phase | Ramp Configuration | Course Range | No. of Blocks | Estimated Working Years |
|---|---|---|---|---|
| Phase 1 | 16 Ramps (12 up/4 down) | 0–9 | 317000 | 0.56 |
| Phase 2 | 8 Ramps (6 up/2 down) | 10–20 | 315000 | 1.12 |
| Phase 3a | 4 Ramps (3 up/1 down) | 21–183 | 1677000 | 11.95 |
| Phase 3b | 2 Ramps (2 up/down) | 184–198 | 3300 | 0.04 |
| Phase 3c | 1 Ramp (1 up/down) | 199–203 | 170 | 0.004 |
| Total | Adaptive | 0–203 | 2317000 | 13.67 |

Monte Carlo simulations figures:

Capacity:

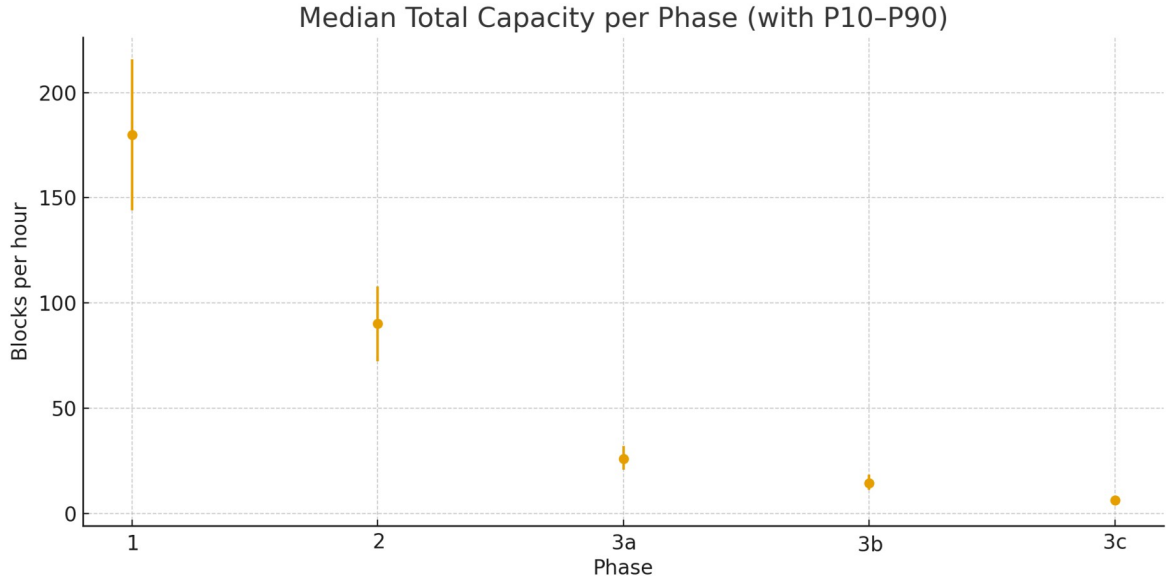

Fig. S9.1. Total ramp capacity by phase (median and P10–P90). Median realized throughput per construction phase; shaded bands indicate the 10th–90th percentile across Monte Carlo trials (N = 10,000, seed=42), use μ = 0.2 and 7° ramp inclination. Phases follow the staged parallel IER configurations (Phase 1: 16 ramps; Phase 2: 8 ramps; Phase 3a: 4 Phase 3b:2 Phase 3c:1 ramp). 4-minutes headway, baseline speeds 0.15 m/s (ramps) and 0.20 m/s (terraces); lognormal corner delay (median 2.8 min, σ = 0.35). Whiskers show P2.5–P97.5; simulation details in Methods.

Different scenarios:

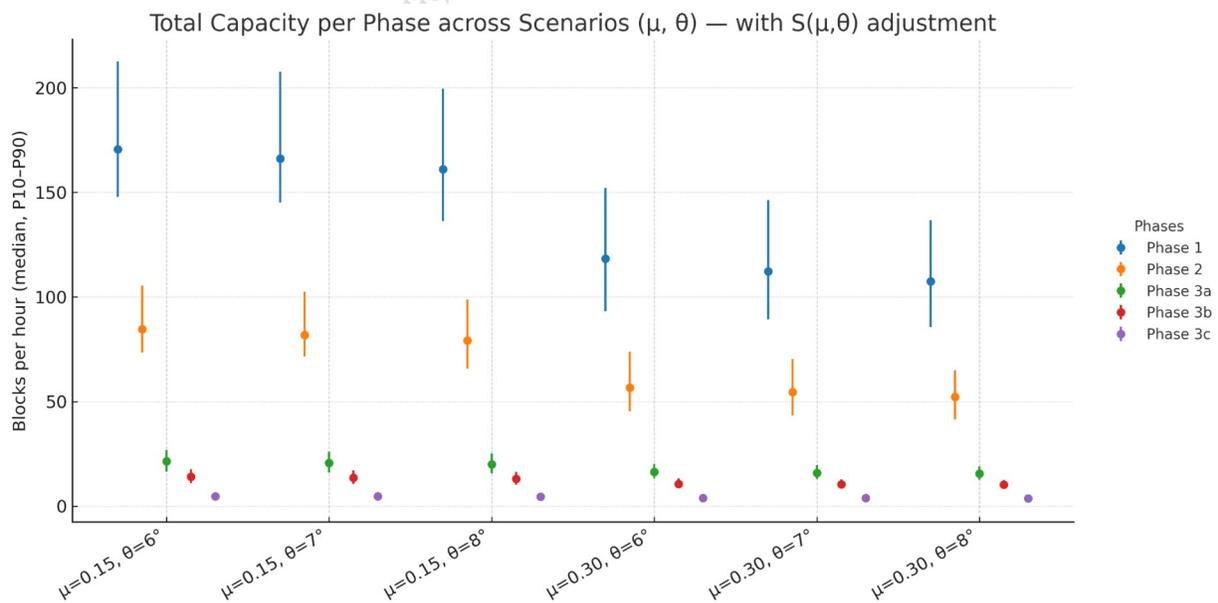

Fig. S9.2. Phase capacity under friction–slope variation. Axes: x = phase (ramp configuration), y = realized throughput (units as in Methods). Statistics: boxes show P10–P90 (operational range); center line = median. Method: capacities are adjusted by $S(\mu,\theta_r)$ from Monte Carlo (N = 10,000, seed=42) sampling ($\mu$ = 0.15–0.30, $\theta_r$ = 6–8°). Reading: lower phases (16 and 8 ramps) operate below saturation, whereas upper phases (4–2–1) run near capacity. 4-minutes headway, baseline speeds 0.15 m/s (ramps) and 0.20 m/s (terraces); lognormal corner delay (median 2.8 min, $\sigma$ = 0.35). Whiskers show P2.5–P97.5. Variations in friction and slope shift the levels modestly but do not change the phase ordering.

On site duration with $\mu$=0.15–0.30 and $\theta_r$=6–8° and different headways:

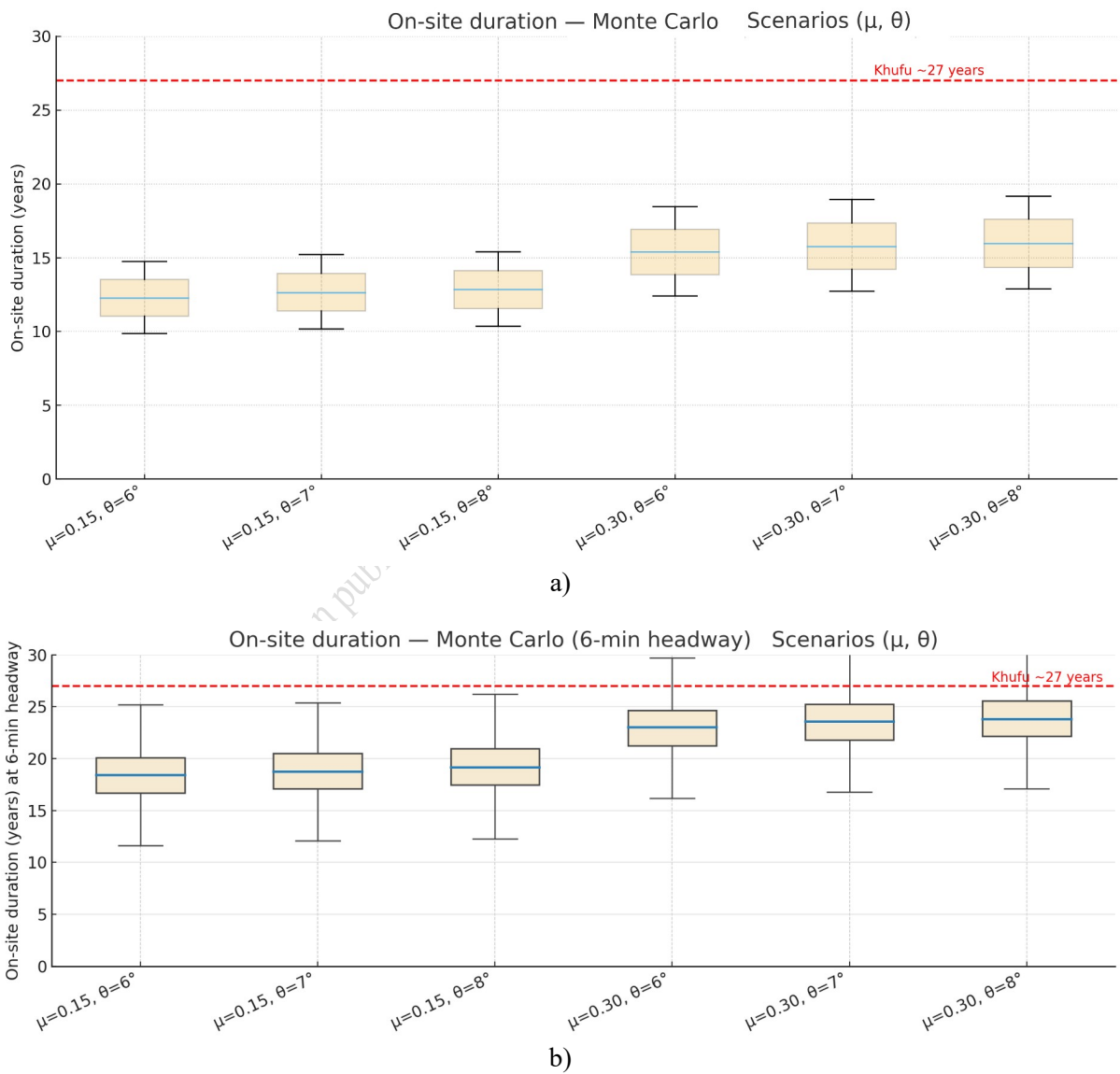

a)

b)

Fig. S9.3. On-site construction duration from 10,000-trial Monte Carlo ($\mu$ = 0.15–0.30; $\theta_r$ = 6–

8°). Dots mark medians; whiskers show P2.5–P97.5 (≈95% percentile interval). The red dashed line indicates Khufu's ~27-year horizon. a) baseline 4-minutes headway b) conservative 6-minutes headway; baseline geometry and calendars as in Methods. All remaining below Khufu's ~27-year historical window.

Headway, a continuous variable, is adjusted per course to match the ramp length. Since each terrace begins after the previous terrace, the work partitions naturally into phases. Headways from 2–13 minutes mapped to the ramp length yield the per course times (Fig. S9.3), with an average time of ~4 minutes, which is the adopted baseline.

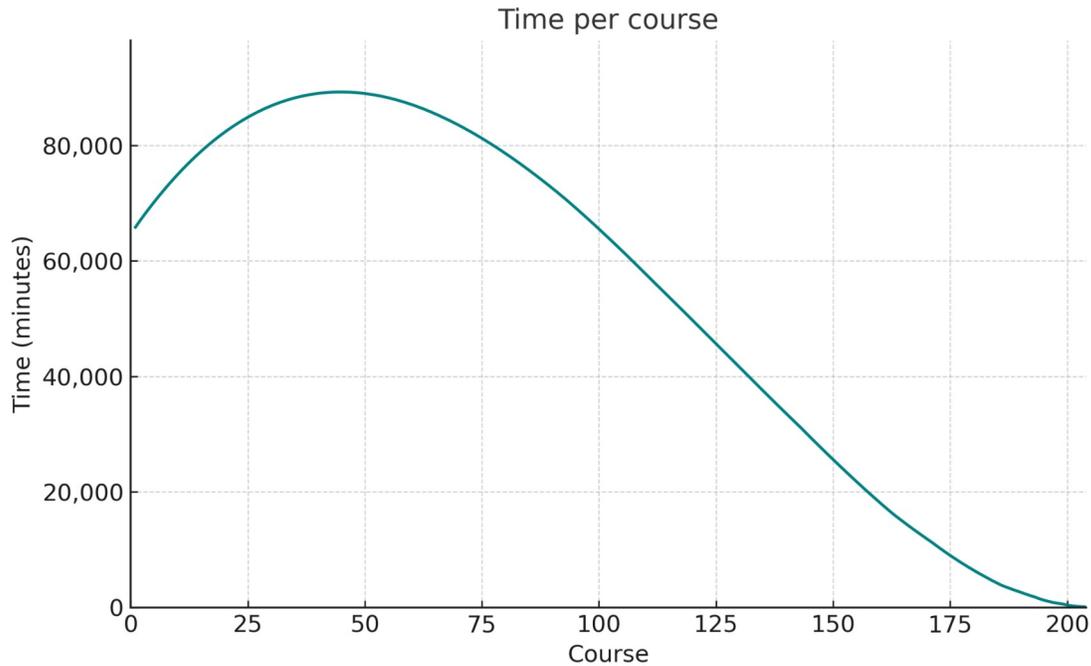

Fig. S9.4. Per-course time (minutes) with variable per-ramp headway (2–13 min). Axes: x = course number; y = total time for that course. Headway is treated as a continuous function of ramp length, bounded 2–13 min and centered on the 4-min baseline. Per-course time equals the blocks assigned per active ramp × realized headway, accounting for parallel ramps. Baseline speeds: 0.15 m/s (ramp) and 0.20 m/s (terrace); corner maneuvers add a lognormal delay (median 2.8 min, σ = 0.35). (Vertical dashed lines, if shown, mark stage transitions.)

Friction coefficient sledge and sand with water:

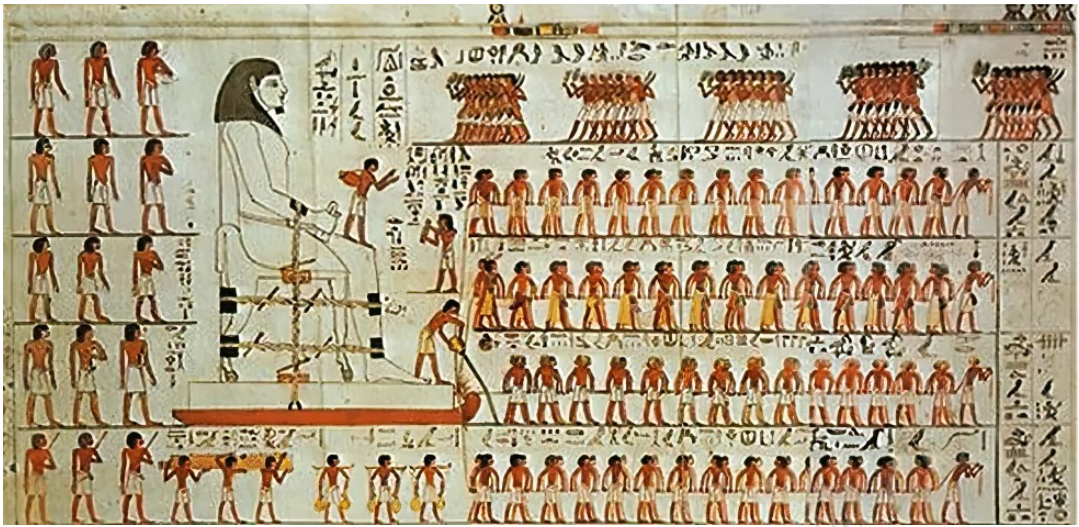

Fig. S9.5. Mural painting from 1880 BC in the tomb of Djehutihotep. A colossal statue on a wooden sledge. The figure at the front of the sled is pouring water onto the sand. The scene provides historical context for sled-on-sand transport and the wet-sand friction assumptions used in this study.

Pyramid plant with quarries:

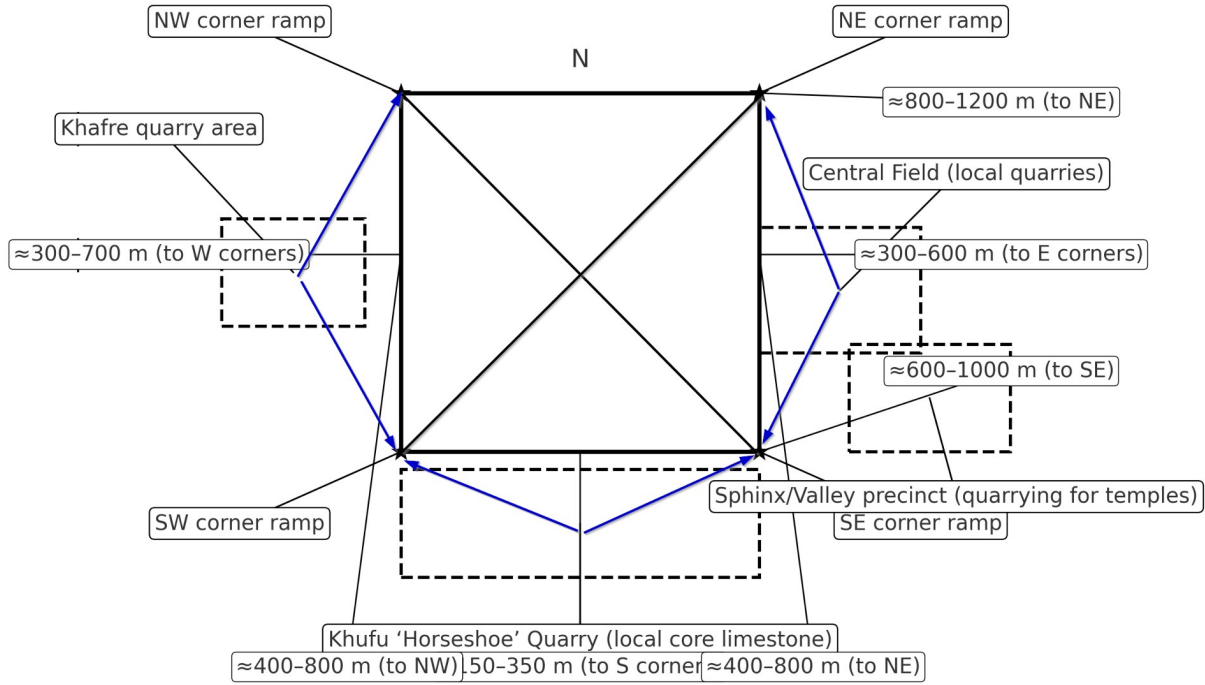

Fig. S9.6. Schematic (not to scale) of local Giza quarries and feasible haul routes to corner ramps. Stars mark corners; blue arrows paths; dashed boxes mark quarry areas. Labels show indicative one-way hauls (hundreds of meters to ~1.2 km) routed around the perimeter.

## S10. Granite megaliths project

*Mezzanine haul using capstans*

We report $F_{incline}$, $T_{line}$, and required pullers (200–300 N per puller), plus capstan holding $T_{tail}$ for $\mu_{capstan}$=0.25, 0.35 and θ=π, 2π. Note: capstan wrap reduces $T_{tail}$; $T_{line}$ is unchanged.

Table S10.1. Mezzanine haul — m = 60–80 t; $\theta_r$ = 3–4°; $\mu_k$ = 0.15–0.30.

| Mass (ton) | Ramp angle $\theta_r$ (deg) | $\mu_r$ | $F_{incline}$ (kN) | $T_{line}$ (kN) | Pullers @200 N | Pullers @300 N | $T_{tail}$ (kN) [$\mu_{capstan}$=0.25, θ=π] | $T_{tail}$ (kN) [$\mu_{capstan}$=0.25, θ=2π] | $T_{tail}$ (kN) [$\mu_{capstan}$=0.35, θ=π] | $T_{tail}$ (kN) [$\mu_{capstan}$=0.35, θ=2π] |
|---|---|---|---|---|---|---|---|---|---|---|
| 60 | 3 | 0.15 | 119.0 | 119.0 | 595 | 397 | 54.24 | 24.73 | 39.62 | 13.19 |
| 60 | 3 | 0.2 | 148.4 | 148.4 | 742 | 495 | 67.64 | 30.84 | 49.41 | 16.45 |
| 60 | 3 | 0.25 | 177.8 | 177.8 | 889 | 593 | 81.04 | 36.95 | 59.2 | 19.71 |
| 60 | 3 | 0.3 | 207.1 | 207.1 | 1036 | 691 | 94.44 | 43.06 | 68.98 | 22.97 |
| 60 | 4 | 0.15 | 129.1 | 129.1 | 646 | 431 | 58.88 | 26.84 | 43.0 | 14.32 |

| | | | | | | | | | |
|---|---|---|---|---|---|---|---|---|---|
| 60 | 4 | 0.2 | 158.5 | 158.5 | 793 | 529 | 72.26 | 32.95 | 52.78 | 17.58 |
| 60 | 4 | 0.25 | 187.9 | 187.9 | 940 | 627 | 85.65 | 39.05 | 62.56 | 20.83 |
| 60 | 4 | 0.3 | 217.2 | 217.2 | 1087 | 725 | 99.03 | 45.15 | 72.33 | 24.09 |
| 70 | 3 | 0.15 | 138.8 | 138.8 | 695 | 463 | 63.29 | 28.85 | 46.22 | 15.39 |
| 70 | 3 | 0.2 | 173.1 | 173.1 | 866 | 577 | 78.92 | 35.98 | 57.64 | 19.2 |
| 70 | 3 | 0.25 | 207.4 | 207.4 | 1037 | 692 | 94.55 | 43.11 | 69.06 | 23.0 |
| 70 | 3 | 0.3 | 241.7 | 241.7 | 1209 | 806 | 110.2 | 50.24 | 80.48 | 26.8 |
| 70 | 4 | 0.15 | 150.7 | 150.7 | 754 | 503 | 68.69 | 31.32 | 50.17 | 16.71 |
| 70 | 4 | 0.2 | 184.9 | 184.9 | 925 | 617 | 84.31 | 38.44 | 61.58 | 20.51 |
| 70 | 4 | 0.25 | 219.2 | 219.2 | 1096 | 731 | 99.92 | 45.56 | 72.98 | 24.3 |
| 70 | 4 | 0.3 | 253.4 | 253.4 | 1268 | 845 | 115.5 | 52.68 | 84.39 | 28.1 |
| 80 | 3 | 0.15 | 158.6 | 158.6 | 794 | 529 | 72.33 | 32.98 | 52.83 | 17.59 |
| 80 | 3 | 0.2 | 197.8 | 197.8 | 990 | 660 | 90.19 | 41.12 | 65.88 | 21.94 |
| 80 | 3 | 0.25 | 237.0 | 237.0 | 1186 | 791 | 108.1 | 49.27 | 78.93 | 26.28 |
| 80 | 3 | 0.3 | 276.2 | 276.2 | 1381 | 921 | 125.9 | 57.41 | 91.98 | 30.63 |
| 80 | 4 | 0.15 | 172.2 | 172.2 | 861 | 574 | 78.5 | 35.79 | 57.34 | 19.09 |
| 80 | 4 | 0.2 | 211.3 | 211.3 | 1057 | 705 | 96.35 | 43.93 | 70.37 | 23.44 |
| 80 | 4 | 0.25 | 250.5 | 250.5 | 1253 | 835 | 114.2 | 52.07 | 83.41 | 27.78 |
| 80 | 4 | 0.3 | 289.6 | 289.6 | 1449 | 966 | 132.0 | 60.2 | 96.45 | 32.12 |

A repetitive cycle of "mezzanine haul" of 30m for each course (≈15m ramp at 3–4° + ≈15m horizontal transfer), repeating it course by course up to the level of the King's Chamber and, above, for the unloading chambers. The stress equations and parameters were the same as those used in the physical model of the article (ramp and horizontal sledge; ) and the operating assumptions were the same (velocities of 0.15 m/s on the ramp and 0.20 m/s on the terrace, later treated in sensitivity), with local ramps of 3–4° for 60–80 t megaliths and $\mu_k$=0.15–0.30.

The strategy of short, reusable ramps (3–4°), sized hauling equipment, and the use of bollards/winches to reduce holding (but not pulling) forces are documented, as well as the fact that the entire lot could be raised in 1–2 days by working in batches and recycling the ramp by course. Furthermore, the text places the bulk of the granite work between courses ~60–85, consistent with the King's Chamber horizon and the spillways.

Table S10.2. The number of pullers is calculated based on line tension (without regulators) at 300–200 N/person; (ii) with $\mu$=0.15, the teams fall into the range ≈300–700; with $\mu$=0.30, the maximums approach ~1,000 in the worst cases, consistent with the text. Forces are reported in kN (1 kN = 1000 N); μ is dimensionless; angles in degrees.

| Case | $F_{incline}$ (kN) | $F_{horizontal}$ (kN) | Work W (MJ) | Pullers (300–200 N/person) |
|---|---|---|---|---|
| 60 ton, 3° | 148.4 | 117.7 | 3.99 | 495–742 |
| 60 ton, 4° | 158.5 | 117.7 | 4.14 | 529–793 |
| 80 ton, 3° | 197.8 | 157.0 | 5.33 | 660–990 |
| 80 ton, 4° | 211.3 | 157.0 | 5.52 | 705–1057 |

*Block summations (reference):*

From course ~60 to ~85 (≈25 steps):

- 60 t, 3–4°, μ=0.20: 100–104 MJ per block; 75–114 min of pure towing time (2.9–4.6 min/step).
- 80 t, 3–4°, μ=0.20: 133–138 MJ; 75–114 min (the time per step barely changes; what increases is the crew/safety).

This fits with the "batch" operation in one or two days (ramp assembly 2–4 h per course, reusable for several megaliths; wedge maneuvers and metrological verification).

From the base to the level of the Chamber (~60 steps of 0.71 m per course, if starting very low):

- 60 t: ≈240–248 MJ per block; 175–275 min of effective hauling (≈3–4.6 h).
- 80 t: ≈319–331 MJ; same times per step (more demanding on equipment).

Coverage of courses 1–85: from the base course to the end of the megalithic horizon (including the start of the King's Chamber).

From courses 61 to 85: the factor decreases linearly from 1.0 → 0.0 to reflect that, as the beams are placed (King's Chamber and lightening), there are fewer and fewer beams left to raise further. This 60–85 range is consistent with the manuscript itself, which places the greatest granitic activity between these courses.

*Granite megalith transport — time calculations*

Assumptions & constants

- Path per lift step: 15 m ramp + 15 m terrace (total 30 m).
- Ramp grades considered: 3–4°. Baseline kinetic friction $\mu_k = 0.20$ (range 0.15–0.30).
- Speeds: ramp 0.15 m/s and terrace 0.20 m/s (baseline); conservative heavy-load case uses 0.12/0.18 m/s.
- Per-course setup overhead: 3.0 h (mid; range 2.0–4.0 h).
- Seating & wedging per block: 10 min (range 5–15 min).
- Phase 1 blocks: 6 (3 shutters + 3 "plugs").
- Phase 2 blocks: 45 (5 course × 9 beams/course).

Totals by phase (one row rise):

- Phase 1 — Shutters + Plugs (6 blocks):
  - Base: 4.29 hours per course.
  - Conservator: 4.35 hours per course.
- Phase 2 — King's Chamber and Loading:
  - Per row (9 beams): 4.94 hours (base) | 5.02 hours (cons.).
  - "Batch of 5 courses" at the same elevation (uncommon, useful for accounting purposes): 24.69 hours (base) | 25.10 hours (cons.) ⇒ approx. 1.0–1.05 clock days.
- Combined phases (Phase 1 + Phase 2 for one row; 6 + 45 = 51 blocks):
  - Base: 14.00 hours per row (including 3 hours of assembly)
  - Conservative: 14.45h per row.
- Cumulative 26 courses: 128.38–130.54 h ≈ 12.8–13.1 working days.

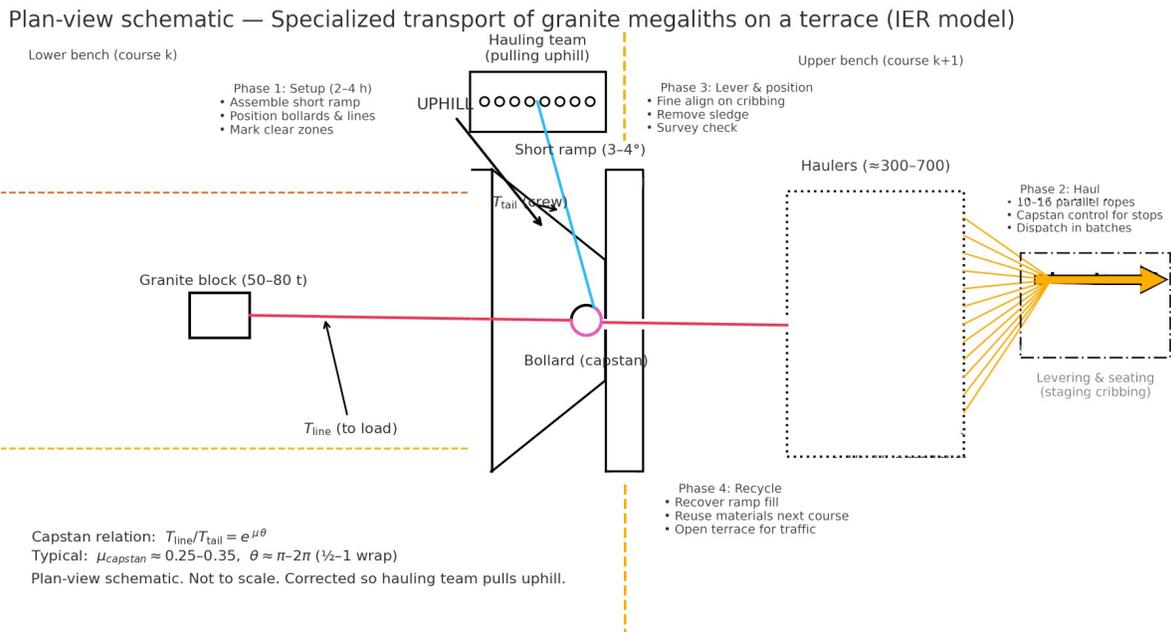

Fig. S10.2. Plan-view schematic — Specialized transport of granite megaliths on a working terrace (IER model). A short, reusable 3–4° slip-ramp bridges adjacent benches while limestone delivery continues along the reserved perimeter corridor. Twelve (typ. 10–16) parallel rope lanes pull a granite beam on a wooden sledge; terrace-mounted wooden bollards ("cabestrants") provide capstan wrap control (θ≈½–1 turn), reducing the tail holding force without changing the drawbar line tension. Phases: 1-Setup (2–4 h): assemble the short ramp, place bollards, lay out rope lanes and clear zones. 2-Haul: batch moves using 10–16 parallel ropes, with capstan-controlled stops/holds. 3-Lever & position: fine alignment on staging cribbing, remove sledge, survey check. 4-Recycle: recover ramp fill and reuse materials for the next course, freeing the terrace. Scale bar 20 m; schematic, not to scale for all features. Operating lane ~3.8 m; crew sizing and friction ranges follow Methods.

Optional cumulative time in the granite activity range (courses 60–85, 26 courses):
- Time per course (one row; 9 beams + assembly): 4.94–5.02 h.
- Cumulative 26 courses: 128.38–130.54 h ⇒ 16.05–16.32 8-hour days.

Summarize:
- The first 60 courses: 30 working days;
- The 60–85 horizon: 13 working days;
- Total maximum: 43 working days

By introducing 5–7 ramp relocations at 2–4 h each course, the preparation term dominates and brings the total to ≈1–2 days, while maintaining limestone traffic in the rest of the perimeter (occupied zone ≈1.35–2.28% of the terrace at that elevation).

Another advantage of this method is that it can be worked in parallel, so each team could prepare a ramp and raise a group of blocks without affecting the other group's position, as they are in parallel positions. This means that work times could be even shorter.

## S11. Other geometric correspondences

A quantitative comparison between the helical trajectory predicted by the IER model and the variations in the height of the pyramid's courses indicates a consistent architectural pattern.

To quantitatively assess the correlation between theoretical ramp models and empirical construction features, a direct comparison was performed (Table S11.1). We tested 20 ramp models, each defined by a constant slope ranging from 6° to 8° (increment 0.1). For each model, the 10 sequential turn locations were calculated as specific course numbers. These theoretical turn courses were then compared against a pre-defined set of 16 architecturally significant course ranges identified from the pyramid's structure. The analysis involved a strict sequential mapping: the first calculated turn for each slope was compared against the first significant range, the second turn against the second range, and so forth. A "match" was recorded only when a calculated turn course height fell precisely within its corresponding significant range 1.5 m (2 courses) below the thickness range change. This method allows for a quantitative scoring of each ramp model, identifying the slope whose geometric progression best aligns with the selected constructional anomalies.

Table S11.1: Correspondence between Calculated Ramp Turn Courses and Pre-defined Significant Course Ranges. The table displays the calculated course height for the 10 sequential turns of each ramp model (course, defined by slope). Each turn is compared against a specific, pre-defined course or range of interest (columns). A value is shown only if it falls within a strict proximity of 1.5 m (2 courses) below the thickness range change. The 'Total Matches' column sums these direct hits, providing a correlation score for each ramp model.

| Height range | 6.0 | 6.1 | 6.2 | 6.3 | 6.4 | 6.5 | 6.6 | 6.7 | 6.8 | 6.9 | 7.0 | 7.1 | 7.2 | 7.3 | 7.4 | 7.5 | 7.6 | 7.7 | 7.8 | 7.9 | 8.0 |
|---|---|---|---|---|---|---|---|---|---|---|---|---|---|---|---|---|---|---|---|---|---|
| [15.365,16.865] | | | | | | | | | | | | | | | | | | | | | |
| [17.515,19.015] | | | | | | | | | | | | | | | | | | | | | |
| [27.055,28.555] | | | | | | | | | | | 27.25 | 27.61 | 28.0 | 28.35 | | | | | | | |
| [35.305,36.805] | | | | | | | | | | | | | | | | | | | | | |
| [38.065,39.565] | | | | | | | | | | | | | | | | | | | | | |
| [51.975,53.475] | | | | | | | | | | | | | | | | | 52.05 | 52.74 | 53.43 | | |
| [57.135,58.635] | | 57.58 | 58.34 | | | | | | | | | | | | | | | | | | |
| [67.34,68.84] | | | | | | | | | | | 67.68 | 68.58 | | | | | | | | | |
| [73.35,74.85] | | | | | | 73.61 | 74.7 | | | | | | | | | | | | | | |
| [81.23,82.73] | | 81.9 | | | | | | | | | | 81.42 | 82.51 | | | | | | | | |
| [87.465,88.965] | | | | | | | | 88.07 | 88.62 | | | | | | | | | | | | 88.49 |
| [95.025,96.525] | | | | | | 95.16 | 96.29 | | | | | | | 95.9 | | | | | | | |
| [99.835,101.335] | 100.0 | 100.86 | | | | | | | | 100.71 | | | | | | | | | | | 101.2 |
| [103.42,104.92] | | | | | | 103.47 | 104.53 | | | | | | | 104.51 | | | | | | | |
| [107.155,108.655] | | 107.91 | | | | | | | | 107.74 | | | | | | 107.94 | | | | | |
| [123.85,125.35] | | | | | | | | | | 123.89 | 124.89 | | | 124.55 | | | | 123.87 | 124.87 | | |

| Total | | 2 | 3 | 1 | 0 | 3 | | 3 | | 0 | 1 | 2 | 2 | 1 | | 0 | 0 | 1 | 5 | | 3 | 1 | 2 | | 2 | 2 | 3 |
| --- | --- | --- | --- | --- | --- | --- | --- | --- | --- | --- | --- | --- | --- | --- | --- | --- | --- | --- | --- | --- | --- | --- | --- | --- | --- | --- | --- |

Based on the validated data, a very clear conclusion emerges. The 7.4° ramp models emerge as the strongest candidates, achieving five direct matches to areas of construction interest. This result provides focused quantitative support for a ramp hypothesis with a slope very close to 7.4–7.5° (Fig. 10 and Fig. S11.1).

The angle sweep exhibits a narrow maximum at 7.4° with 5 post-turn matches out of 10 tiers (closely matched by 7.5°), whereas adjacent angles are $\leq 4$ (Table S11.1; Fig. 11, Fig. S11.1, and Fig. S11.2.e). In 2,000 Monte Carlo realizations (seed = 42) of 203 courses drawn from a truncated N(0.722,0.235) within [0.495, 1.50] m, no simulation reached the observed total (max = 4), yielding an empirical $p < 5 \times 10^{-4}$ at MC resolution. We interpret this as a geometric correspondence—consistent with re-leveling/compensation immediately after face changes—that is hypothesis-generating, not confirmatory. Full tier-by-tier counts, window/detrending sensitivities.

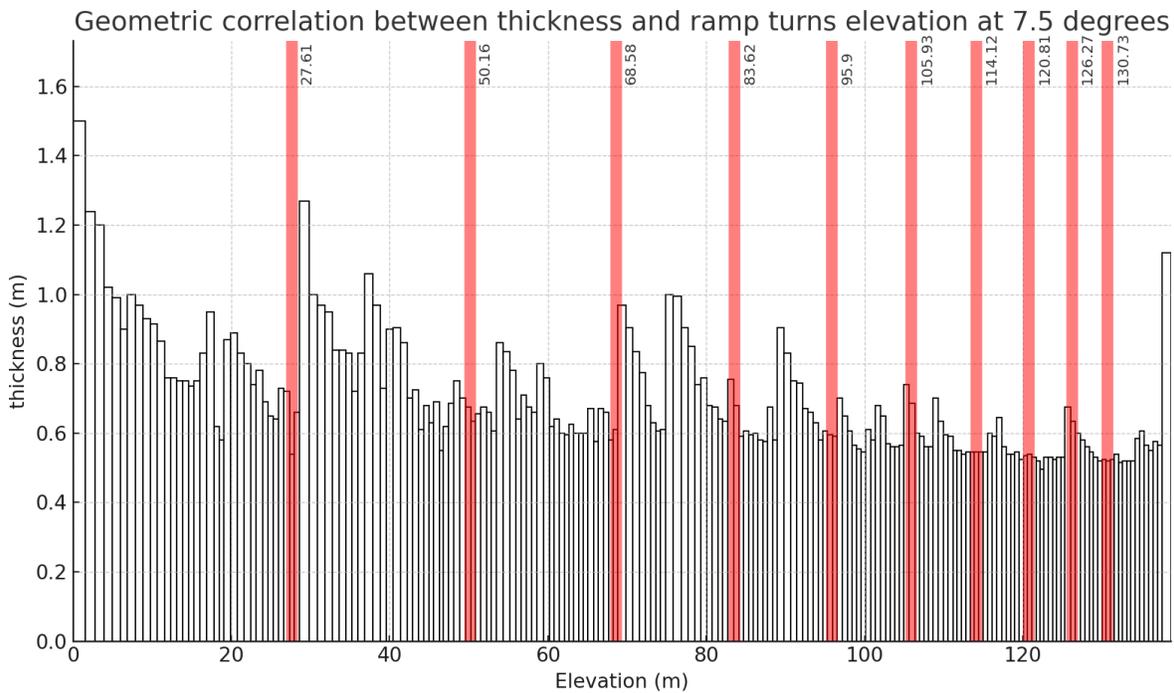

Fig. S11.1. Course thickness vs elevation (m) ($\theta_r = 7.5°$). White-filled bars with black borders show thickness per course (m) versus elevation (m). The red bands indicate the turning nodes predicted by the IER scenario with a 7.5° slope in courses 39 (27.61 m), 97 (68.58 m), 135 (95.9 m); The turning elevation positions are highlighted in red. The figure is descriptive.

Other prominent thickness peaks do not follow IER turning nodes. We read these as survey-driven leveling/stiffening belts or operational phase boundaries (e.g., platforms near the King's Chamber horizon), not evidence for/against the IER haul path. In other words, IER constrains logistics, not the cadence of geometric re-leveling, and belts can arise independently of turns. Falsification route: targeted GPR/ERT along predicted edge-bands, endoscopy, joint micro-topography, and independent re-measurement of course heights at the specified elevations should reveal (or fail to reveal) rubble-rich, heterogeneous fills and corner wear/anchor traces; a systematic absence would falsify this aspect of the model.

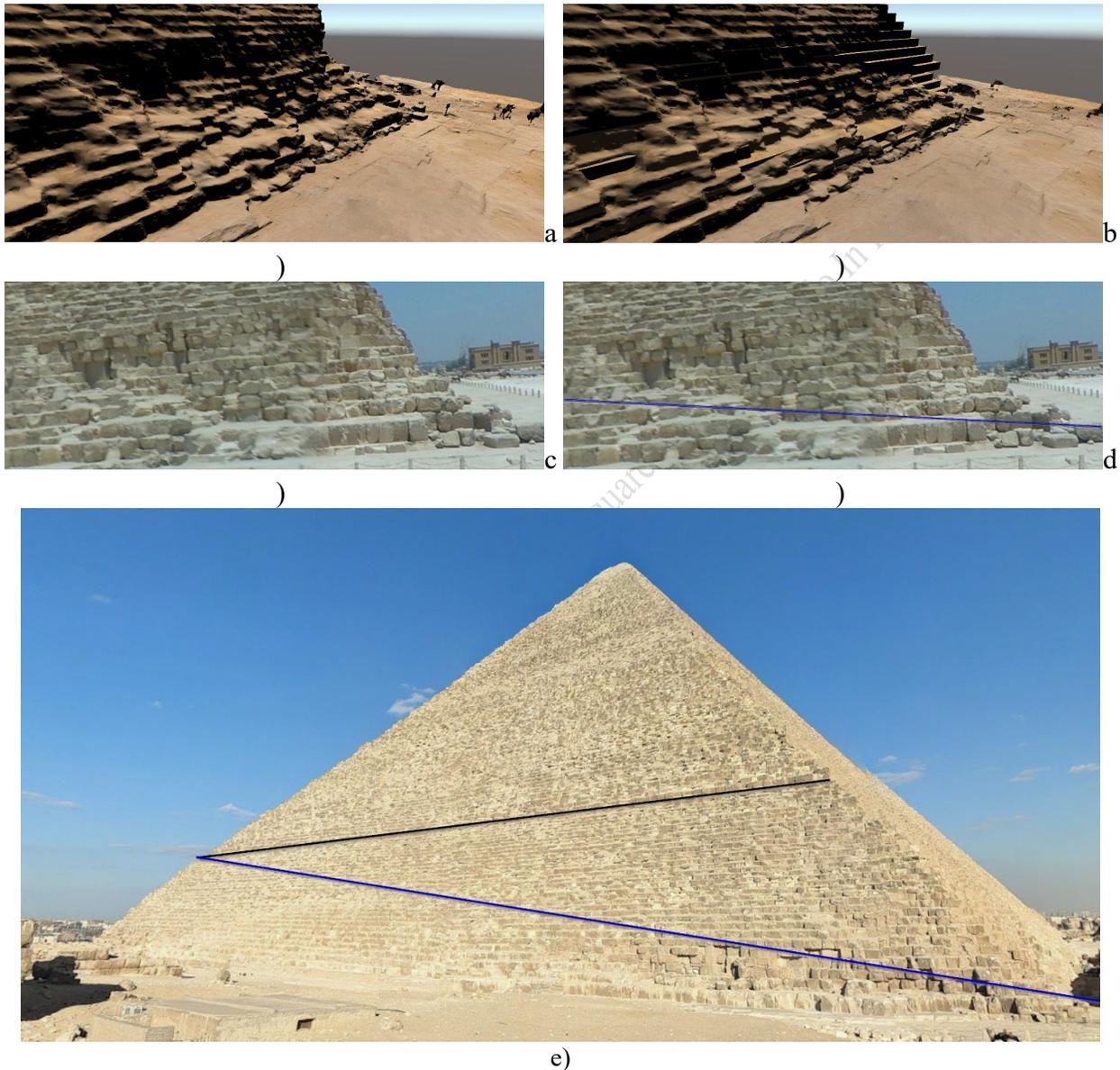

Fig. S11.2. Parametric 3D modeling (Unity) shows a geometric correspondence that warrants further investigation: the predicted northeastern (NE) corner access for the Integrated Edge-Ramp (IER) corresponds to a zone of substantial structural loss on the pyramid's face, likely due to stone removal or natural decay. a) 3D photogrammetry reconstruction of the corner, b) 3D photogrametry superposed to the 3D rendered model, c) original photo of the northeastern corner, d) blue line drawn over the original photo of the corner with 7.5° ramp slope, e) black

line below the first thickness change (27 m) and blue line ramp slope with 7.4°. See the correlation between IER and its possible ramp entrance at the corner.

## S12. ScanPyramids Mission Findings figures

This study conducted a comprehensive sensitivity analysis of the Integrated Edge-Ramp (IER) model by systematically varying the ramp inclination angle ($\theta_r$) from 6.0° to 8.0°. For each angle, the recursive geometric algorithm was executed to calculate the cumulative height at each of the ten helical turns. The results, compiled in a complete dataset of turn elevations, reveal a clear and predictable relationship: steeper ramp angles yield higher turn elevations for a given turn number, effectively "compressing" the helical path vertically. Crucially, this analysis allowed us to identify the optimal ramp angles that best align the model's predictions with the muographic anomalies reported by the ScanPyramids mission. The angles of 7.4°, 7.5°, and 7.6° were found to produce turn heights most consistent with the observed cavities at ~52.5 m, ~83 m, and ~105 m, with 7.5° providing the best overall fit. This quantitative assessment not only quantifies the geometric consistency (hypothesis-generating, not evidentiary) of the IER model with external empirical evidence but also defines a narrow and archaeologically plausible range for the operational parameters of the proposed construction system.

Table S12.1. Cumulative height (m) at each helical turn for different ramp inclination angles ($\theta_r$) from 6.0° to 8.0° in the Integrated Edge-Ramp (IER) model.

| Turn | 6.0 | 6.25 | 6.5 | 6.75 | 7.0 | 7.25 | 7.3 | 7.4 | 7.5 | 7.6 | 7.75 | 8.0 |
|---|---|---|---|---|---|---|---|---|---|---|---|---|
| 1 | 22.08 | 22.97 | 23.91 | 24.81 | 25.79 | 26.70 | 26.89 | 27.25 | 27.61 | 28.00 | 28.53 | 29.52 |
| 2 | 40.86 | 42.36 | 43.91 | 45.39 | 47.04 | 48.55 | 48.85 | 49.51 | 50.16 | 50.82 | 51.71 | 53.43 |
| 3 | 56.82 | 58.72 | 60.67 | 62.52 | 64.56 | 66.43 | 66.78 | 67.68 | 68.58 | 69.44 | 70.60 | 72.80 |
| 4 | 70.38 | 72.53 | 74.70 | 76.77 | 79.00 | 81.06 | 81.42 | 82.51 | 83.62 | 84.63 | 85.99 | 88.49 |
| 5 | 81.90 | 84.18 | 86.45 | 88.62 | 90.90 | 93.03 | 93.37 | 94.62 | 95.90 | 97.02 | 98.53 | 101.20 |
| 6 | 91.69 | 94.01 | 96.29 | 98.48 | 100.71 | 102.82 | 103.13 | 104.51 | 105.93 | 107.13 | 108.75 | 111.49 |
| 7 | 100.00 | 102.30 | 104.53 | 106.68 | 108.79 | 110.83 | 111.10 | 112.58 | 114.12 | 115.38 | 117.08 | 119.82 |
| 8 | 107.06 | 109.29 | 111.43 | 113.50 | 115.45 | 117.38 | 117.61 | 119.17 | 120.81 | 122.11 | 123.87 | 126.57 |
| 9 | 113.06 | 115.19 | 117.21 | 119.17 | 120.94 | 122.74 | 122.92 | 124.55 | 126.27 | 127.60 | 129.40 | 132.04 |
| 10 | 118.15 | 120.17 | 122.05 | 123.89 | 125.46 | 127.13 | 127.26 | 128.94 | 130.73 | 132.08 | 133.91 | 136.47 |

Table S12.2. Alignment assessment: Absolute differences (m) between IER model predictions and ScanPyramids anomaly heights across ramp angles.

| Angle | Diff N3 (52.5 m) | Diff C2 (83 m) | Diff C1 (105 m) | Mean |
|---|---|---|---|---|
| 7.5 | 2.34 | 0.62 | 0.93 | 1.30 |
| 7.4 | 2.99 | 0.49 | 0.49 | 1.32 |
| 7.6 | 1.68 | 1.63 | 2.13 | 1.81 |
| 7.3 | 3.65 | 1.58 | 1.87 | 2.37 |
| 7.75 | 0.79 | 2.99 | 3.75 | 2.51 |
| 7.25 | 3.95 | 1.94 | 2.18 | 2.69 |
| 7.0 | 4.46 | 4.00 | 4.29 | 4.25 |
| 8.0 | 0.93 | 5.49 | 6.49 | 4.30 |
| 6.75 | 6.11 | 6.23 | 6.52 | 6.29 |
| 6.5 | 7.59 | 8.30 | 8.71 | 8.20 |
| 6.25 | 9.14 | 10.47 | 10.99 | 10.20 |
| 6.0 | 10.64 | 12.62 | 13.31 | 12.19 |

Table S12.3. Comparison between the predicted heights of ramp notches for a 7.5° edge-ramp and the observed elevations of the three known northeast notches (N1–N3). Predicted and measured values differ by $<\sim 2.5$ m. We report observational uncertainties for both data and model registration; within these ranges, alignments are treated as geometric consistency and a testable hypothesis, not diagnostic evidence.

| Ramp | Iteration height | Total height | Notches height | Known notches and cavities |
|---|---|---|---|---|
| 1 | 27.61 | 27.61 | | |
| 2 | 22.55 | 50.16 | 52.5 | N3 |
| 3 | 18.42 | 68.58 | | |
| 4 | 15.04 | 83.62 | 83.00 | N2 (C2) |
| 1 | 12.28 | 95.90 | | |
| 2 | 10.03 | 105.93 | 105.00 | N1 (C1) |
| 3 | 8.19 | 114.12 | | |
| 4 | 5.46 | 120.81 | | |
| 1 | 4.46 | 126.27 | | |

If Khufu employed an integrated edge ramp, the near-contemporary pyramid of Khafre may have used the same method. A concrete, falsifiable test is to compute a Khafre-specific list of predicted edge-band elevations from Khafre's face angle $\theta_{Khafre}$ and course module $h_c$ (Table S12.4), assuming the same ramp grade ($\approx 7.5°$) and corner-turn scheduling, and to target these heights in muographic and endoscopic surveys. A statistically significant one-to-one correspondence (within registration tolerance and after correcting for casing loss) between predicted bands and observed notches/voids on Khafre would provide convergent support for a shared construction method; a systematic mismatch would argue against generalization.

Table S12.4. Predicted heights of ramp notches for a 7.5° edge-ramp in the Khafre's pyramid

| Ramp | 1 | 2 | 3 | 4 | 1 | 2 | 3 | 4 | 1 | 2 |
|---|---|---|---|---|---|---|---|---|---|---|
| Height | 25.52 | 46.51 | 63.78 | 77.99 | 89.68 | 99.30 | 107.21 | 113.72 | 119.07 | 123.47 |

These values are mathematical predictions derived from the recursive algorithm using the pyramid's ideal geometric parameters. It is important to note that the actual construction of the pyramid involved courses of varying height and irregularities in the core masonry. Therefore, when testing these predictions against physical evidence (e.g., through geophysical surveys or muography), the actual turn elevations may vary by several meters from these calculated values. The model provides a precise theoretical framework, but its validation must account for the inherent construction tolerances of Old Kingdom pyramid building.

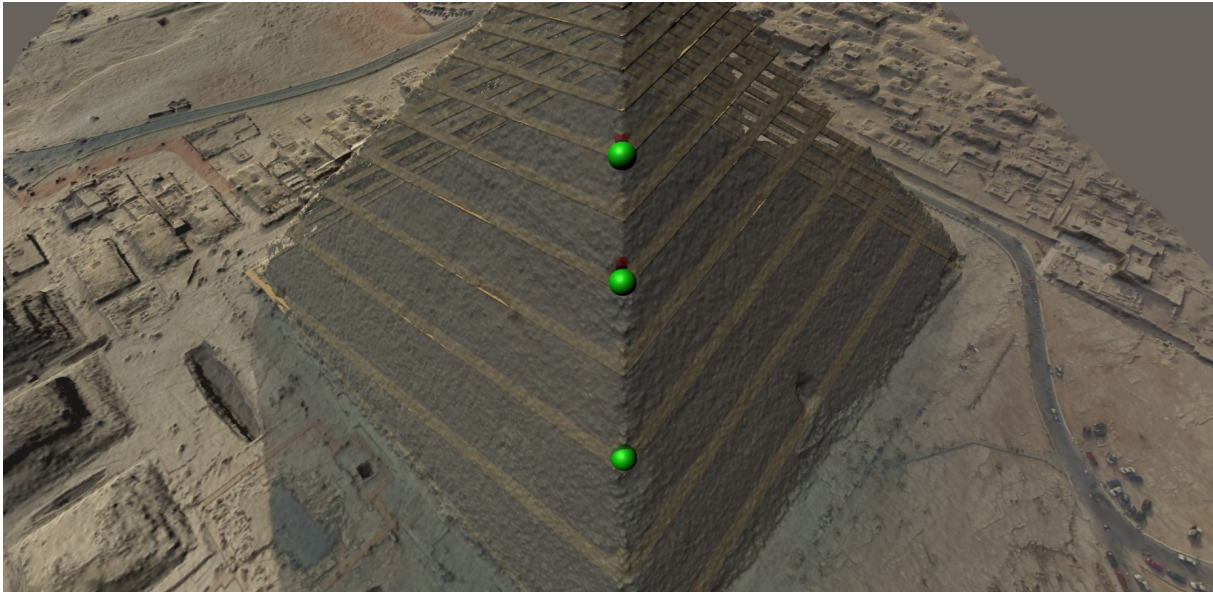

Fig. S12.1. Geometric consistency of the IER with ScanPyramids anomalies. 3-D parametric reconstruction of the four edge-integrated helical channels at 7.5° with photogrammetric overlaid. Published control points; green spheres = observed notches, red cubes = reported muographic cavities. The overlay is consistent with the predicted edge helices passing adjacent to these anomaly zones.

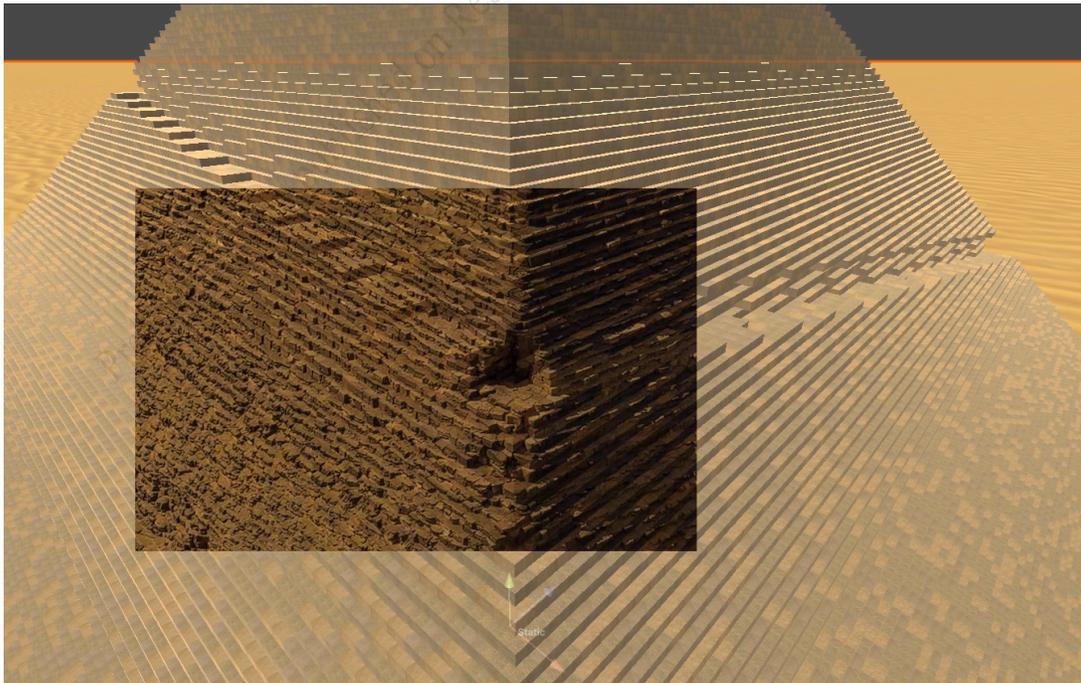

Fig. S12.2. Corner notch photograph registered to the IER 3-D model. Field photograph of a Pyramid corner notch overlaid on the parametric edge-integrated ramp (IER) geometry. The overlay is rigidly registered to the model using visible corner/edge control points; colored polylines show the predicted helical channel centerlines and platform extents at that elevation. The match illustrates how a notch occurs adjacent to the modeled edge path without tuning parameters.

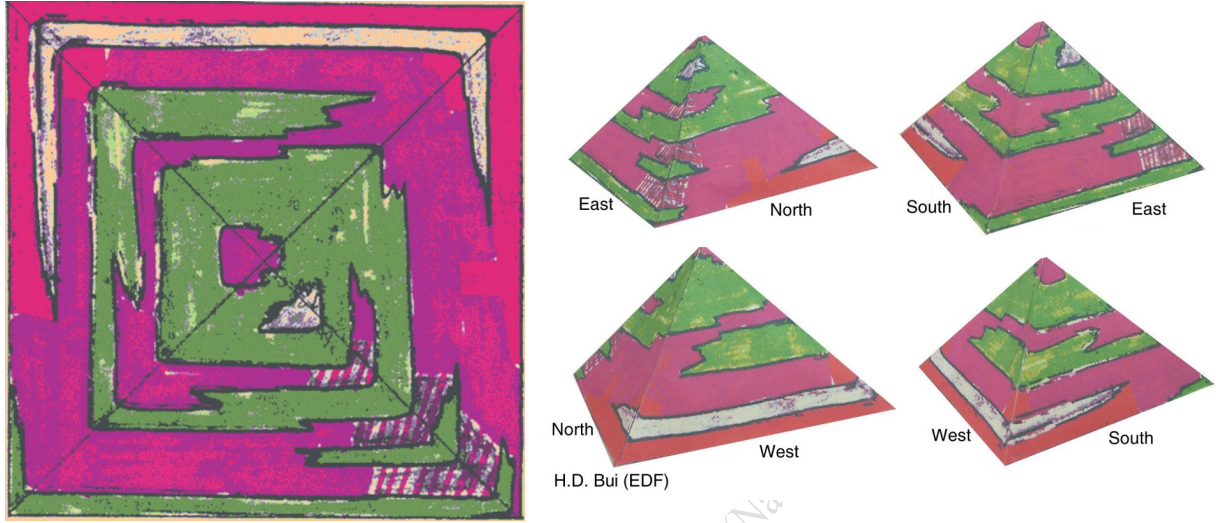

Fig. S12.3. In 1986/1987, a microgravimetric edge-density map was obtained thanks to the EDF Foundation (Redrawn after Dormion. Orientation per Houdin; scale bar and acquisition lines indicated)

**S13. Comparison of pyramid ramp construction theories.**

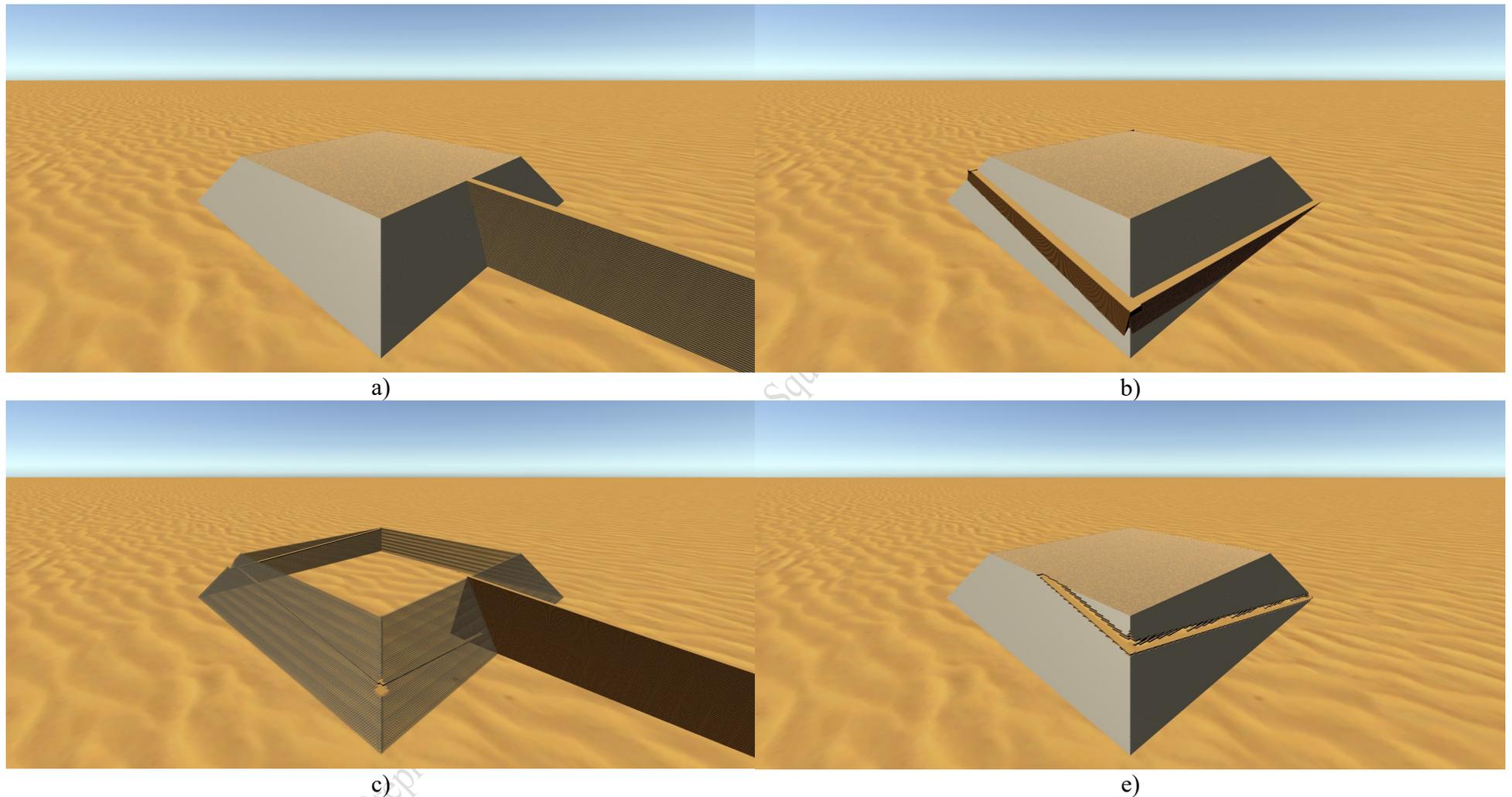

Fig. S13.1. Comparative visualization of the four ramp configurations implemented in the same computational framework up to course 56, where the Houdin model transitions from the external to the internal ramp. (a) Straight external ramp at 4°, showing the long single frontal embankment. (b) Spiral external ramp (4°, 5 m offset, 6.35 m lane) wrapping around the pyramid faces. (c) Dual-phase Houdin model combining a short external

ramp to ~40 m (4°) with an emerging internal spiral (4°) (transparent pyramid). (d) Integrated Edge-Ramp (7°) confined along the edges and progressively backfilled. All renders share identical geometric and logistic parameters ($\mu = 0.20$, 0.15 m/s, one-cell separation rule). Up to this height, the straight and spiral variants are distance-dominated and require the largest external fills; Houdin reduces early earthwork but remains single-lane.

The IER achieves the shortest cumulative paths and the lowest auxiliary volume, illustrating the efficiency gains of edge confinement.

This supplementary note extends the open geometric–logistic framework (Zenodo dataset and code archive) to four representative ramp hypotheses:

- Straight external ramp (4°) – single frontal ramp; side-slope angle for embankment 60° slope angle (from ground); 6.35 m wide (5 blocks).
- Spiral external ramp (4°) – helical lane offset 5 m from the pyramid edge; 6.35 m wide.
- Houdin dual-phase system (4–5°) – straight external ramp to ~40 m (4°); 6.35 m wide. followed by an internal spiral (~5°); 2.54 m wide.
- Integrated Edge-Ramp (IER, 7°) – single line; 3.8 m apertures (3 blocks).

All were simulated using identical physical and logistical assumptions: kinetic friction $\mu=0.20$ (wet sand), block mass 2.27 t, gravity $g=9.81$ m s$^{-2}$, traction 300 N per worker, ramp velocity 0.15 m/s, and a one-cell separation rule (team + 3 m sled + 15 m buffer). Each course (rows 0–99) follows empirical block counts from the base upward. Outputs include per-course ramp and terrace distances, work, vertical/horizontal force components, and cumulative energy. We evaluated all models up to course 100 (~71 m) (Fig. S13.1). We used these results to extend the comparison to the entire pyramid. All runs share identical parameters, enabling like-for-like comparison and FEA export.

Using identical parameters, the framework reproduces and contrasts four major ramp hypotheses (Table 5). The straight and spiral models are distance- and work-dominated; the Houdin dual-phase system reduces early earthwork but remains single-lane and slow; the IER integrates geometry and logistics into a reproducible design. The data files in the Zenodo record contain the full per-course metrics for replication or parameter sensitivity.

This rubric compares published ramp theories without new simulations. For each model we assign a short qualitative label and a one-line, source-referenced note. Any numeric entry is marked "as reported" and scope-tagged (on-site vs. integrated). When a comparable figure is unavailable, we write ns. We do not rank models.

Criteria:

- Material efficiency (ME) — auxiliary volume fraction (ramp + platforms). ME = $1 - V_{aux}/V_{pyr}$: Qualitatively gauges how much temporary work (ramps, fills, platforms) the method demands beyond the pyramid core. We compute the Embankment volume in our framework for the straight ramp.
- Archaeological footprint (external remains expected): Assesses the extent and likelihood of persistent, visible external remains the method would leave.
- Survey visibility of edges/corners during build: Rates the ability to keep edges and corners observable for sighting, alignment, and casing control throughout construction.
- Corner-turn feasibility & safety (turning stations/platforms): Judges the practicality and safety of block rotation at corners given platforms, clearances, rigging, and procedures.

- Ramp width & clearances: Reviews operational lane width and safety clearances for sleds, teams, anchorage, passing, and guard features.
- Structural risk/novelty required: Appraises the degree of structural hazard or reliance on untested/novel temporary works to make the method function.
- Construction time potential (median on-site vs 20–27 years): Evaluates whether the on-site assembly implied by the method plausibly fits within a 20–27-year overall horizon.
- Throughput/headway potential (parallel lanes, switchbacks, two-way conflicts): Looks at the capacity to sustain stable minute-scale dispatch intervals via parallel lanes and conflict-free routing.
- Granite megalith handling (space, slopes, staging): Checks that space, gradients, and intermediate staging are adequate to haul and place large granite elements.
- Muography/void evidence consistency (ScanPyramids):Considers geometric compatibility between the model's predicted voids/fills and ScanPyramids findings, noting non-diagnostic limits.

This comparison summarizes published evaluations for other ramp theories, while the IER values are derived from our computational model. The heterogeneity of sources and methods prevents a direct quantitative comparison, so the table should be interpreted as a contextual qualitative assessment highlighting how the IER comprehensively addresses multiple constraints.

## Table S13.2. Comparison of pyramid ramp construction theories by quantitative/qualitative values explained

| Feature | Integrated Edge Ramp (this work) | Frontal Ramp | External Spiral Ramp | Internal Spiral Ramp | Step-core construction with short ramps | Integrated Ramp |
|---|---|---|---|---|---|---|
| Description | Helical channel along edges formed by temporarily omitted blocks. (Fig. 3) | Straight external ramp ascending one face from the ground (Fig. S1.1a) | Exterior ramp spiraling around the pyramid perimeter to ascend (Fig. S1.1c) | Helical haul tunnel inside the core volume, turning at corners.(Fig. S1.1d) | Stepped-core build using short tangential ramps between benches (Fig. S1.1f) | Embedded external helical ramp rising from opposite corners (Fig. S1.1e) |
| Material efficiency (auxiliary volume) | Very low external works; Skipped blocks: ≈0.5% (single-ramp) → ≈1.8% (four ramps). Filled: 0.063% ramp zero-footprint | Very high temporary works; ~20–40% of pyramid volume. | Continuous wrapping fill/berms around the monument (ns). | Low external mass —internal excavation/roofing and low straight ramp for megaliths self-consumed | Short ramps repeatedly built/removed on terraces. | Embedded/inner ramp concept; external mass low, internal fabric added. |
| Archaeological footprint | | Foundations, berms, scarps would persist at high courses. | Helical scar/berm traces expected. | Internal traces/voids would dominate. | Short ramps in rubble/adobe dismantled or re-used. | Zero-footprint. Embedded features mainly internal. |
| Survey visibility of edges/corners | Edges/corners remain observable | Ramp occludes the working face. | Continuous face occlusion along the helix. | Faces open, but internal haul path is not visible. | Edges visible | Depends on how/where the integrated ramp is exposed. |
| Corner-turn feasibility & safety | Feasible on open-air corner platforms; short, repeatable maneuvers with simple rigging. | Few turns on the main run. | Frequent turns on exposed ledges | Turning "rooms" required special tools. | Short, staged transfers (as reported) [14]. | Turning difficult, not expected corner platforms" → Turning at embedded stations; access/ventilation constraints (as reported) [14]. |
| Ramp width & clearances | Adequate lane + safety margins along arrises. | Very wide lane; heavy retaining works. | Narrow ledges; limited passing/refuge. | Constrained by tunnel cross-section. | Adequate locally; width resets each terrace. | Constrained unless large integrated sections are engineered. |
| Structural risk/novelty required | Open-air, back-filled lanes. | Tall embankments; stability risks. | Continuous external ledge stability. | Long self-supporting roofed corridors. | Many small, low-rise works; localized risk | Open-air, embedded ramps require stable roofing/support. |
| Construction time | on-site (this work; Monte Carlo, N=10 000): median 13.79 years; conservative median 20.59 years. | 40-50 years (this work; baseline) | ≥45–55 years (this work; baseline) | 35–45 years (this work; baseline) | on-site (as reported): 20–22 years [14] | on-site (as reported) [14]: ≈ 54 years for Khufu under a single upward path at 5-min takt. |
| Throughput/ headway | Parallel lanes; ~4–6 min per active ramp. | Single lane. Headway >3.7 min. (this work; baseline) | Single helical lane with recurrent crossing/merging. Headway >3.7 min. (this work; baseline) | One-way internal corridors; station bottlenecks at turns. >3.7 min. (this work; baseline).Turn times are not evaluated. | Multiple short lanes. 5 minutes headway (as reported) [14] | Internal lanes possible but stationing/egress constrain flow. ~5 min per stone per transport path (assumed; as reported) |
| Granite megalith | Viable with gentle grades and staged platforms inside the footprint (lever-assisted moves). | Viable but demands very wide, gently-graded approach and heavy temporary works. | Constrained by narrow helical ledges and curvature. | Viable with frontal straight and Grand Gallery ramps. | Viable on short gentle ramps. | Constrained because embedded ramps must provide generous sections. |
| Muography/void evidence consistency | Geometrically consistent (non-diagnostic): the predicted edge bands pass near the NFC, C1-2 and N1-3. | Not expected to create internal voids. | Not expected to create internal voids | Model claims to explain the Big Void; current muography does not show a continuous internal spiral (non-diagnostic). | Consistent with absence of large internal voids. | Non-diagnostic: embedded paths might not yield clear muon signatures. |

Note: Numeric entries are as reported and scope-tagged (on-site vs integrated). We do not recompute rival models or harmonize missing parameters; ns = not specified in a scope-matched form.